\documentclass[longauth]{aa}  

\usepackage{graphicx}
\usepackage{txfonts}

\usepackage{natbib}
\bibpunct{(}{)}{;}{a}{}{,}

\defcitealias{Lusso_2010}{L10}
\defcitealias{Stern_2015}{S15}
\defcitealias{Duras_2020}{D20}

\usepackage{booktabs}
\usepackage{rotating}
\usepackage{subfig}
\usepackage{caption}
\usepackage{float}
\usepackage{array}
\usepackage{tabularx}
\usepackage{longtable}
\usepackage{pdflscape}
\usepackage{multirow}
\usepackage{colortbl}
\usepackage{anyfontsize}

\usepackage{soul}

\usepackage{amsmath}
\usepackage{amssymb}

\usepackage{orcidlink}

\usepackage{hyperref}
\hypersetup{
    colorlinks=true,
    citecolor=blue,
    linkcolor=black,      
    urlcolor=black}

\begin{document}

   \title{The WISSH quasar project }
   \subtitle{XII. X-ray view of the most luminous quasi-stellar objects at Cosmic Noon}

   \author{C. Degli Agosti\inst{1}\fnmsep\inst{2}\fnmsep\thanks{E-mail: cdagosti@mpifr-bonn.mpg.de \\ Member of the International Max Planck Research School (IMPRS) for Astronomy and Astrophysics at the Universities of Bonn and Cologne.}\orcidlink{0009-0003-6383-4950}
          \and
          C. Vignali\inst{2}\fnmsep\inst{3}\orcidlink{0000-0002-8853-9611}
          \and
          E. Piconcelli\inst{4}\orcidlink{0000-0001-9095-2782}
          \and
          L. Zappacosta\inst{4}\orcidlink{0000-0002-4205-6884}
          \and
          E. Bertola\inst{5}\orcidlink{0000-0001-5487-2830}
          \and
          R. Middei\inst{4}\fnmsep\inst{6}\orcidlink{0000-0001-9815-9092}
          \and
          I. Saccheo\inst{4}\fnmsep\inst{7}\orcidlink{0000-0003-1174-6978}
          \and
          G. Vietri\inst{8}\orcidlink{0000-0001-9155-8875}
          \and
          F. Vito\inst{3}\orcidlink{0000-0003-0680-9305}
          \and
          A. Bongiorno\inst{4}\orcidlink{0000-0002-0101-6624}
          \and
          M. Bischetti\inst{9}\fnmsep\inst{10}\orcidlink{0000-0002-4314-021X}
          \and
          G. Bruni\inst{11}\orcidlink{0000-0002-5182-6289}
          \and
          S. Carniani\inst{12}\orcidlink{0000-0002-6719-380X}
          \and
          G. Cresci\inst{5}\orcidlink{0000-0002-5281-1417}
          \and
          C. Feruglio\inst{10}\orcidlink{0000-0002-4227-6035}
          \and
          F. Salvestrini\inst{10}\fnmsep\inst{13}\orcidlink{0000-0003-4751-7421}
          \and
          A. Travascio\inst{14}\orcidlink{0000-0002-8863-888X}
          \and
          M. Gaspari\inst{15}\orcidlink{0000-0003-2754-9258}
          \and
          E. Glikman\inst{16}\orcidlink{0000-0003-0489-3750}
          \and
          E. Kammoun\inst{17}\orcidlink{0000-0002-0273-218X}
          \and
          G. Lanzuisi\inst{3}\orcidlink{0000-0001-9094-0984}
          \and
          M.\,Laurenti\inst{4}\fnmsep\inst{18}\fnmsep\inst{19}\orcidlink{0000-0001-5762-6360}
          \and
          G. Miniutti\inst{20}\orcidlink{0000-0003-0707-4531}
          \and
          C. Pinto\inst{21}\orcidlink{0000-0003-2532-7379}
          \and
          V. Testa\inst{4}\orcidlink{0000-0003-1033-1340}
          \and
          F. Tombesi\inst{4}\fnmsep\inst{13}\fnmsep\inst{19}\orcidlink{0000-0002-6562-8654}
          \and 
          A. Tortosa\inst{4}\orcidlink{0000-0003-3450-6483}
          \and 
          F. Fiore\inst{10}\fnmsep\inst{14}\orcidlink{0000-0002-4031-4157}
          }

   \institute{Max-Planck-Institut für Radioastronomie, Auf dem Hügel 69, 53121 Bonn, Germany\\
        \and
            Dipartimento di Fisica e Astronomia “Augusto Righi”, Università degli Studi di Bologna, Via P. Gobetti 93/2, 40129 Bologna, Italy\\
        \and
            INAF -- Osservatorio di Astrofisica e Scienza dello Spazio, Via P. Gobetti 93/3, 40129 Bologna, Italy\\
        \and
            INAF -- Osservatorio Astronomico di Roma, Via Frascati 33, 00040 Monte Porzio Catone, Italy\\
        \and
            INAF--OAA, Osservatorio Astrofisico di Arcetri, largo E. Fermi 5, 50127, Firenze, Italy\\
        \and
            Space Science Data Center, Agenzia Spaziale Italiana, Via del Politecnico snc, 00133 Roma, Italy\\
        \and
            Dipartimento di Matematica e Fisica, Università Roma Tre, Via della Vasca Navale 84, 00146 Roma, Italy\\
        \and
            INAF –- Istituto di Astrofisica Spaziale e Fisica Cosmica Milano, Via A. Corti 12, 20133 Milano, Italy\\
        \and
            Dipartimento di Fisica, Università di Trieste, Via Alfonso Valerio 2, 34127 Trieste, Italy\\
        \and
            INAF--OAT, Osservatorio Astronomico di Trieste, Via Tiepolo 11, 34131 Trieste, Italy\\
        \and
            INAF –- Istituto di Astrofisica e Planetologia Spaziali, Via Fosso del Cavaliere 100, 00133 Roma, Italy\\
        \and
            Scuola Normale Superiore, Piazza dei Cavalieri 7, 56126 Pisa, Italy\\
        \and
            IFPU -- Institute for Fundamental Physics of the Universe, via Beirut 2, 34151 Trieste, Italy\\
        \and
            Dipartimento di Fisica “G. Occhialini”, Università degli Studi di Milano-Bicocca, Piazza della Scienza 3, 20126 Milano, Italy\\
        \and
            Department of Physics, Informatics and Mathematics, University of Modena and Reggio Emilia, 41125 Modena, Italy\\
        \and
            Department of Physics, Middlebury College, Middlebury, VT 05753, USA\\
        \and
            Cahill Center for Astronomy and Astrophysics, California Institute of Technology, 1200 California Boulevard, Pasadena, CA 91125, USA\\
        \and
            Physics Department, Tor Vergata University of Rome, Via della Ricerca Scientifica 1, 00133 Rome, Italy\\
        \and
            INFN - Rome Tor Vergata, Via della Ricerca Scientifica 1, 00133 Rome, Italy\\
        \and 
            Centro de Astrobiolog\'ia (CAB), CSIC-INTA, Camino Bajo del Castillo s/n, 28692 Villanueva de la Ca\~nada, Madrid, Spain\\  
        \and
            Istituto Nazionale di Astrofisica INAF IASF Palermo, Via Ugo La Malfa 153, Palermo, 90146, Italy\\
             }

   \newpage
 
   \abstract{}{To improve our knowledge of the nuclear emission of luminous quasi-stellar objects (QSOs) at Cosmic Noon, we studied the X-ray emission of the WISE/SDSS-selected hyper-luminous (WISSH) QSO sample. It consists of 85 broad-line active galactic nuclei (AGN) with bolometric luminosities $\rm L_{bol} > few\times 10^{47}\,\,erg\,s^{-1}$ at $z \rm \approx 2-4$. Our goal is to characterise their X-ray spectral properties and investigate the relation between the X-ray luminosity and the energy output in other bands. To this end, we compared the nuclear properties of powerful QSOs with those derived for the majority of the AGN population.}{We were able to perform X-ray spectral analysis for about one-half of the sample. For 16 sources, we applied the hardness ratio analysis, while for the remaining sources we estimated their $\rm 2 - 10\,\,keV$ intrinsic luminosity $\rm L_{2-10}$; only 8 sources were not detected.}{We report a large dispersion in $\rm L_{2-10}$ despite the narrow distribution in $\rm L_{bol}$, $\rm 2500\,\,\AA$ intrinsic luminosity $\rm L_{2500\,\AA}$, and $\rm 6\,\,\mu m$ intrinsic luminosity $\rm \lambda L_{6\,\mu m}$ of WISSH QSOs (approximately one-third of the sources classified as X-ray-weak QSOs). This suggests that the properties of the X-ray corona and inner accretion flow in hyper-luminous QSOs can be significantly different from those of typical less powerful AGN. The distribution of the X-ray spectral index does not differ from that of AGN at lower redshift and lower $\rm L_{bol}$, and does not depend on the Eddington ratio ($\rm \lambda_{Edd}$) and X-ray weakness. The majority of WISSH QSOs, for which it was possible to estimate the presence of intrinsic absorption ($\rm \approx 65\%$ of the sample), exhibit little to no obscuration (i.e. column density $\rm N_H \le 5 \times 10^{22}\,\,cm^{-2}$). Among the X-ray obscured sources, we find some blue QSOs without broad absorption lines (BALs) that fall within the `forbidden region' of the $\rm Log(N_H) - Log(\lambda_{Edd})$ plane, which is typically occupied by dust-reddened QSOs and is associated with intense feedback processes. Additionally, we confirm a significant correlation between $\rm L_{2-10}$ and velocity shift of the {\sc Civ} emission line, a tracer of nuclear ionised outflows.}{Multi-wavelength observations of the broad-line WISSH quasars at Cosmic Noon and, in particular, their complete X-ray coverage, allow us to properly investigate the accretion disk$-$corona interplay to the highest luminosity regime. The distribution of bolometric corrections $\rm k_{bol}$ and X-ray$-$to$-$optical indices $\rm \alpha_{OX}$ of the WISSH quasars is strikingly broad, suggesting that caution should be exercised when using $\rm L_{bol}$, $\rm L_{2500\,\AA}$, and $\rm \lambda L_{6\,\mu m}$ to estimate the  X-ray emission of individual luminous QSOs.}

   \keywords{galaxies: active -- 
                galaxies: high-redshift --
                QSOs: general --
                QSOs: supermassive black holes --
                X-rays: galaxies
            }

   \maketitle

\section{Introduction}\label{sec:intro}
The X-ray emission from an active galactic nucleus (AGN) carries valuable information on the physical properties of the material distributed over sub-parsec scales around the central supermassive black hole \citep[SMBH; e.g.][]{Turner_2009}, due to inverse Compton scattering of UV accretion disk photons with electrons in the corona. The X-ray spectrum of broad-line Type 1 AGN in the $\rm \approx 0.3 - 10\,\,keV$ band can be described by a power law with a slope of $\rm \Gamma \approx 1.7 - 2$. It typically exhibits absorption features due to the presence of ionised outflowing material along the line of sight to the nucleus, with a broad distribution in velocity, distance from the SMBH, ionisation state, and column density \citep[e.g.][and references therein]{Krongold_2003, McKernan_2007, Yamada_2024}. The vast majority of AGN show an extra-continuum emission below 2\,\,keV, called the soft excess, likely related to warm Comptonisation or relativistically blurred reflection from the inner accretion disk \citep[e.g.][]{Miniutti_2004, Kubota_2018}. Moreover, emission features, such as Fe K emission lines and the broad Compton hump component, which result from fluorescence and reflection off the accretion disk and other sub-parsec material, are also commonly detected above 6\,\,keV \citep[e.g.][]{Matt_1991, Patrick_2012}. This picture has basically emerged from  an extensive study of local, X-ray-bright, low- to moderate-luminosity AGN with a bolometric luminosity $\rm L_{bol} < 10^{46}\,\,erg\,s^{-1}$ \citep[e.g.][]{Reynolds_1995, Ricci_2017}. 

Our knowledge of the X-ray properties of luminous $\rm (L_{bol} \geq 10^{47}\,\,erg\,s^{-1})$, highly accreting quasi-stellar objects (QSOs) shining at Cosmic Noon (i.e. $\rm \textit{z}\approx 2-4$) is less accurate due to the lack of high-quality X-ray spectra for these distant sources. Nonetheless, during the past two decades \textit{XMM-Newton} and \textit{Chandra} observations have provided a large amount of high-$z$ QSOs detected in X-rays. The combination of X-ray data from local AGN and distant QSOs highlights several trends:
(i) the intensity of the Fe K line and Compton hump in luminous AGN is weaker than in low-luminosity sources; (ii) the relative contribution of the luminosity in the $\rm 2 - 10\,\,keV$ band $\rm (L_{2-10})$ to $\rm L_{bol}$ (which approximately corresponds to the UV luminosity in these luminous AGN) progressively decreases as a function of $\rm L_{bol}$; (iii) the ratio of the mid-infrared (MIR) luminosity (typically measured at $\rm 6\,\,\mu m$) to X-ray luminosity diminishes for increasing $\rm L_{bol}$; (iv) a correlation exists between the strength and blueshift of the {\sc Civ} emission line at 1550\,\,\AA, and the X-ray$-$to$-$optical index, defined as $\rm \alpha_{OX} = 0.3838 \times Log(L_{2\,keV}/L_{2500\,\AA})$ \citep{Tananbaum_1979}, where $\rm L_{2\,keV}$ and $\rm L_{2500\,\AA}$ are the monochromatic luminosities at 2\,\,keV and 2500\,\,\AA, respectively; (v) a sizeable fraction $\rm (\approx 30\%)$ of luminous QSOs exhibit intrinsic X-ray weakness, i.e. the difference between the measured $\rm \alpha_{OX}$ and that expected from the $\rm \alpha_{OX} - Log(L_{2500\,\AA})$ relation is $\rm \Delta(\alpha_{OX}) \lesssim - (0.2 - 0.3)$ \citep[e.g.][]{Vignali_2003, Bianchi_2007, Just_2007, Stern_2015, Zappacosta_2018, Nardini_2019, Duras_2020, Timlin_2020, Zappacosta_2020}.
These findings highlight the importance of strengthening our understanding of the X-ray properties of luminous high-$z$  QSOs, as well as their complex relationships with other multi-wavelength nuclear parameters. This holds significant potential for uncovering largely unexplored aspects of the accreting SMBH phenomenon, which have remained elusive due to the challenges in obtaining reliable systematic X-ray and multi-band data for such distant AGN populations.

It is widely recognised that luminous QSOs (i.e. highly accreting, billion-solar-mass SMBHs) play a crucial role in the evolution of massive galaxies during Cosmic Noon. In these sources, strong AGN-driven feedback is expected to significantly influence both SMBH growth and star formation activity within the host galaxy \citep[e.g.][]{Silk_1998, Fiore_2017, Choi_2018, Gaspari_2020, Byrne_2024}. Motivated by these results, we have embarked on a comprehensive multi-band investigation on a large sample of broad-line WISE/SDSS-selected hyper-luminous (WISSH) QSOs at $\rm \textit{z} \approx 2-4$ \citep{Saccheo_2023}. Our goal is to shed light on their nuclear properties, the presence and power of multi-scale multi-phase outflows generated by their huge bolometric radiative output (i.e. $\rm L_{bol} \approx 10^{47-48}\,\,erg\,s^{-1}$), and their host galaxies. The study of these QSOs has revealed the common presence of ionised gas winds, typically with very high velocities relative to the bulk of AGN, at various distances from the central SMBH. Their host galaxies appear to be in a growth phase, with star formation rates of $\rm SFR \approx 10^{2-3}\,\,M_\odot /yr$, disturbed kinematics, and that preferentially reside in high-density environments surrounded by companions. The molecular gas content of these host galaxies is typically lower than that of main-sequence galaxies with the same IR luminosity. This  combination of a small cold gas reservoir and high SFR suggests that host galaxies of WISSH QSOs may be the progenitors of giant quiescent elliptical galaxies \citep[e.g,][]{Bischetti_2017, Bischetti_2018, Bischetti_2021, Duras_2017, Vietri_2018, Bruni_2019, Vietri_2022}. 

In this paper we extend the initial results on the X-ray properties of a sub-sample of WISSH QSOs reported in \citet{Martocchia_2017} and \citet{Zappacosta_2020}, by presenting the analysis of the X-ray observations that are available for all  85 sources in the sample. The paper is organised as follows. Section \ref{sec:obs_and_dataReduction} describes the X-ray observations used in our investigation and the data reduction. Section \ref{sec:Xray_analysis} presents the X-ray data analysis techniques used depending on the number of collected X-ray photons. The main results are presented in Section \ref{sec:results}. In Section \ref{sec:discussion} we discuss the fraction of intrinsically X-ray-weak sources among QSOs, the relation between some X-ray spectral properties and the Eddington ratio (i.e. the ratio of $\rm L_{bol}$ to the Eddington luminosity $\rm \lambda_{Edd} = L_{bol}/L_{Edd}$), and the comparison with the X-ray properties of QSOs at $\rm \textit{z} > 6$. Finally, the summary and our conclusions are given in Section \ref{sec:conclusions}. Throughout this work we adopt $\rm H_0 = 70$, $\rm \Omega_m = 0.27$, and $\rm \Omega_\Lambda = 0.73$ \citep{Bennett_2013}.

\section{Observations and data reduction}\label{sec:obs_and_dataReduction}

\subsection{Data presentation}
\begin{figure*}
    \centering
    \includegraphics[width=17cm]{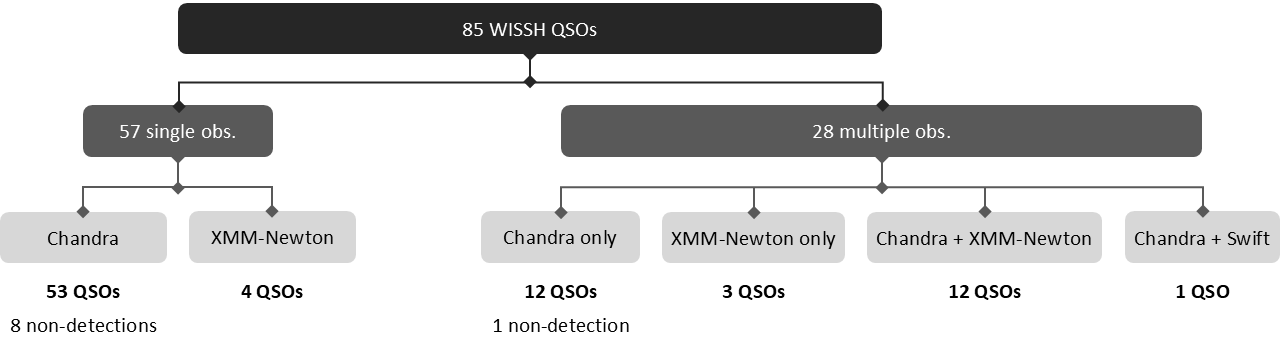}
    \caption{X-ray coverage of the WISSH sample.}
    \label{data_scheme}
\end{figure*}

All 85 WISSH sources have been observed by \textit{Chandra} and/or \textit{XMM-Newton}. Specifically, 53 and 4 objects have single \textit{Chandra} and \textit{XMM-Newton} observations, respectively, while 28 have been targeted multiple times (see Figure \ref{data_scheme}). The complete dataset is presented in Table \ref{tab:logObs}. In Sections \ref{sec:Xray_analysis} and \ref{sec:results}, we consider for each source only the observation with the highest number of X-ray photons.

We present new proprietary \textit{Chandra} observations for 44 sources (PI: E. Piconcelli; Prop. ID: 23700190; Cycle 23), which complete the WISSH sample X-ray coverage. We also made use of two proprietary \textit{XMM-Newton} observations of WISSH59 (PI: C. Pinto), in which the QSO is not the target. \textit{Chandra} and \textit{XMM-Newton} archival observations for the remaining 41 sources \citep{Martocchia_2017} were re-analysed to provide a homogeneous analysis of the entire WISSH sample. Finally, for WISSH57, showing peculiar properties in the \textit{Chandra} spectrum, a \textit{Swift}-XRT follow-up observation is also presented.
 
\subsection{\textit{Chandra} observations}\label{subsec:Chandra_reduction}
We downloaded the data from the \textit{Chandra} Data Archive\footnote{\url{https://cda.harvard.edu/chaser/}} and reprocessed them through the \texttt{chandra\_repro} task using {\sc CIAO}\footnote{\url{https://cxc.cfa.harvard.edu/ciao/}} {\sc 4.14} package with {\sc CALDB 4.9.8}. 
Applying the \texttt{dmcopy} tool, we filtered the data and produced an image in the $\rm 0.3 - 7\,\,keV$ energy range. We inspected the image with {\sc SAOImageDS9}\footnote{\url{https://sites.google.com/cfa.harvard.edu/saoimageds9}} to define the extraction regions, selecting circular regions of radii $\rm \approx 2$ arcsec and $\rm \approx 11 - 42$ arcsec for the source and background, respectively. The source extraction region was defined to correspond to $\rm \approx 90\%$ of the encircled energy fraction, while the background one was selected to be free of sources and close enough to the target to be representative of the background in the source extraction region. Once the source and background regions were identified, spectral extraction was performed with {\sc CIAO} task \texttt{specextract}, which also produces the Auxiliary Response File (ARF) and the Redistribution Matrix File (RMF). 

\subsection{\textit{XMM-Newton} observations}\label{subsec:XMM_reduction}
EPIC pn Observation Data Files (ODFs) were downloaded from \textit{XMM-Newton} Science Archive\footnote{\url{https://www.cosmos.esa.int/web/xmm-newton/xsa}} (XSA). We reprocessed them using version {\sc 20.0.0} of the {\sc \textit{XMM-Newton} SAS} \citep{Gabriel_2004} and produced the event files through the \texttt{epproc} task. Extracting the light curve in the $\rm 10-12\,\,keV$ energy range, we checked for possible presence of flaring background, typically attributed to soft protons. Accordingly, the event files were filtered with either \texttt{espfilt} or \texttt{gtigen} tool, producing Good Time Interval (GTI) tables, and light curves were re-extracted. Finally, we created cleaned event files in the energy range $\rm 0.3 - 10\,\,keV$. A circular region of radius $\rm \approx 10-15$ arcsec (corresponding to $\rm \approx 70\%$ of the encircled energy fraction) was used for the source. This choice limits the background contribution and is well suited for relatively faint X-ray sources. A circular region of radius $\rm \approx 15 - 55$ arcsec was used for the background, instead. Spectra were extracted through the {\sc \textit{XMM-Newton} SAS} meta-task \texttt{especget}, which runs also the \texttt{arfgen}, \texttt{rmfgen} and \texttt{backscale} tasks. Thus, ARF and RMF matrices are calculated, and the source and background extraction region are re-scaled to obtain the spectra. 

\subsection{\textit{Swift}-XRT observation}\label{subsec:Swift_reduction}
The {\it Swift}-XRT observations of WISSH57 have been performed in photon counting (PC) readout mode. Data were first reprocessed thanks to the on-the-fly facilities developed by the ASI-SSDC and included in the {\sc NASA-HEASARC HEASoft} package\footnote{\url{https://heasarc.gsfc.nasa.gov/docs/software/heasoft/}} (version \texttt{v6.31.1}). The data processing relied on the {\sc XRTDAS} software package and the standard calibration, filtering processing steps were taken, and the calibration files available {\sc CALDB} (version {\sc 20220803}) used. For each of the three exposure available for WISSH57, the source spectrum was extracted using a circle of 20-pixel ($\rm \approx 47$ arcsec) radius centred on the target. The background was computed adopting a circular region with a radius of 40 pixels centred on a blank area. The resulting spectra were stacked.

\section{Data analysis}\label{sec:Xray_analysis}
Depending on the number of detected X-ray photons, the sources were divided into three groups as follows:
\begin{itemize}
    \item 39 QSOs with $\rm \ge 20$ net (i.e. background-subtracted) counts;
    \item 16 QSOs with $\rm 5 < \text{net counts} < 20$;
    \item 30 QSOs with $\rm \le 5$ net counts (8 of them are undetected).
\end{itemize}
Their X-ray properties are presented in the following Sections. For all parameters, errors and upper limits are reported at 1$\rm \sigma$ and $\rm 90\%$ confidence level, respectively.

\subsection{Analysis of the HC-WISSH sub-sample}
In this paper we refer to $\rm > 5$ counts sources as the High-Counts WISSH (HC-WISSH) sub-sample.

\subsubsection{Analysis of QSOs with $\rm \ge 20$ counts: spectral fitting}\label{subsubsec:20_counts}
For the QSOs corresponding to $\rm \ge 20$ counts (maximum value: $\rm \approx 2500$ counts; median value: $\rm \approx 140$ counts), we performed basic/moderate-quality spectral analysis. Thus, we derived the continuum X-ray properties ($\rm \Gamma$ and the intrinsic column density $\rm N_H$), along with the observed $\rm 0.5 - 10\,\,keV$ flux $\rm (F_{0.5-10})$ and the intrinsic $\rm 2 - 10\,\,keV$ luminosity, through the X-ray spectral fitting package {\sc Xspec} \citep{Arnaud_1996}. Errors on $\rm F_{0.5 - 10}$ and $\rm L_{2-10}$ were calculated using the \texttt{cflux} and \texttt{clumin} tools in {\sc Xspec}. These are convolution models to calculate the flux of all the components included in the best fit spectral model. 

\begin{figure*}
    \centering
    \subfloat[][\label{fig:spec_eg}]{\includegraphics[height=190pt]{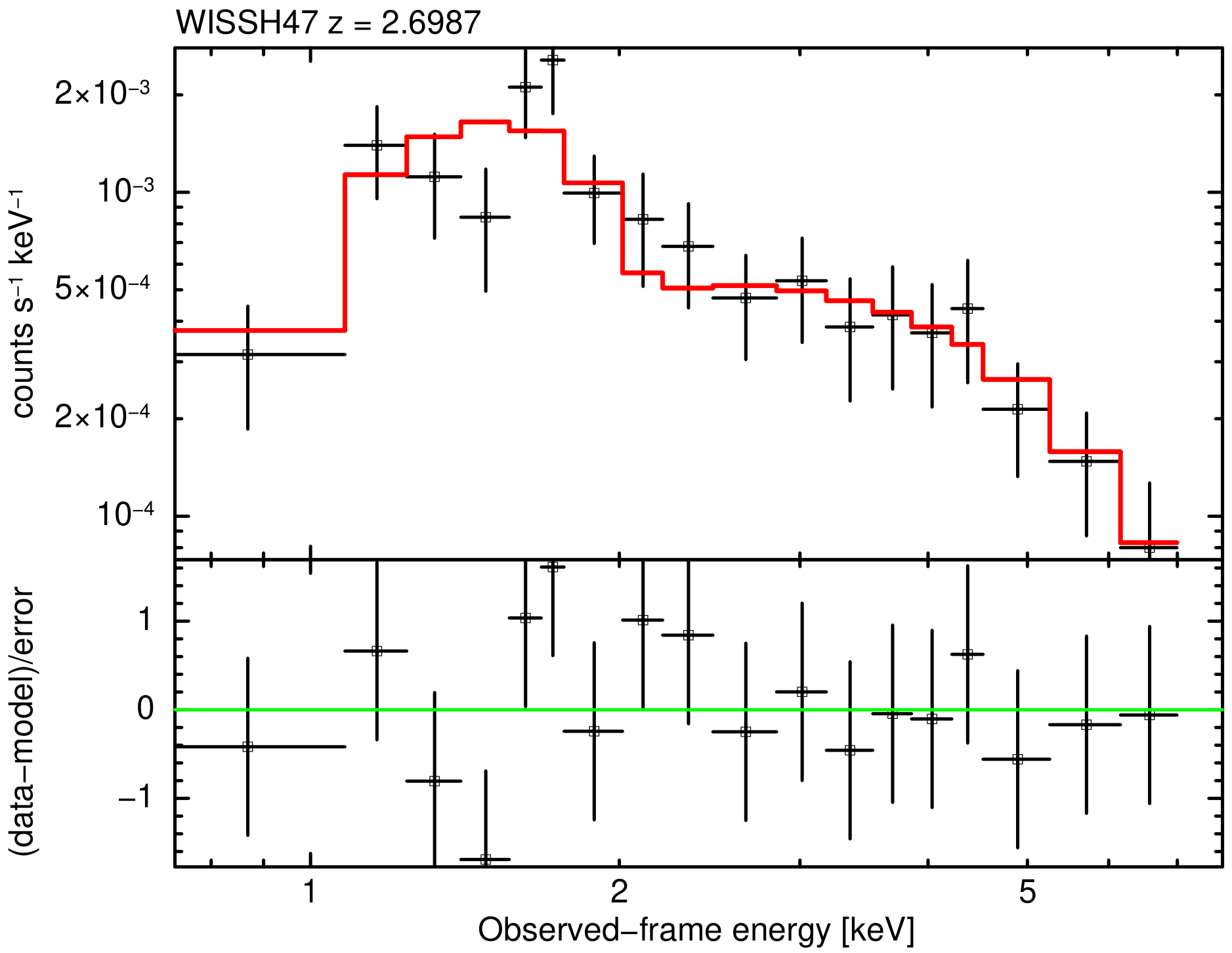}} \quad
    \subfloat[][\label{fig:contour_plot}]{\includegraphics[height=190pt]{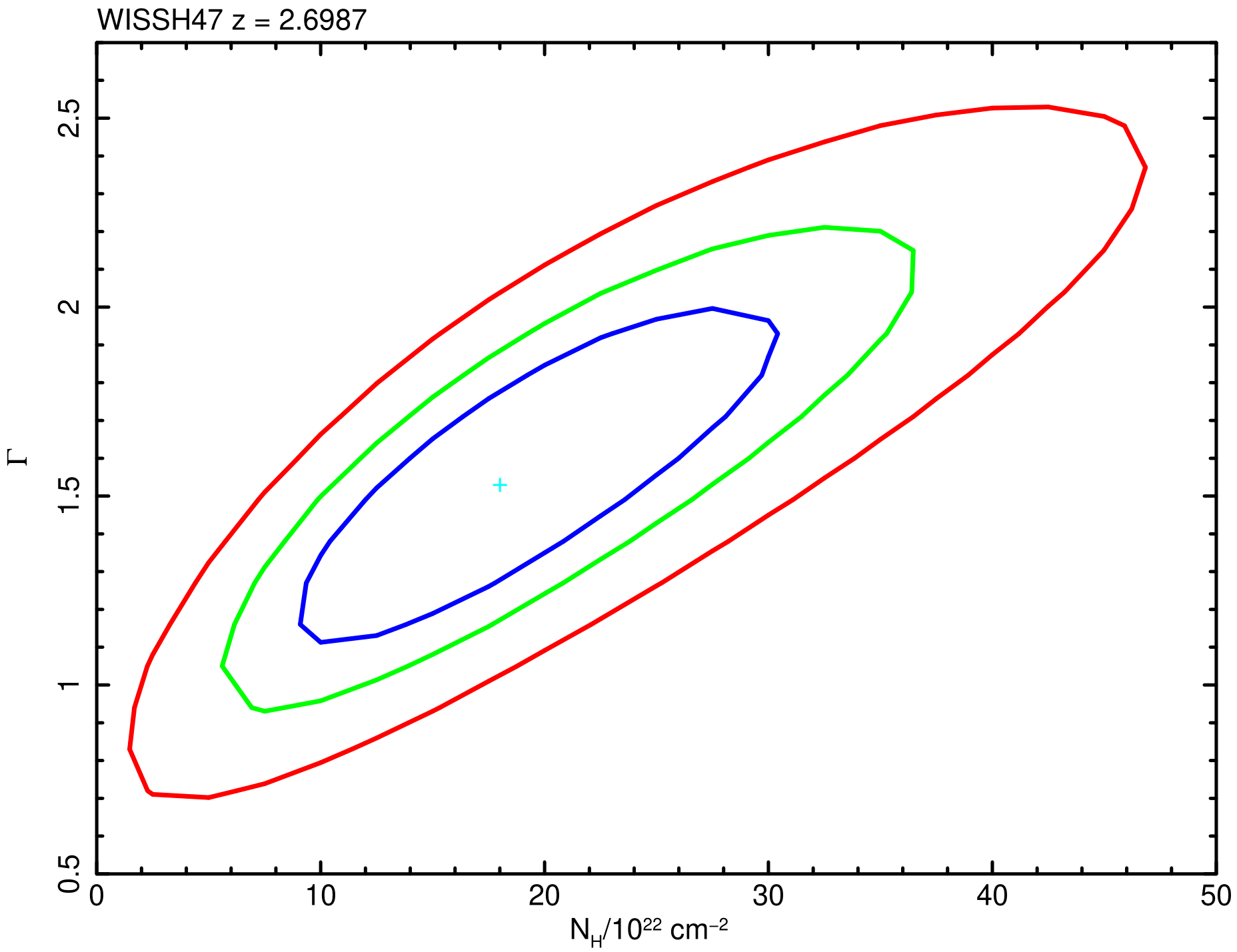}}
    \caption{(a) \textit{Chandra} spectrum (re-binned for display purposes) of WISSH47 ($z \rm = 2.6987$). We detect about 130 counts for this source and measure significant intrinsic absorption. The residuals are defined as \texttt{(data - model)} in units of $\rm \sigma$. (b) $\rm N_H - \Gamma$ contour plot for the best fit model of WISSH47. The blue, green, and red curves represent the 68\%, 90\%, and 99\% confidence levels, respectively.}
\end{figure*}

We grouped spectral points to one count/bin with the \texttt{grppha} tool of the {\sc FTOOLS} package.\footnote{\url{https://heasarc.gsfc.nasa.gov/ftools/}} Cash statistics with direct background subtraction \citep[C-stat in {\sc Xspec};][]{Cash_1979, Wachter_1979} was used for the spectral fitting. We initially adopted a power law model modified by Galactic absorption (see Table \ref{tab:20counts}), which translates into an {\sc Xspec} model of the form \texttt{phabs*po}. We also tested the presence of intrinsic absorption by adding a \texttt{zphabs} multiplicative component. For about one-third of the analysed QSOs, this addition yielded to an improvement in the fit quality with respect to the power law model by $\rm > 95\%$ confidence level using an F-test. We therefore included an absorption component in the best fit model used to calculate fluxes and luminosities reported in Table \ref{tab:20counts}. If an additional intrinsic obscuring component was not significantly required, an upper limit is reported in the $\rm N_H$ column of Table \ref{tab:20counts}. As an example, in Figure \ref{fig:spec_eg} we report the \textit{Chandra} spectrum of WISSH47. We detect $\rm \approx 130$ counts for this source in the $\rm 0.3 - 7\,\,keV$ energy range, which is representative of the median number of counts for the $\rm \ge 20$ net counts sub-sample, and find a significant $\rm N_H$ (i.e. $\rm > 95\%$ confidence). Figure \ref{fig:contour_plot} shows the $\rm N_H - \Gamma$ confidence contours using the best fit model.

\subsubsection{Analysis of QSOs with $\rm 5 < counts < 20$: hardness ratio analysis}\label{subsubsec:619_counts}
For the QSOs detected with $\rm 5 - 20$ photons, we derived the number of soft ($\rm 0.5 - 2\,\,keV$, S) and hard ($\rm 2 - 10\,\,keV$, H) source and background photons. These values, normalised by the different source and background extraction areas, were required to derive the hardness ratio (HR) 
\begin{equation}\label{eq:HR}
    \rm HR = \frac{H - S}{H + S}
\end{equation}
through the Bayesian Estimator of Hardness Ratio ({\sc BEHR}) \citep{Park_2006}. {\sc BEHR} is especially useful in Poissonian regimes, where the background subtraction is not a good solution, as correctly deals with the non-Gaussian nature of the error propagation, whether or not the source is detected in both bands. 
Then, we constrained intrinsic $\rm N_H$ comparing the measured HR with the values from simulated absorbed power law models with fixed $\rm \Gamma = 1.8$ \citep{Piconcelli_2005}, which also take into account the sources redshift. 
$\rm N_H$ was considered to be significant when its 1$\rm \sigma$ lower boundary was not consistent with zero, otherwise upper limits on $\rm N_H$ were derived. One-fourth of the objects in the sub-sample exhibit a significant $\rm N_H$.

We calculated $\rm F_{0.5-10}$ through \textit{Chandra} {\sc WebPIMMS},\footnote{\url{https://cxc.harvard.edu/toolkit/pimms.jsp}} assuming $\rm \Gamma = 1.8$ and the Galactic absorption. 
$\rm L_{2-10}$ was derived as 
\begin{equation}\label{eq:hardL}
    \rm L_{2 - 10} = 4\pi d_L^2 \,F_{2-10}^{corr} \,(1 + z)^{\Gamma - 2} \,\,erg\,s^{-1},
\end{equation}
where $\rm d_L$ is the luminosity distance,\footnote{Calculated from \url{https://astro.ucla.edu/~wright/CosmoCalc.html}.} $\rm F_{2-10}^{corr}$ (derived using {\sc WebPIMMS}) is the $\rm 2 - 10\,\,keV$ flux corrected for Galactic and intrinsic absorption, $z$ is the source redshift and $\rm \Gamma = 1.8$. Results are reported in Table \ref{tab:520counts}.

\subsection{Analysis of QSOs with $\rm \le 5$ counts and non-detections}\label{subsubsec:05_counts}
We adopted the binomial method of \citet{Weisskopf_2007} (see their Appendix A3) and derived the probability distribution function of net counts, to test whether the sources having $\rm \le 5$ counts could be considered detected or not. The nominal value of the net counts coincides with the peak of the probability distribution. As in Section \ref{subsubsec:619_counts}, we derived $\rm F_{0.5-10}$ through {\sc WebPIMMS}, assuming $\rm \Gamma = 1.8$ and the Galactic absorption. $\rm L_{2-10}$ were obtained from Equation \ref{eq:hardL}, assuming $\rm N_H = 5 \times 10^{22}\,\,cm^{-2}$, which is the median value for the 14 absorbed sources in the HC-WISSH sample;\footnote{We excluded WISSH59, which exhibits an exceptionally high intrinsic absorption.} their values with and without $\rm N_H$ change, on average, by $\rm \approx 7\%$. Results are reported in Table \ref{tab:05counts}. Only 8 sources result to be undetected (detection significance is at 99\% confidence level). 

\subsection{QSOs with multiple observations}\label{subsec:multi_obs}
For the 28 out of 85 QSOs that have multiple observations (see Figure \ref{data_scheme}), we evaluated possible changes in terms of $\rm N_H$, observed $\rm 0.5 - 2 \,\,keV$ flux $\rm (F_{0.5-2})$ and observed $\rm 2 - 10 \,\,keV$ flux $\rm (F_{2-10})$. We did not consider $\rm \Gamma$ variability, as we do not have a spectroscopic estimate of this parameter for all of the observations; moreover, even when available, it has non-negligible uncertainties. We report the most significant cases of variability, set to $\rm > 2 \sigma$ confidence level for what concerns $\rm N_H$ variability (to spot candidate changing look AGN) and to $\rm >3\sigma$ confidence level for flux variability. Results are presented in detail in Appendix \ref{apdx:variability}. In particular:
\begin{itemize}
    \item two QSOs (WISSH59 and WISSH69) show intrinsic $\rm N_H$ variability of up to an order of magnitude (as fast as $\rm \approx 40$ days rest-frame in case of WISSH59);
    \item four QSOs (WISSH13, WISSH33, WISSH35 and WISSH63) show $\rm F_{0.5-2}$ variability (as fast as $\rm \approx 9$ days rest-frame in case of WISSH35);
    \item three QSOs (WISSH70, WISSH82 and WISSH83) show both $\rm F_{0.5-2}$ and $\rm F_{2-10}$ variability (as fast as $\rm \approx 120$ days rest-frame in case of WISSH83).
\end{itemize}

We also checked for possible coincidence of WISSH objects with the \textit{eROSITA} eRASS1 Main catalogue sources \citep{Merloni_2024}. We find seven cross-identifications (i.e. WISSH14, WISSH27, WISSH29, WISSH37, WISSH49, WISSH54, WISSH60) within a 5 arcsec radius. Comparing the $\rm 0.2 - 2.3\,\,keV$ fluxes (\textit{eROSITA} sensitivity is maximum in this band), we find a $\rm > 3\sigma$ variability over $\rm \approx 6$ years rest frame for WISSH60, which has a high detection significance both in \textit{Chandra} and \textit{eROSITA} observation. Even assuming $\rm \Gamma = 1.7$ as for \textit{eROSITA}, only WISSH60 exhibits a flux variation $\rm > 3\sigma$.

Finally, as the WISSH57 \textit{Chandra} observation exhibits a particularly steep $\rm \Gamma = 2.80 \pm 0.32$, we submitted a request for a follow-up observation with \textit{Swift}-XRT. The observation yields a more standard $\rm \Gamma = 2.10_{-0.40}^{+0.35}$. Although possibly tracing a rapid ($\rm \approx 4$ months rest-frame) spectral change, the photon index variability is $\rm < 3\sigma$ due to the large uncertainties affecting both measurements.

\section{Results}\label{sec:results}

\subsection{Intrinsic absorption, X-ray luminosity and X-ray bolometric correction}\label{subsec:kBol}
\begin{figure}
    \resizebox{\hsize}{!}{\includegraphics{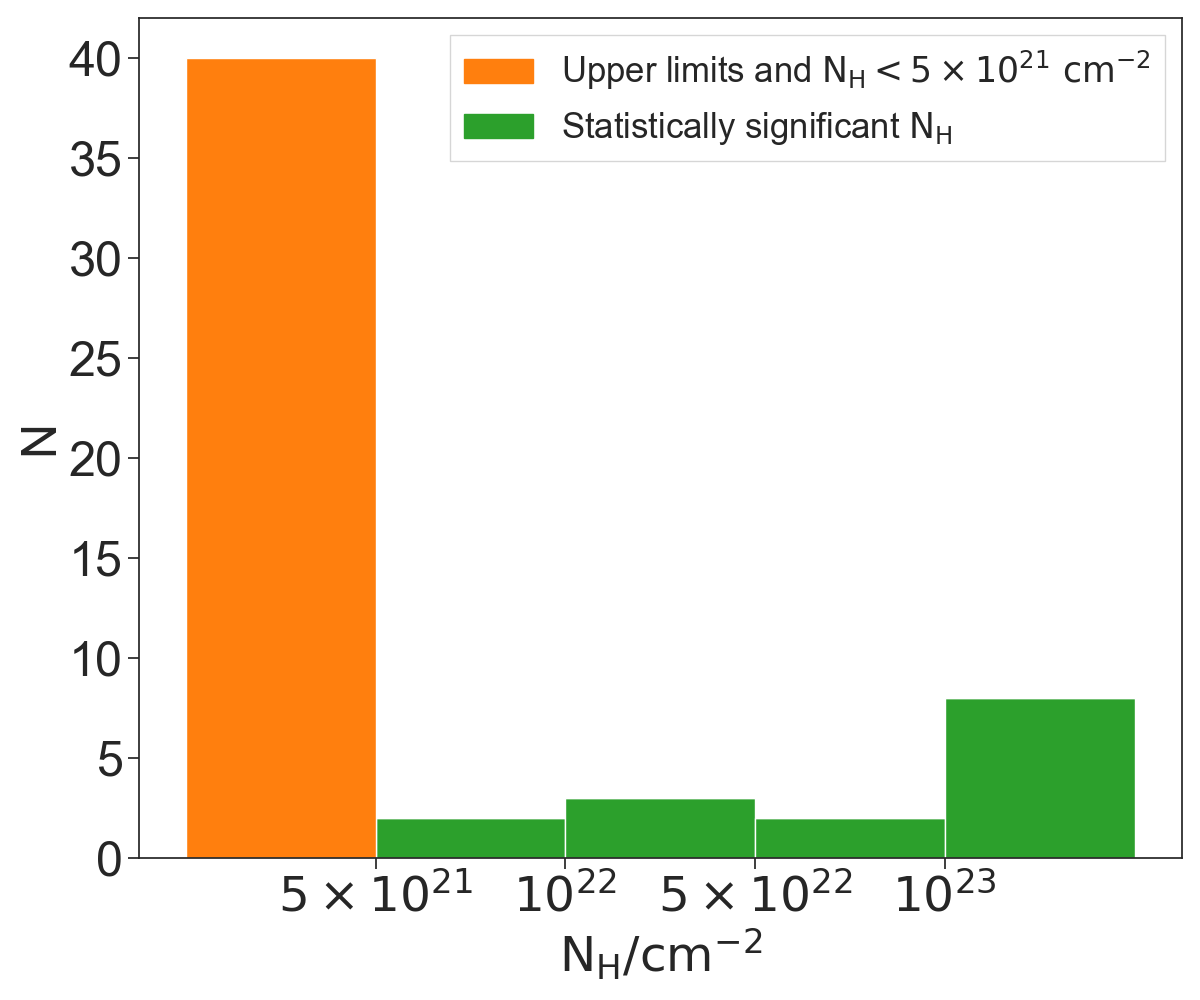}}
    \caption{Intrinsic column density distribution for the HC-WISSH sample. Sources for which the power law model represents the best fit (i.e. unobscured) are included in the bin $\rm N_H < 5 \times 10^{21}\,\,cm^{-2}$.}
    \label{fig:nh_distribution}
\end{figure}

\begin{figure}
    \resizebox{\hsize}{!}{\includegraphics{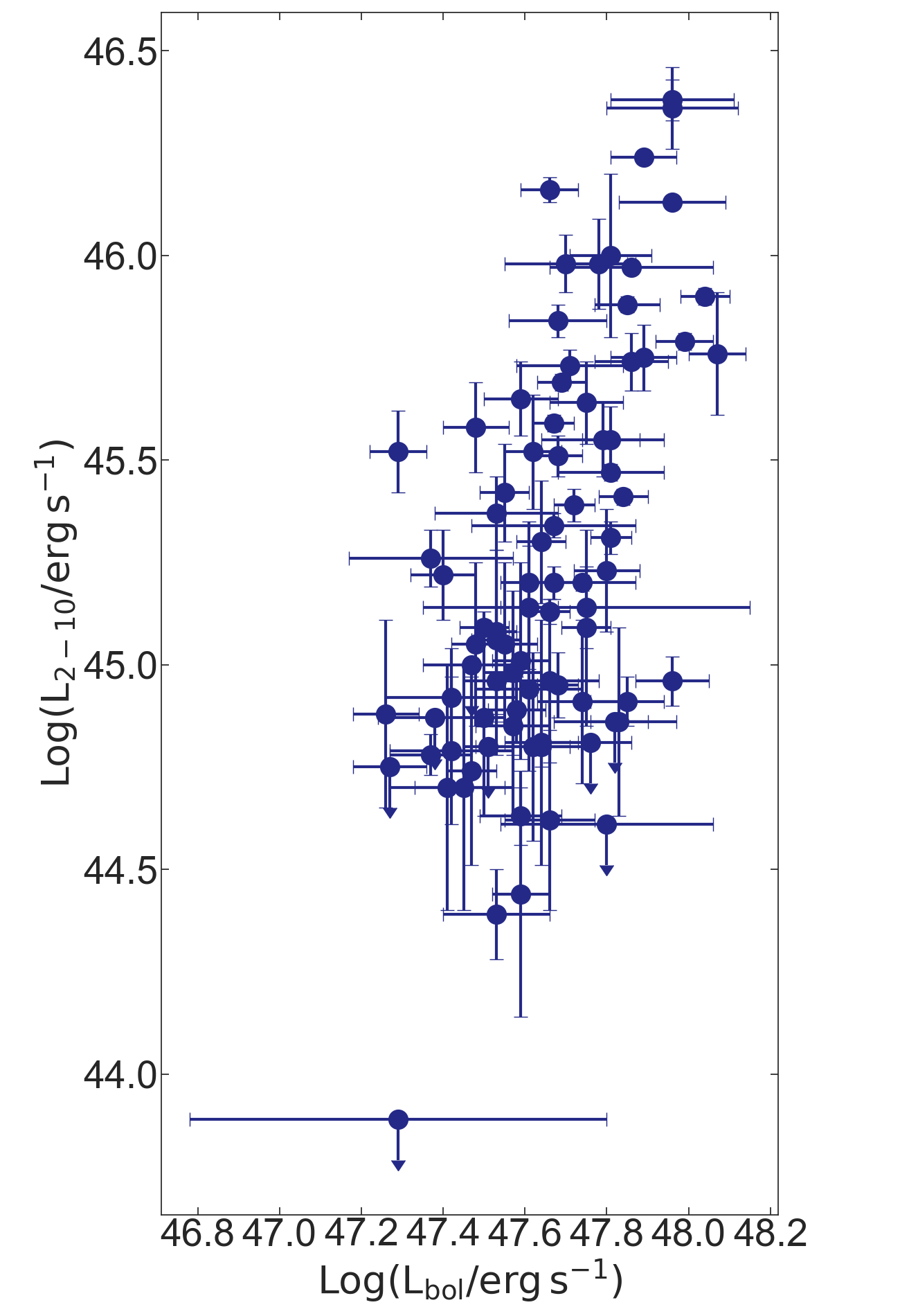}}
    \caption{$\rm Log(L_{2-10})$ as a function of $\rm L_{bol}$ for the entire WISSH sample.}
    \label{fig:Lbol_L210}
\end{figure}

 Figure \ref{fig:nh_distribution} shows the $\rm N_H$ distribution derived for the HC-WISSH sub-sample. The bulk $\rm (\approx 76\%)$ of the sources exhibits low level of intrinsic absorption $\rm (N_H < 10^{22}\,\,cm^{-2})$. We note, however, that $\rm \approx 15\%$ of the QSOs are strongly obscured $\rm (N_H \ge 10^{23}\,\,cm^{-2})$. Sources for which the power law model with no absorption represents the best fit are included in the bin $\rm N_H < 5\times10^{21}\,\,cm^{-2}$ (orange bar). For each QSO, the intrinsic $\rm L_{2-10}$ was derived by correcting for nuclear obscuration $\rm N_H$. In Figure \ref{fig:Lbol_L210}, $\rm L_{2-10}$ is shown as a function of $\rm L_{bol}$, taken from \citet{Saccheo_2023} who performed multi-band SED fitting after correcting for dust extinction for all WISSH sources. The scaled axis highlights the large $\rm L_{2-10}$ distribution compared to the narrow $\rm L_{bol}$ distribution.

\begin{figure}
    \resizebox{\hsize}{!}{\includegraphics{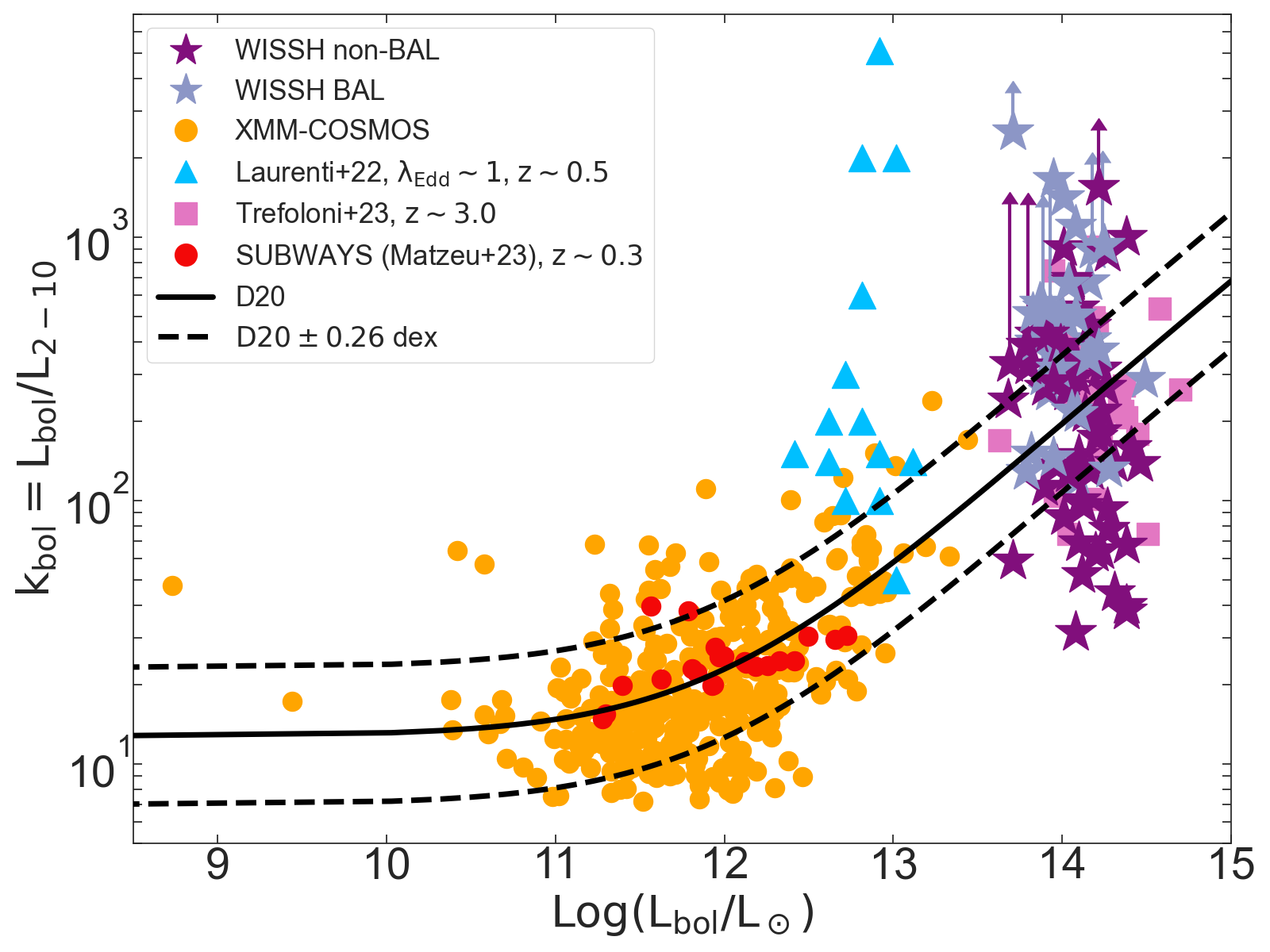}}
    \caption{Bolometric correction as a function of $\rm Log(L_{bol})$. WISSH BAL and non-BAL QSOs are shown as indigo and purple stars, respectively. The XMM-COSMOS sample of 343 Type 1 AGN presented by \citetalias{Lusso_2010} is represented by yellow dots; it covers a wide range of redshifts (0.04 < $z$ < 4.25) and X-ray luminosities $\rm (40.6 \le Log(L_{2-10}/erg\,s^{-1}) \le 45.3)$. The 22 SUBWAYS QSOs and Type 1 AGN at intermediate redshifts $\rm (0.1 \lesssim$ $z$ $\rm \lesssim 0.4)$ from \citet{Matzeu_2023} can be seen as red dots. Light blue triangles correspond to the 14 radio-quiet high-$\rm \lambda_{Edd}$ $\rm (\lambda_{Edd} \gtrsim 1)$ QSOs at $\rm 0.4 \le$ $z$ $\rm \le 0.75$ presented by \citet{Laurenti_2022}. The sample of luminous and intrinsically blue QSOs at $z$ $\rm \approx 3$ presented by \citet{Trefoloni_2023}, for which constrained $\rm L_{2-10}$, $\rm \Gamma$, $\rm \lambda_{Edd}$, and {\sc Civ} line velocity values are available, are shown as pink squares. The black solid and dashed lines correspond to \citetalias{Duras_2020} best fit to Type 1 sources and its 0.26 dex spread, respectively.}
    \label{fig:Lbol_kBol}
\end{figure}

We also calculated the hard X-ray bolometric correction $\rm k_{bol} = L_{2-10}/L_{bol}$. In Figure \ref{fig:Lbol_kBol}, the $\rm k_{bol}$ of WISSH QSOs is reported as a function of $\rm L_{bol}$, along with other low- and high-$\rm L_{bol}$ AGN samples from literature. We distinguished between broad absorption line (BAL) and non-BAL WISSH objects \citep[as listed in][and shown as indigo and purple stars, respectively]{Bruni_2019}, since a sizeable fraction of BAL sources has been found to be intrinsically X-ray weak \citep[e.g.][]{Luo_2014, Vito_2018}. The XMM-COSMOS sample of 343 Type 1 AGN presented by \citet[L10 hereafter]{Lusso_2010}, which constitutes the bulk of low-luminosity AGN, is shown with yellow dots. 
The 22 SUBWAYS QSOs and Type 1 AGN with $\rm L_{bol} > 10^{45}\,\,erg\,s^{-1}$ from \citet{Matzeu_2023} can be seen as red dots. 
Light blue triangles correspond to the 14 radio-quiet high-$\rm \lambda_{Edd}$ $\rm (\lambda_{Edd} \gtrsim 1)$ sources at $z \rm \approx 0.5$ published by \citet{Laurenti_2022}. The sample of luminous and intrinsically blue QSOs at $z \rm \approx 3$ presented by \citet{Trefoloni_2023}, for which constrained $\rm L_{2-10}$, $\rm \Gamma$, $\rm \lambda_{Edd}$, and {\sc Civ} line velocity values are available, is shown as pink squares. The black solid line represents the relation that \citet{Duras_2020} \citepalias[][hereafter]{Duras_2020} found fitting Type 1 sources only, while the black dashed curves correspond to the 0.26 dex spread of the same relation. 

\begin{figure*}
    \centering
    \subfloat[][\label{fig:Lbol_ratioKbol}]{\includegraphics[height=250pt]{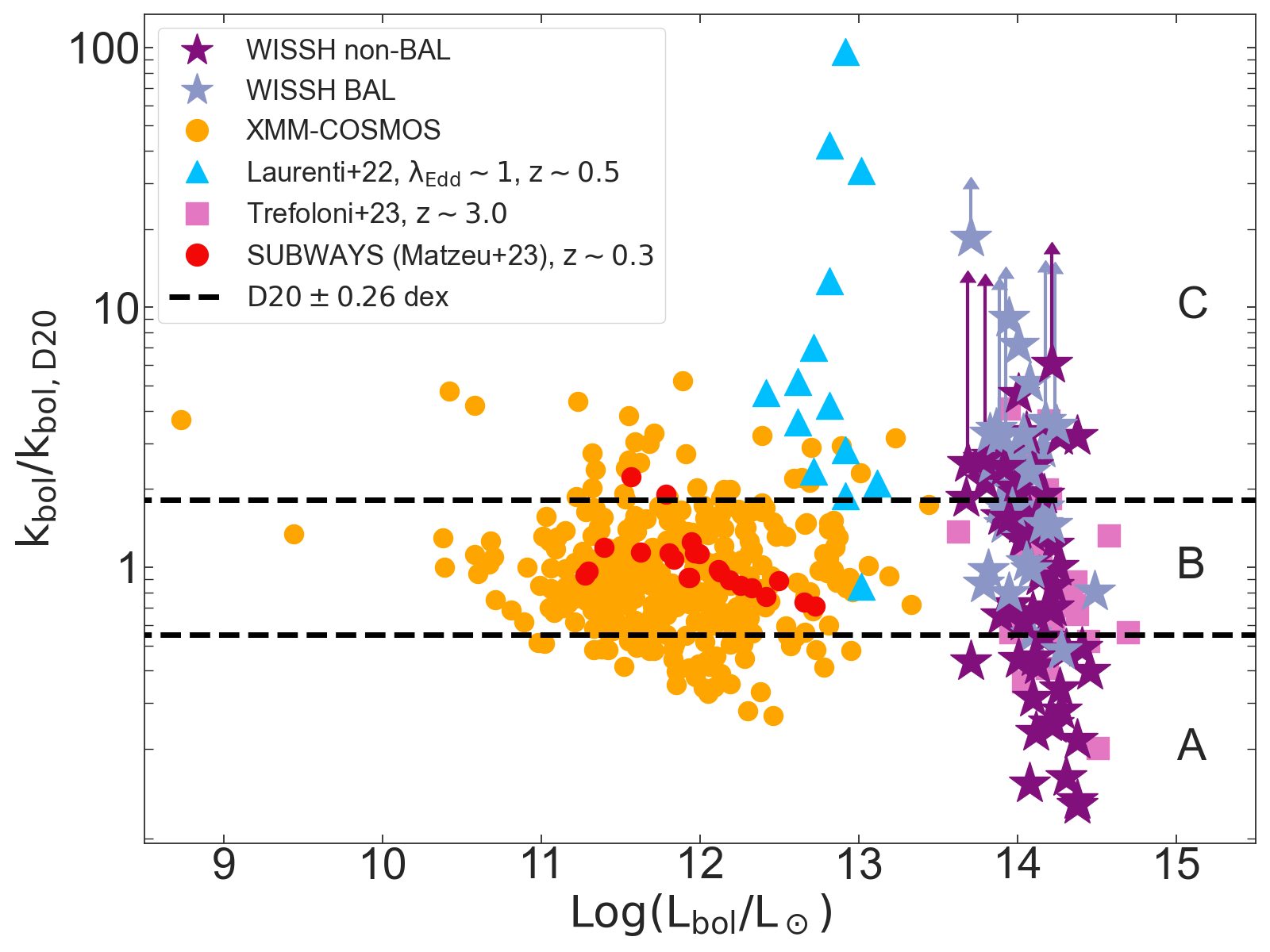}} \quad
    \subfloat[][\label{fig:D20_above_below}]{\includegraphics[height=250pt]{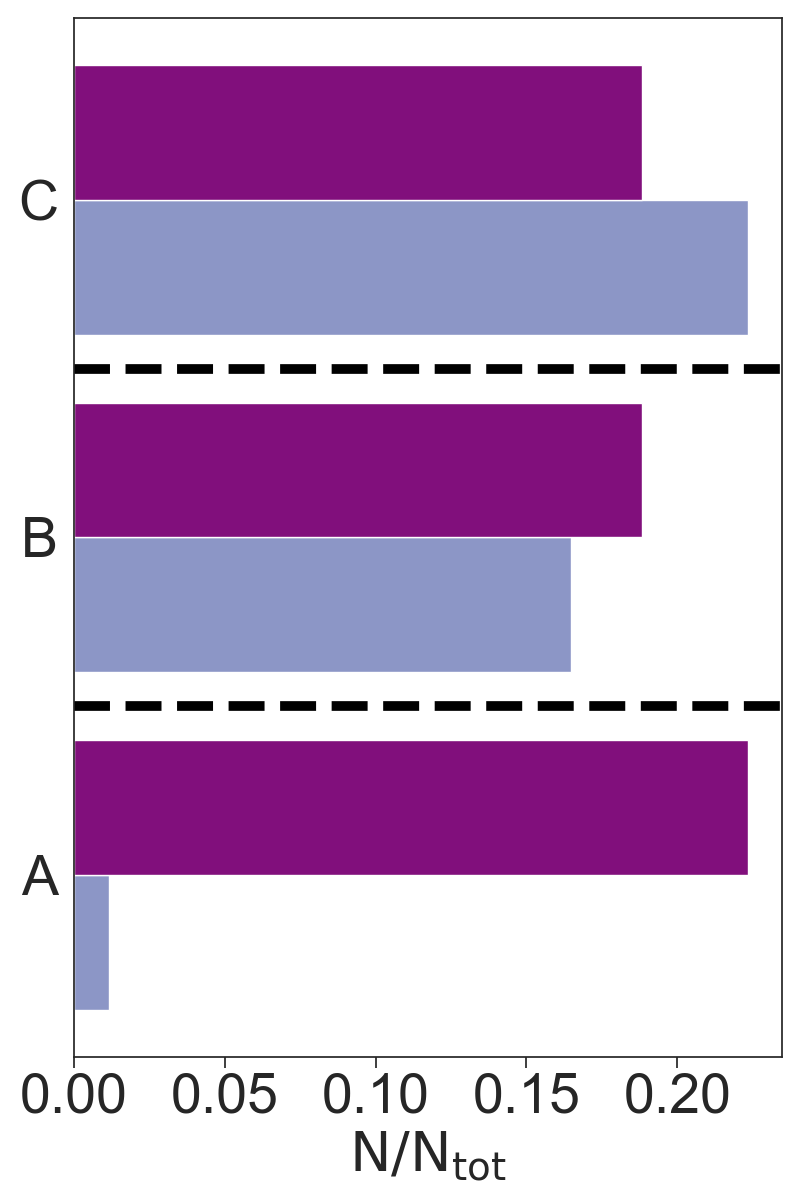}}
    \caption{(a) Ratio of measured $\rm k_{bol}$ values to expected $\rm k_{bol}$ values from \citetalias{Duras_2020} as a function of $\rm Log(L_{bol})$. WISSH QSOs are compared to literature samples; the symbols are the same as in Figure \ref{fig:Lbol_kBol}. The two black dashed lines correspond to the 0.26 dex spread of \citetalias{Duras_2020} best fit to Type 1 sources. (b) Distribution of WISSH QSOs $\rm k_{bol}$ with respect to the \citetalias{Duras_2020} relation, including its spread (black horizontal dashed lines), i.e. being below, within, or above it. BAL and non-BAL sources are represented as indigo and purple bars, respectively.}
    \label{fig:kBol_plots}
\end{figure*}

The huge spread of WISSH QSOs in the $\rm k_{bol} - Log(L_{bol})$ plane was already apparent (although with half of the current sources) in \citet{Martocchia_2017}. This spread looks more evident due to the relatively narrow range of $\rm L_{bol}$ sampled by WISSH objects. Figure \ref{fig:Lbol_ratioKbol} shows the ratio of the observed $\rm k_{bol}$ to the expected values at the same $\rm L_{bol}$ measured using the \citetalias{Duras_2020} relation ($\rm k_{bol}/k_{bol,D20}$; symbols are the same as in Figure \ref{fig:Lbol_kBol}). WISSH BAL QSOs are mainly $\rm (\approx 56\%)$ located above the \citetalias{Duras_2020} relation including its spread: at a given $\rm L_{bol}$, they are more often characterised by relatively weak X-ray emission. Nonetheless, non-negligible fractions of $\rm \approx 41\%$ and $\rm \approx 3\%$ of WISSH BAL QSOs fall within and below \citetalias{Duras_2020} spread, respectively. Conversely, non-BAL objects are more equally distributed: $\rm \approx 31\%$ above, $\rm \approx 30\%$ within and $\rm \approx 39\%$ below the best fit relation. To summarise, only $\rm \approx 34\%$ of the whole WISSH sample falls within the prediction of the \citetalias{Duras_2020} relation (including its spread), while $\rm \approx 66\%$ is distributed above $\rm (\approx 41\%)$ or below $\rm (\approx 25\%)$ it. The complete distribution for WISSH QSOs is visible in Figure \ref{fig:D20_above_below}, where BAL and non-BAL sources are reported as indigo and purple bars, respectively.

\subsection{Distribution of $ \alpha_{OX}$ and the fraction of X-ray-weak WISSH QSOs}\label{subsec:aox}
\begin{figure}
    \resizebox{\hsize}{!}{\includegraphics{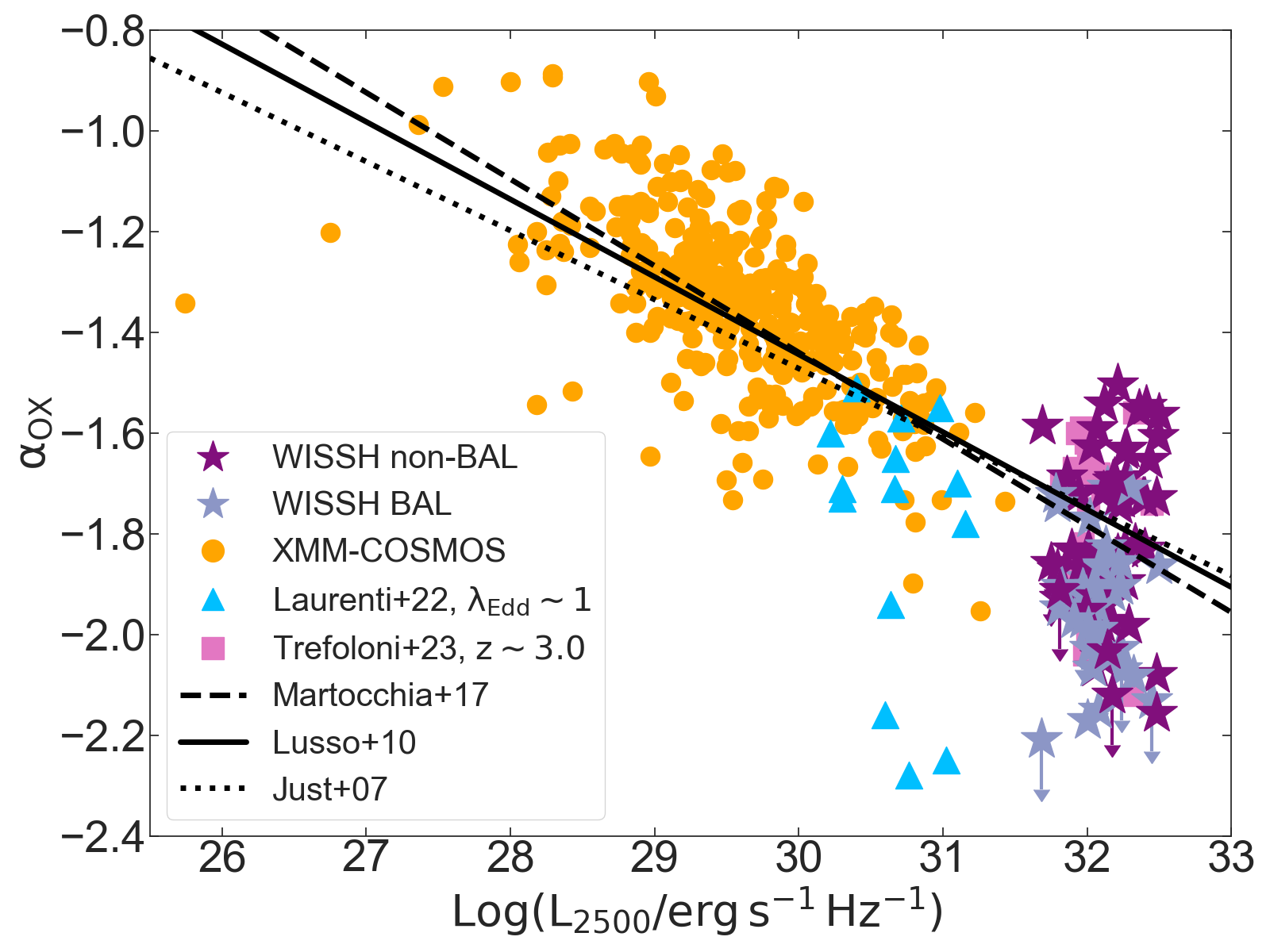}}
    \caption{X-ray$-$to$-$optical index $\rm (\alpha_{OX})$ as a function of $\rm Log(L_{2500\,\AA})$. WISSH QSOs are compared to literature samples; the symbols are the same as in Figure \ref{fig:Lbol_kBol}. The black dashed, solid, and dotted lines are the best fits from \citet{Martocchia_2017}, \citetalias{Lusso_2010}, and \citet{Just_2007}, respectively. In particular, \citetalias{Lusso_2010} fit XMM-COSMOS data only; \citet{Just_2007} consider 34 QSOs of their core sample, 332 sources from \citet{Steffen_2006}, and 14 from \citet{Shemmer_2006}; and \citet{Martocchia_2017} consider XMM-COSMOS objects, 23 optically selected QSOs from the Palomar-Green (PG) Bright QSO Survey of the complete sample by \citet{Laor_1994} and the 41 WISSH QSOs with available X-ray data at the time of the publication.}
    \label{fig:L2500_aox}
\end{figure}

We derived $\rm \alpha_{OX}$ for WISSH QSOs using $\rm L_{2500\,\AA}$ from \citet{Saccheo_2023}. Figure \ref{fig:L2500_aox} displays the $\rm \alpha_{OX} - Log(L_{2500\,\AA})$ plane, where WISSH QSOs are compared to literature samples. Symbols are the same as in Figure \ref{fig:Lbol_kBol}. The relations by \citet{Martocchia_2017}, \citetalias{Lusso_2010} and \citet{Just_2007} (black dashed, solid and dotted line, respectively) are also reported as representative of the numerous literature works. The WISSH sample occupies the bottom-right region of the plane, meaning that luminous QSOs are among the X-ray-weakest sources. This result is expected, given the indications of Figure \ref{fig:Lbol_kBol}, where high $\rm k_{bol}$, corresponding to low $\rm \alpha_{OX}$ values, implies a low coronal X-ray contribution to the overall AGN emission, i.e. most of the accretion-related luminosity comes from the accretion disk in the UV band \citep[but see also models of][]{Kubota_2018}. As in the $\rm k_{bol} - Log(L_{bol})$ plane, WISSH QSOs broadly follow the decreasing $\rm \alpha_{OX}$ trend at increasing $\rm L_{2500\,\AA}$, as proposed by previous papers and already observed by \citet{Martocchia_2017}, though with lower statistics.

\begin{figure}
    \resizebox{\hsize}{!}{\includegraphics{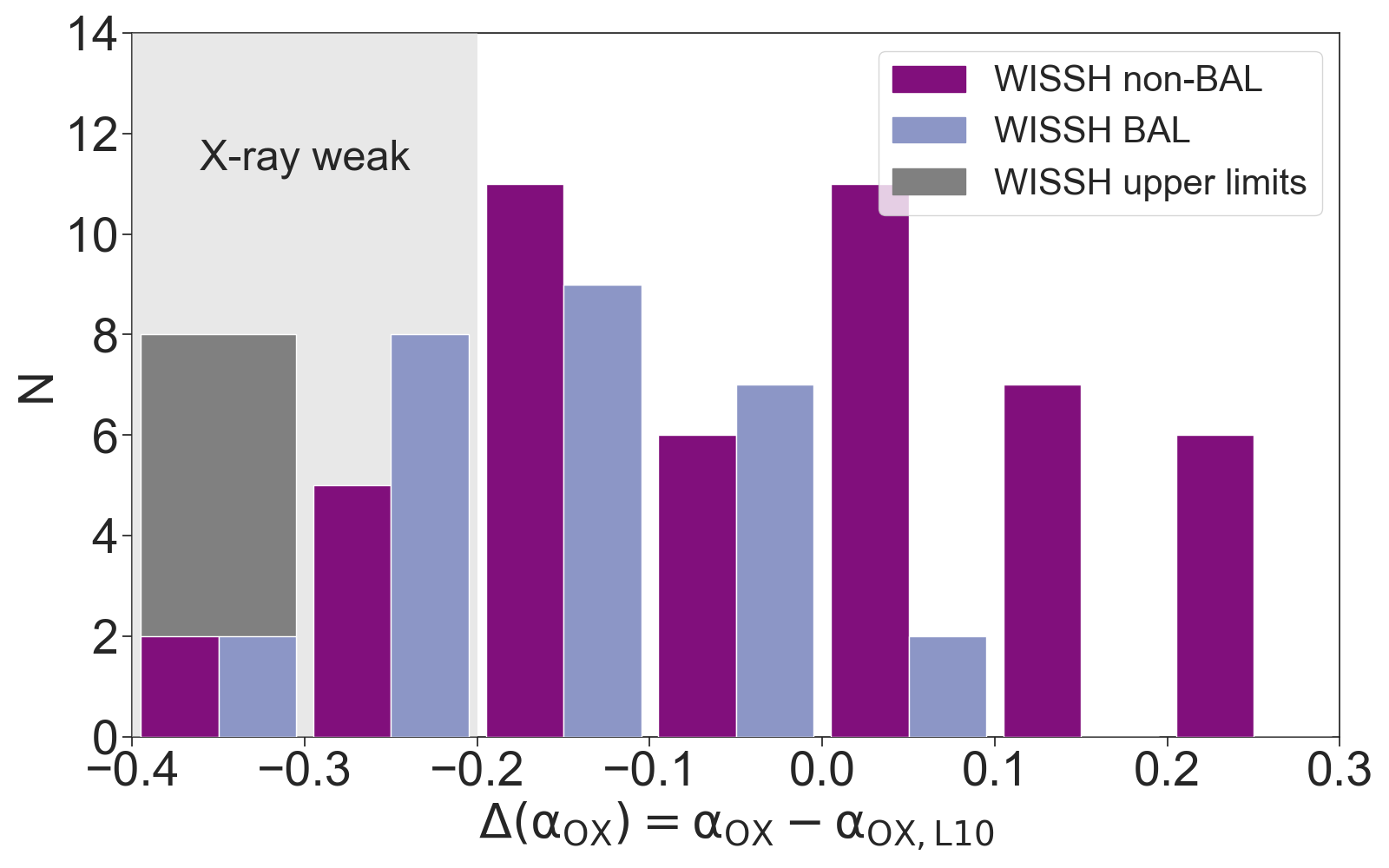}}
    \caption{$\rm \Delta(\alpha_{OX}) = \alpha_{OX} - \alpha_{OX, L10}$ distribution. $\rm \alpha_{OX, L10}$ refers to the value derived from the \citetalias{Lusso_2010} relation at a given $\rm L_{2500\,\AA}$. The BAL and non-BAL QSOs are represented as indigo and purple bars, respectively. The grey bar is for undetected sources, which have all been included in the X-ray-weakest bin. The grey-shaded area highlights the locus of X-ray-weak sources.}
    \label{fig:deltaAOX_distribution}
\end{figure}

Measuring the offset between the observed values $\rm (\alpha_{OX})$ and those expected from \citetalias{Lusso_2010} relation $\rm (\alpha_{OX, L10})$ at a given $\rm L_{2500\,\AA}$, one can define $\rm \Delta(\alpha_{OX})$ as
\begin{equation}\label{eq:deltaAOX}
    \rm \Delta(\alpha_{OX}) = \alpha_{OX} - \alpha_{OX, L10}.
\end{equation}
This quantity can be used to point out X-ray-weak sources, characterised by $\rm \Delta(\alpha_{OX}) \le -0.2$ \citep{Luo_2015}. Figure \ref{fig:deltaAOX_distribution} shows the $\rm \Delta(\alpha_{OX})$ distribution of WISSH QSOs (see also Table \ref{tab:results}). 
About 31\% of them falls under the X-ray-weak category, which is highlighted by the grey-shaded area in Figure \ref{fig:deltaAOX_distribution}. It is evident that BAL and non-BAL QSOs are differently distributed. The Kolmogorov-Smirnov test results in p = 0.003, which rejects the two-sided null-hypothesis probability that the two distributions are derived from the same parent distribution at 3$\rm \sigma$ confidence level. In particular, $\rm \Delta(\alpha_{OX})$ value is below the adopted threshold for $\rm \approx 47\%$ and $\rm \approx 20\%$ of BAL and non-BAL objects, respectively. X-ray-weak sources are therefore much more common among BAL objects than non-BAL ones. In case of QSOs with multiple X-ray observations (see Section \ref{subsec:multi_obs}), we report no X-ray weak to X-ray normal (or vice versa) transitions.

\subsection{Photon index distribution}\label{subsec:gamma_distributions}
To provide a deeper investigation of the possible origin of X-ray weakness in the presence of an absorbing column density, we searched for a possible correlation between $\rm \Gamma$ and $\rm \Delta(\alpha_{OX})$. Indeed, additional absorption, if not included in the X-ray spectral analysis, would result in X-ray-weak AGN to have a flatter photon index. 

\begin{figure*}
    \centering
    \subfloat[][\label{fig:deltaAOX_gammaPL}]{\includegraphics[width=0.48\textwidth]{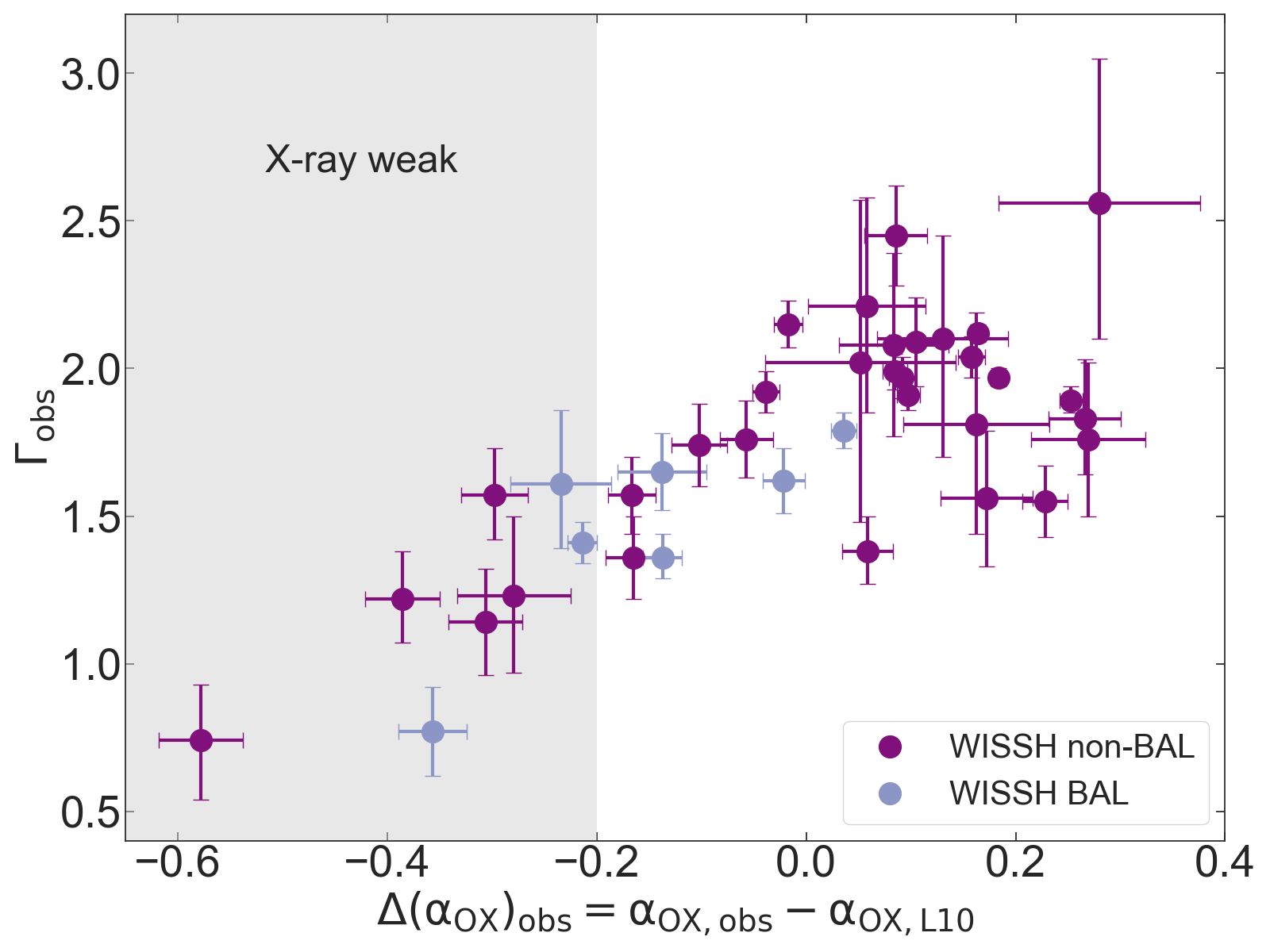}} \quad
    \subfloat[][\label{fig:deltaAOX_gamma}]{\includegraphics[width=0.48\textwidth]{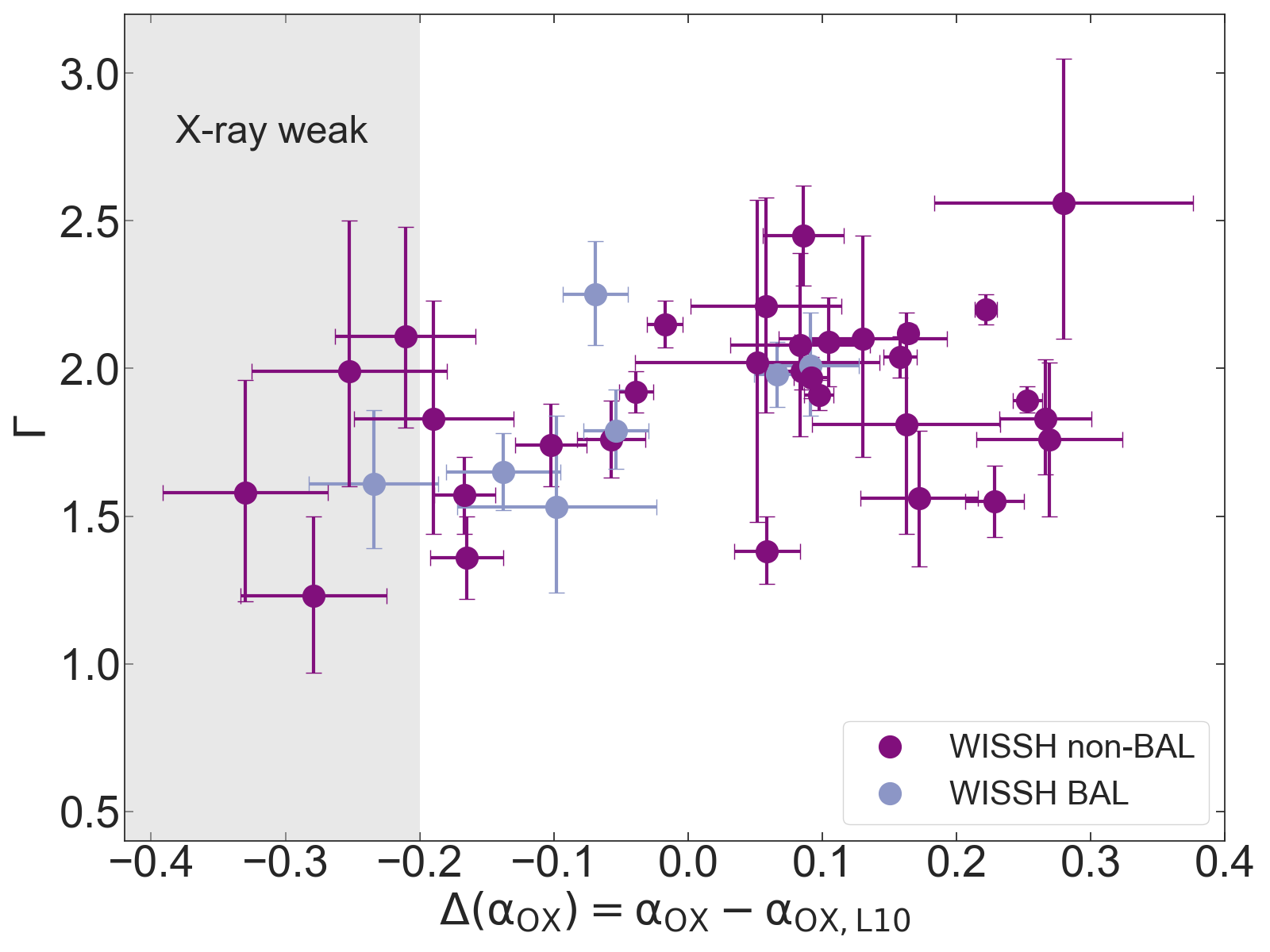}}
    \caption{X-ray photon index as a function of $\rm \Delta(\alpha_{OX})$ for the sources with $\rm \ge 20$ counts. (a) $\rm \Gamma_{obs}$  derived using a power law model modified by Galactic absorption. (b) Cold absorption component  included in the spectral model used to derive $\rm \Gamma$ (see Section \ref{subsubsec:20_counts}). The BAL and non-BAL QSOs are represented as indigo and purple dots, respectively. The grey-shaded areas highlight the locus of X-ray-weak sources.}
    \label{fig:deltaAOX_gamma_both}
\end{figure*}

We only considered the sources with $\rm \ge 20$ photons, for which we derived $\rm \Gamma$ through X-ray spectral analysis. In Figure \ref{fig:deltaAOX_gammaPL}, $\rm \Gamma$ is derived applying a power law model modified by Galactic absorption, while in Figure \ref{fig:deltaAOX_gamma} we also take into consideration the presence of intrinsic absorption (see Section \ref{subsubsec:20_counts}). The main variations occur at the lowest $\rm \Delta(\alpha_{OX})$: the suppression of the soft X-ray emission by intrinsic $\rm N_H$ causes both $\rm \Gamma$ flattening and X-ray weakening with respect to the optical/UV emission. However, once the best spectral model is applied (Figure \ref{fig:deltaAOX_gamma}), the Pearson correlation test only results in $\rm p = 0.01$, i.e. $\rm \Delta(\alpha_{OX})$ and $\rm \Gamma$ correlate at less than $\rm 3\sigma$ significance level. 
Although we cannot rule out the possibility of additional obscuration among the X-ray-weakest sources, it is clear that the roughly uniform spectral shape as a function of $\rm \Delta(\alpha_{OX})$ makes it unlikely that extra-absorption is responsible for the X-ray-weak phenomenon in our sample.

For 36 sources ($\rm \approx 42\%$ of WISSH QSOs), a black hole mass $\rm (M_{BH})$ estimate is available \citep[][Vietri et al. in prep.]{Bischetti_2017, Vietri_2018}. $\rm M_{BH}$ were derived through a single epoch virial method relation \citep{Bongiorno_2014}, which depends on the H$\rm \beta$ line Full Width at Half Maximum $\rm (FWHM_{H\beta})$ and the continuum luminosity at 5100\,\,\AA\ $\rm (\lambda L_\lambda)$:
\begin{equation}\label{eq:singleEpoch}
    \rm Log(M_{BH}/M_\odot) = 6.7 + 2\,Log\bigg(\frac{FWHM_{H\beta}}{10^3\,\,km\,s^{-1}}\bigg) + 0.5\,Log\bigg(\frac{\lambda L_\lambda}{10^{44}\,\,erg\,s^{-1}}\bigg).
\end{equation}
The systematic uncertainty in the $\rm Log(M_{BH})$ determination is estimated to be about 0.3 dex \citep{Bongiorno_2014}. From the $\rm M_{BH}$ values, we calculated the Eddington luminosity, which spans the range $\rm L_{Edd} \approx 3\times10^{47} - 3\times10^{48} \,\,erg\,s^{-1}$. Having both $\rm L_{Edd}$ and $\rm L_{bol}$, we derived the Eddington ratios $\rm \lambda_{Edd}$. This results in a range between 0.2 and 2.9, with a median value 0.6. These numbers indicate a high-accretion regime, as expected for the most luminous QSOs at Cosmic Noon \citep[e.g.][]{Hopkins_2007, Merloni_2008, Delvecchio_2014}.

\begin{figure}
    \resizebox{\hsize}{!}{\includegraphics{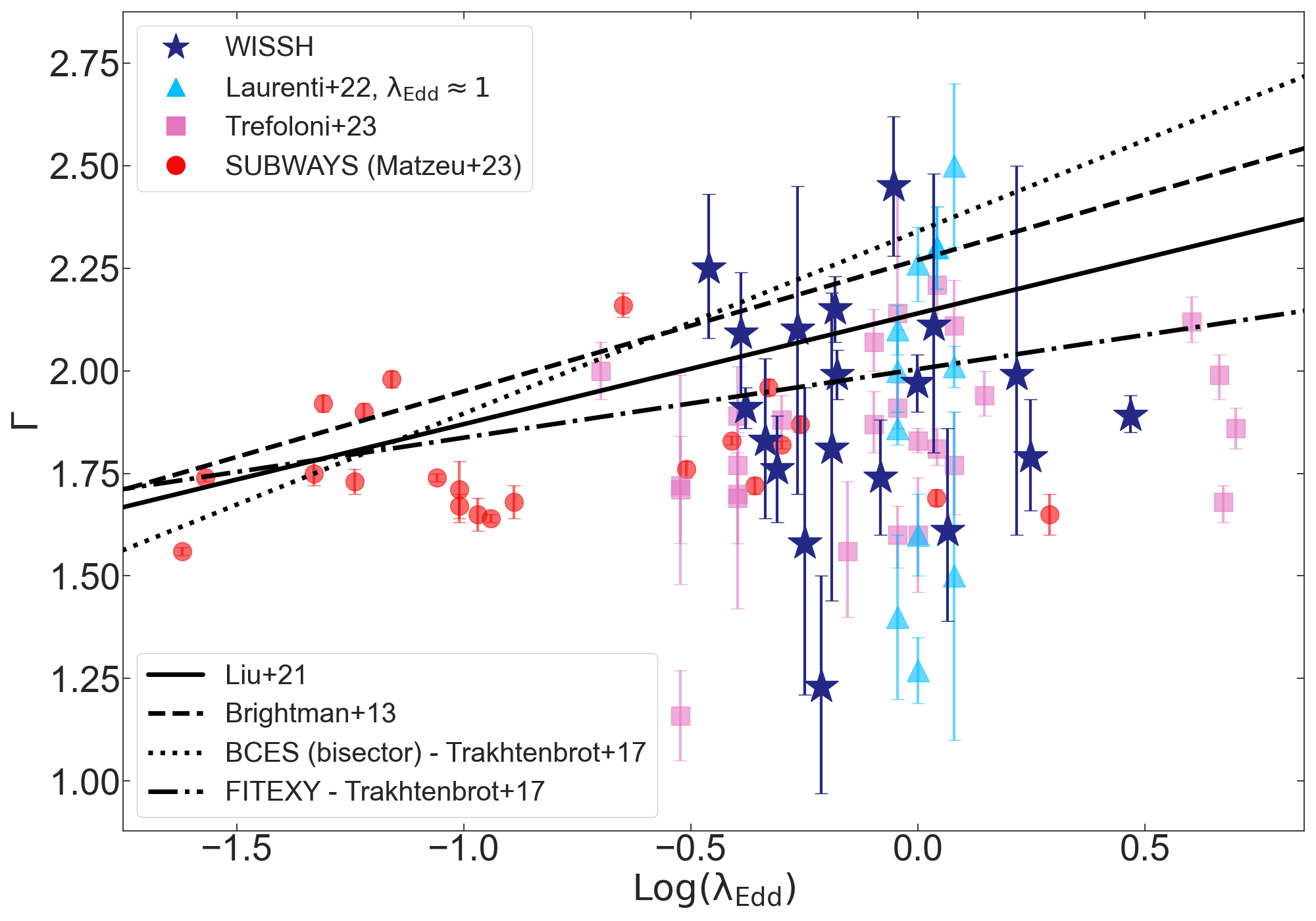}}
    \caption{X-ray photon index as a function of $\rm Log(\lambda_{Edd})$. Only WISSH QSOs with $\rm \ge 20$ photons (i.e. with X-ray spectral analysis), for which an estimate of $\rm M_{BH}$ is available are considered. The symbols are the same as in Figure \ref{fig:Lbol_kBol}, with the exception of WISSH sources, which are represented as blue stars. The black solid, dashed, dotted, and dash-dotted lines correspond to the \citet{Liu_2021} and \citet{Brightman_2013} relations, and two models (BCES bisector and FITEXY method) from \cite{Trakhtenbrot_2017}, respectively. Specifically, \citet{Liu_2021} consider 47 AGN with both super- and sub-Eddington accretion rates from the sample compiled by the SEAMBH collaboration \citep{Du_2018}. \citet{Brightman_2013} use 69 broad-line AGN from the extended \textit{Chandra} Deep Field South and COSMOS surveys. \citet{Trakhtenbrot_2017} refer to 228 hard X-ray selected AGN at 0.01 < $z$ < 0.5, drawn from the \textit{Swift}/BAT AGN Spectroscopic Survey \citep[BASS,][]{Baumgartner_2013}.}
    \label{fig:eddRatio_gamma}
\end{figure}

For the 19 WISSH sources which have both an available $\rm M_{BH}$ and $\rm \ge 20$ photons to perform X-ray spectral analysis, we also studied $\rm \Gamma$ as a function of $\rm \lambda_{Edd}$. This relation may be expected as higher $\rm \lambda_{Edd}$ corresponds to more intense disk emission, thus a more efficient coronal Compton cooling, which leads to a softer (i.e. steeper) photon index \citep[e.g.][]{Haardt_1991, Haardt_1993, Pounds_1995, Fabian_2015, Cheng_2020}. The results are shown in Figure \ref{fig:eddRatio_gamma}, where symbols for literature samples are the same as in Figure \ref{fig:Lbol_kBol}. WISSH sources are represented as blue stars, while the black solid, dashed, dotted and dash-dotted lines correspond to \citet{Liu_2021} and \citet{Brightman_2013} relations, and two models (BCES bisector and FITEXY method) from \citet{Trakhtenbrot_2017}, respectively. 
In order to quantify the distribution in the $\rm \Gamma -  Log(\lambda_{Edd})$ plane, we fitted a linear model to the data in a Bayesian framework using the python package \texttt{linmix} \citep{Kelly_2007}. 
The Pearson correlation test results in $\rm p = 0.07$, suggesting no significant correlation. This highlights the importance of populating the plane with high-$\rm \lambda_{Edd}$ QSOs.

\subsection{The $Log(N_H) - Log(\lambda_{Edd})$ diagram}\label{subsec:NH_Edd}
\begin{figure*}
    \centering
    \includegraphics[width=12cm]{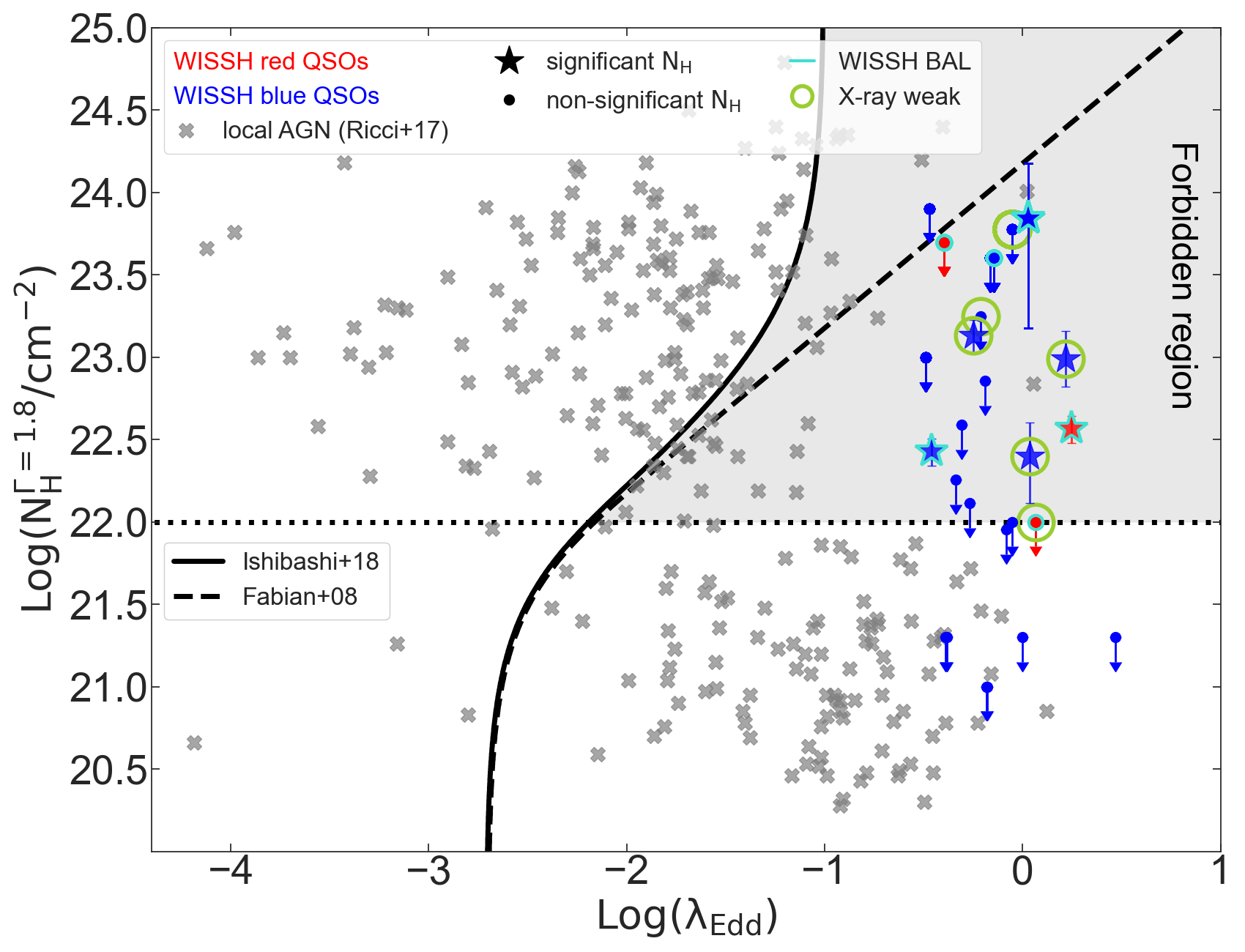}
    \caption{Intrinsic column density (derived assuming $\rm \Gamma = 1.8$) as a function of $\rm Log(\lambda_{Edd})$. QSOs from the HC-WISSH sample with available $\rm M_{BH}$ are plotted. WISSH BAL objects are contoured with a turquoise line. X-ray-weak QSOs $\rm (\Delta(\alpha_{OX}) < -0.2)$ are surrounded with a green circle. Blue and red QSOs are shown in the corresponding colour. Stars correspond to sources for which intrinsic absorption is significantly detected (see Sections \ref{subsubsec:20_counts} and \ref{subsubsec:619_counts}). QSOs from \citet{Ricci_2017} are shown as grey crosses. Black solid and dashed lines are \citet{Ishibashi_2018} and \citet{Fabian_2008} relations, derived from a maximal- and no-photon trapping model, respectively. The black horizontal dotted line indicates $\rm N_H = 10^{22}\,\,cm^{-2}$.}
    \label{fig:eddRatio_nh}
\end{figure*}

In the context of studying AGN evolution from an early, dust-reddened, obscured phase to a later, blue
and unobscured one, \citet{Fabian_2008} introduced the $\rm Log(N_H) - Log(\lambda_{Edd})$ plane. Specifically, they identified a region of this plane which can be associated with a transitional blow-out phase in AGN
evolution, during which the nuclear source, previously obscured, is set free from absorbing gas thanks to
AGN-driven feedback via nuclear winds. In this scenario, the high-$\rm N_H$/high-$\rm \lambda_{Edd}$ conditions correspond to a `forbidden region' in the plane (grey-shaded in Figure \ref{fig:eddRatio_nh}). Following \citet{Fabian_2008} and \citet{Ishibashi_2018}, they assumed a dusty gas of partially ionised hydrogen, and took into consideration the trapping of reprocessed radiation by dusty gas. In Figure \ref{fig:eddRatio_nh}, maximal- and no-photon trapping regimes are represented as a black solid and dashed line, respectively.

We populated the $\rm Log(N_H) - Log(\lambda_{Edd})$ plane combining the available information on $\rm M_{BH}$ and $\rm N_H$ of the sources from the HC-WISSH sample. To better constrain $\rm N_H$, we fixed the X-ray continuum slope to $\rm \Gamma = 1.8$ \citep[e.g.][]{Piconcelli_2005}, since it does not vary significantly within its uncertainties, as shown in Figure \ref{fig:deltaAOX_gamma}. Specifically, WISSH sources exhibit an average photon index $\rm \Gamma = 1.88 \pm 0.29$ (where the reported error is the dispersion of the distribution). The vast majority of WISSH objects are optically classified as blue sources, while only seven QSOs (i.e. WISSH25, WISSH34, WISSH40, WISSH49, WISSH58, WISSH66, WISSH74) exhibit a colour excess $\rm E(B-V) \ge 0.15$, accordingly to \citet{Saccheo_2023}, and can be classified as dust-reddened broad-line sources. Figure \ref{fig:eddRatio_nh} reports WISSH BAL objects contoured with a turquoise line. X-ray-weak QSOs $\rm (\Delta(\alpha_{OX}) < -0.2)$ are surrounded with a green circle. Blue and red QSOs are shown in the corresponding colour. Stars correspond to sources for which intrinsic absorption is significantly detected (see Sections \ref{subsubsec:20_counts} and \ref{subsubsec:619_counts}). Sources from the large sample of local AGN presented by \citet{Ricci_2017} are also reported (grey crosses) and typically populate the $\rm Log(N_H) - Log(\lambda_{Edd})$ plane outside the forbidden region. 
All but one of WISSH sources with significantly detected $\rm N_H$ located in the forbidden region are blue QSOs. Surprisingly, among them, non-BAL QSOs are X-ray-weak sources, while BAL QSOs show $\rm \Delta(\alpha_{OX}) > -0.2$. Finally, unlike what is commonly found for red QSOs \citep[e.g.][]{Lansbury_2020}, WISSH40 falls outside the forbidden region.

\subsection{X-ray versus mid-infrared luminosity}\label{subsec:MIR}
The X-ray and MIR emissions are strictly connected in QSOs. The former is coronal emission of accretion disk Comptonised photons, and the latter is due to reprocessed accretion disk emission being thermalised by the dusty torus. Thus, a positive correlation is expected in the X-ray $-$ MIR luminosity plane \citep[e.g.][]{Gandhi_2009}. By construction, all 85 WISSH QSOs have been detected in the WISE 3.3$\rm \,\,\mu m$ band, from which the luminosity at $\rm 6\,\,\mu m$ ($\rm \lambda L_{6\,\mu m}$ in Table \ref{tab:results}) can be recovered applying a correction derived from the mean SED of the sample \citep{Saccheo_2023}. This allowed us to study the X-ray$-$MIR relation in the highest luminosity regime, while most of previous works focused on the low-luminosity regime (\citealp[e.g.][]{Lutz_2004, Fiore_2009, Lanzuisi_2009, Mateos_2015}; \citealp[but see also][S15 hereafter,]{Stern_2015} \citealp[and][]{Chen__2017}).

\begin{figure*}
    \centering
    \subfloat[][\label{fig:Lmir_LX}]{\includegraphics[height=6cm]{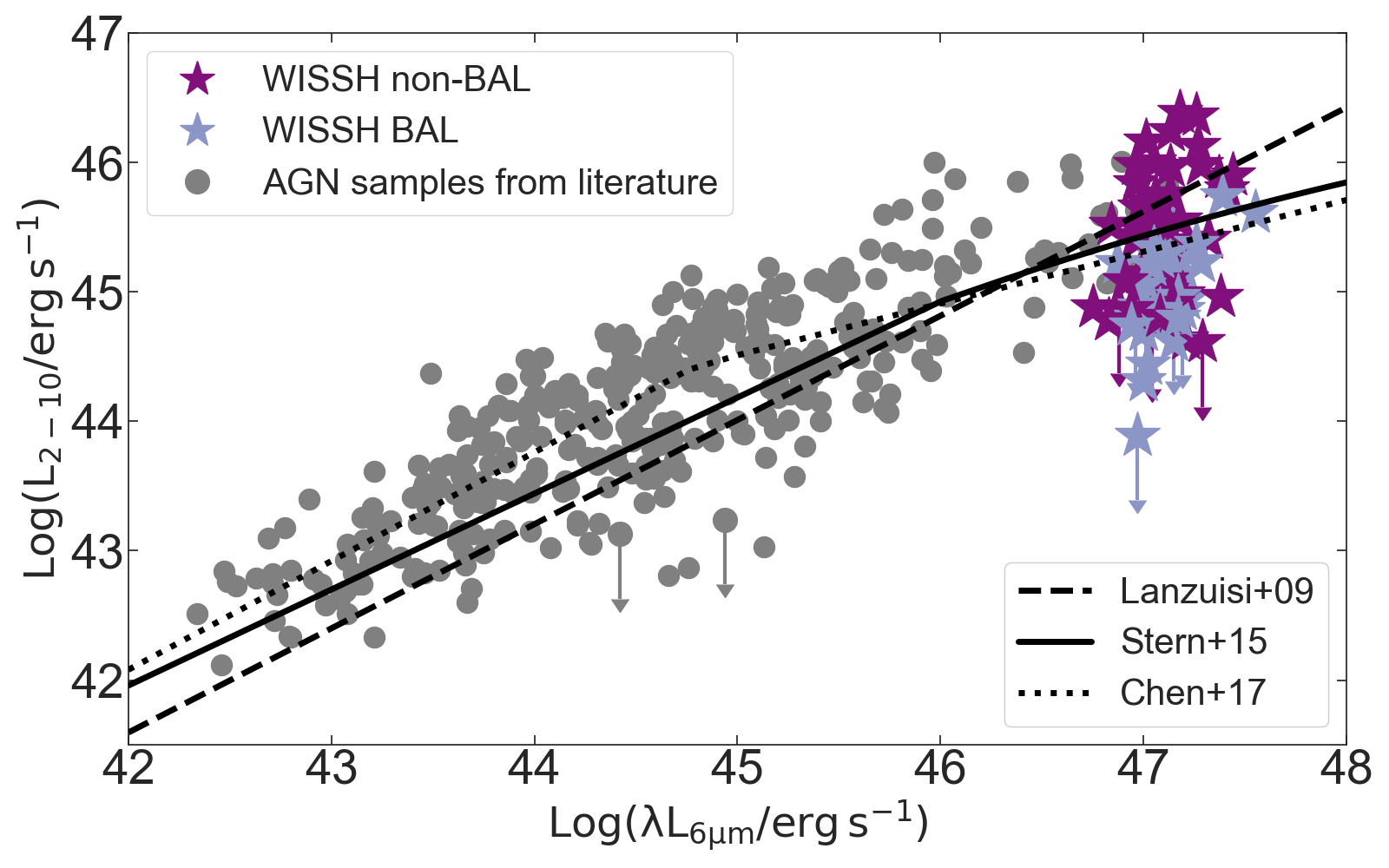}} \quad
    \subfloat[][\label{fig:deltaMIR_deltaAOX}]{\includegraphics[height=6cm]{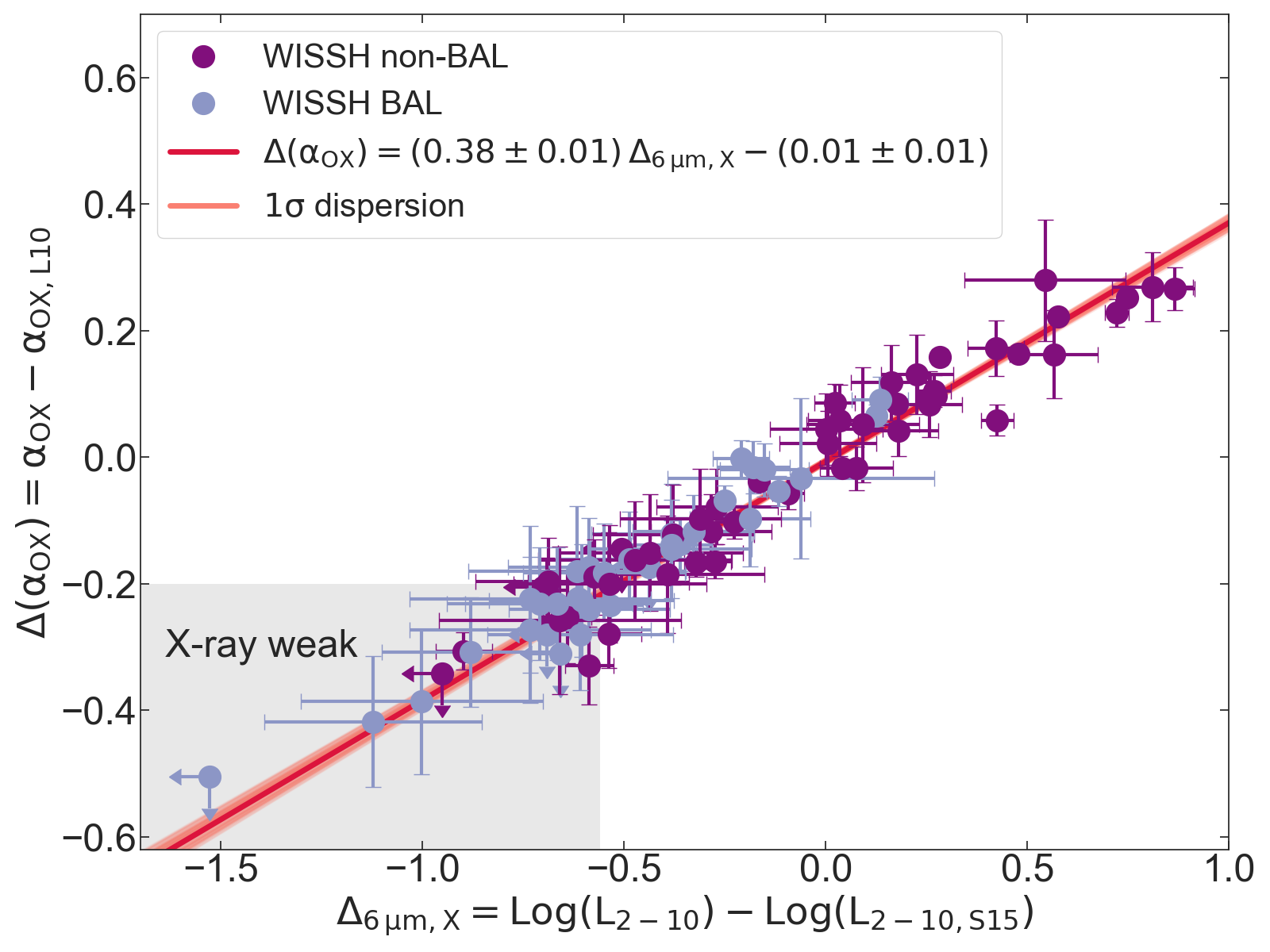}}
    \caption{(a) Intrinsic $\rm 2 - 10\,\,keV$ luminosity as a function of $\rm \lambda L_{6\,\mu m}$. WISSH BAL and non-BAL objects are shown as indigo and purple stars, respectively, while the grey dots represent various comparison samples from \citet{Lanzuisi_2009}, \citet{Mateos_2015}, and \citetalias{Stern_2015}. The black solid, dashed, and dotted lines correspond to the relation by \citetalias{Stern_2015}, \citet{Lanzuisi_2009}, and \citet{Chen__2017}, respectively; (b) $\rm \Delta(\alpha_{OX})$ as a function of the $\rm L_{2-10}$ offset of WISSH QSOs from the relation by \citetalias{Stern_2015} $\rm (\Delta_{6\,\mu m, X})$. WISSH BAL and non-BAL objects are indigo and purple dots, respectively. The grey-shaded area highlights the X-ray-weak region of the plane. The dispersion around the best fit (solid red line) is given by plotting $\rm \approx 200$ realisations considering the values of slope and intercept within $\rm 1\sigma$ of the sampled marginalised posterior distribution.}
    \label{fig:MIR_figures}
\end{figure*}

Figure \ref{fig:Lmir_LX} shows the intrinsic $\rm L_{2-10}$ as a function of $\rm \lambda L_{6\,\mu m}$. WISSH BAL and non-BAL objects are indigo and purple stars, respectively, while grey dots are AGN from the samples of \citet{Lanzuisi_2009}, \citet{Mateos_2015} and \citetalias{Stern_2015}. The black solid, dashed and dotted lines are \citetalias{Stern_2015}, \citet{Lanzuisi_2009} and \citet{Chen__2017} relations, respectively. Despite the limited range in terms of $\rm Log(\lambda L_{6\,\mu m})$ due to the sample selection, WISSH sources cover a wide range of $\rm Log(L_{2 - 10})$. Nonetheless, they seem to agree with the luminosity-dependent trends found by \citetalias{Stern_2015} and \citet{Chen__2017}, who report a flattening at the highest MIR luminosities. Since the MIR luminosity in AGN is strongly linked to the reprocessing of UV accretion disk emission, this behaviour has been interpreted within the framework of the $\rm \alpha_{OX} - L_{UV}$ relation \citep[e.g.][]{Chen__2017}.

Similarly to the derivation of $\rm \Delta(\alpha_{OX})$, we studied the offset of WISSH QSOs measured $\rm Log(L_{2-10})$ from the one expected from \citetalias{Stern_2015} relation $\rm Log(L_{2-10, S15})$:
\begin{equation}\label{eq:deltaMIR}
    \rm \Delta_{6\,\mu m, X} = Log(L_{2-10}) - Log(L_{2-10, S15}).
\end{equation}
In Figure \ref{fig:deltaMIR_deltaAOX}, $\rm \Delta_{6\,\mu m,X}$ is compared to $\rm \Delta(\alpha_{OX})$. WISSH BAL and non-BAL objects are indigo and purple dots, respectively. We fitted the positive correlation with a first order model using the hierarchical Bayesian model \texttt{linmix} (see Section \ref{subsec:gamma_distributions}), resulting in the following tight relation:
\begin{equation}\label{eq:deltaAOX_deltaMIR}
    \rm \Delta(\alpha_{OX}) = (0.38 \pm 0.01)\,\Delta_{6\,\mu m,X} - (0.01 \pm 0.01).
\end{equation}
This robust relation ($\rm p = 3 \times 10^{-50}$; Pearson correlation coefficient $\rm r_P = 0.97$; see Figure \ref{fig:deltaMIR_deltaAOX_corner} and Table \ref{tab:relations_parameters}) makes it possible to translate the $\rm \Delta(\alpha_{OX})$-based definition of X-ray-weak sources into the $\rm \Delta_{6\,\mu m, X}$ parameter, resulting in $\rm \Delta_{6\,\mu m, X} \le -0.52$. The X-ray-weak region of the $\rm \Delta(\alpha_{OX}) - \Delta_{6\,\mu m, X}$ plane is highlighted by the grey-shaded area. We notice that once $\rm L_{2500\,\AA}$ and $\rm \lambda L_{6\,\mu m}$ are known, both Equation \ref{eq:deltaAOX} and \ref{eq:deltaMIR} can be expressed as a function of $\rm L_{2-10}$ (assuming a photon index, e.g. $\rm \Gamma \approx 1.8 - 2$, to estimate $\rm L_{2-10}$ from $\rm L_{2\,\,keV}$). Therefore, $\rm L_{2-10}$ can be derived through Equation \ref{eq:deltaAOX_deltaMIR}. This would provide less ambiguous results, thanks to the narrow distribution of the sources in the $\rm \Delta(\alpha_{OX}) - \Delta_{6\,\mu m, X}$ plane compared to the spread exhibited by WISSH QSOs in the $\rm Log(L_{2-10}) - Log(\lambda L_{6\,\mu m})$ plane. In addition, the comparison between $\rm L_{2-10}$ and the observed $\rm 2 - 10\,\,keV$ luminosity can provide an estimate of the intrinsic $\rm N_H$.

\subsection{The $\rm Log(L_{2-10}) - v_{CIV}$ and $\rm \Delta(\alpha_{OX}) - v_{CIV}$ relations}\label{subsec:CIV_shift}
A weak but significant correlation has been reported between $\rm \alpha_{OX}$ and $\rm \Delta(\alpha_{OX})$, and {\sc Civ} emission line blueshift (tracing the velocity of BLR-scale ionised winds) \citep[e.g.][]{Kruczek_2011, Richards_2011, Ni_2018, Vietri_2018, Timlin_2020}, suggesting a link between the shape of the AGN SED and the capability of accelerating  winds from the nuclear region. This is not surprising, as it is worth noting that a substantial ionising flux, such as QSO X-ray emission, can overionise the gas surrounding the accreting SMBH, thereby preventing the launch of strong nuclear winds \citep[e.g,][and references therein]{Proga_2007}. Nonetheless, given their large radiative output and high accretion rate, the most luminous AGN $\rm (L_{bol} \ge 10^{47}\,\,erg\,s^{-1})$ are expected to give rise to the most powerful winds \citep[e.g.][]{Faucher-Giguère_2012, Giustini_2019, Ward_2024} as supported by a wealth of observations over the last decade \citep[e.g.][]{Fiore_2017, Vietri_2018, Meyer_2019, Perrotta_2019, Musiimenta_2023}. 

\begin{figure}
    \resizebox{\hsize}{!}{\includegraphics{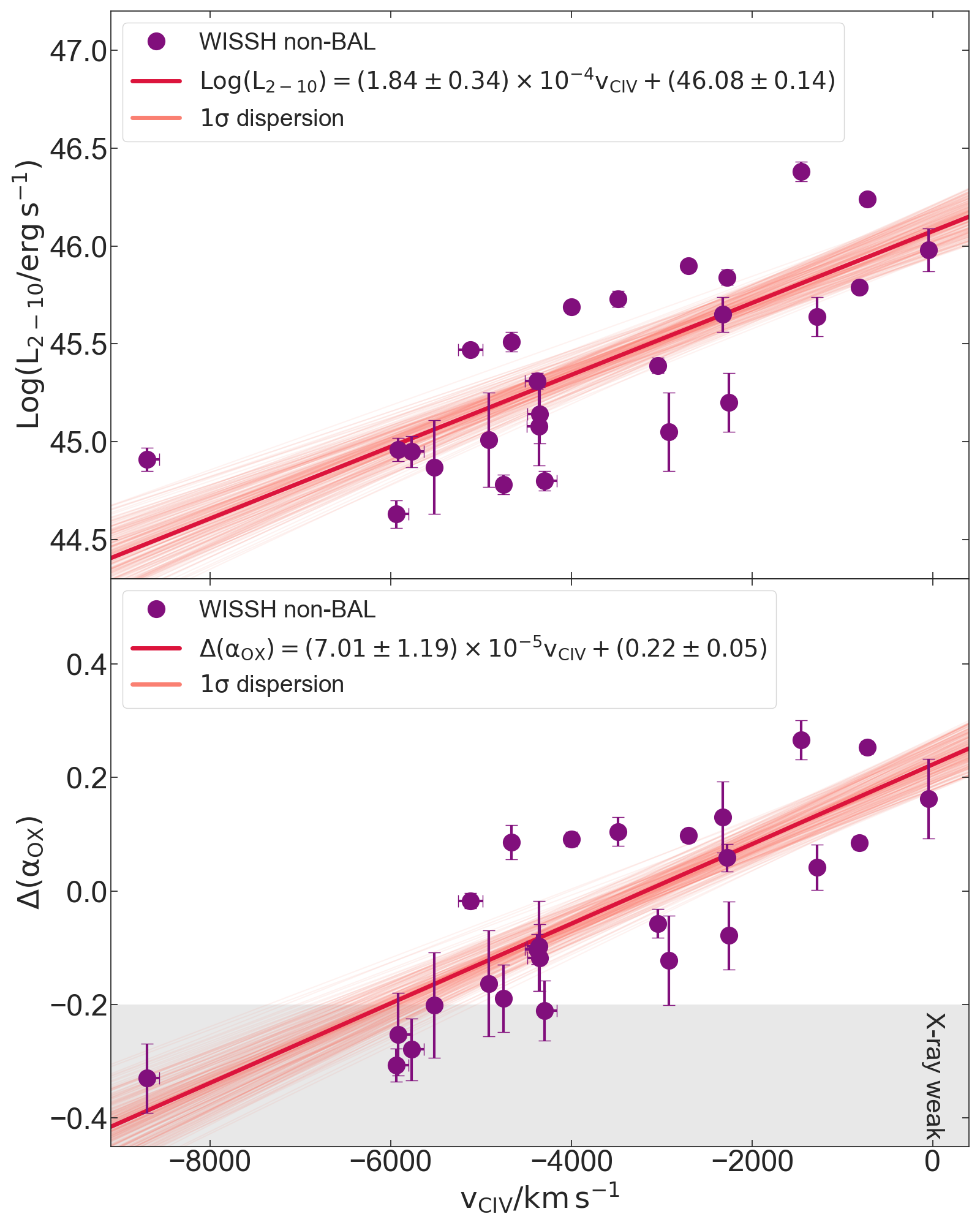}}
    \caption{Intrinsic $\rm 2 - 10\,\,keV$ luminosity (top panel) and $\rm \Delta (\alpha_{OX})$ (bottom panel) as a function of {\sc Civ} velocity $\rm v_{CIV}$. WISSH non-BAL objects are shown as purple dots. The dispersion around the best fits (solid red lines) is given by plotting $\rm \approx 200$ realisations considering the values of slope and intercept within $\rm 1\sigma$ of the sampled marginalised posterior distribution (see Figures \ref{fig:vCIV_L210_corner} and \ref{fig:vCIV_deltaAOX_corner}, and Table \ref{tab:relations_parameters} for further details).}
    \label{fig:vCIV_plots}
\end{figure}

Interestingly, \citet{Zappacosta_2020} reported the discovery of a relation between $\rm L_{2-10}$ and {\sc Civ} blueshift, based on a small sub-sample of WISSH QSOs. Due to  the narrow UV luminosity range covered by the WISSH sources, this supports a scenario in which faster nuclear outflows are unequivocally associated with high-luminosity QSOs that exhibit weaker X-ray emission. We further investigated this relation by taking advantage of the wide X-ray coverage of WISSH QSOs presented in Section \ref{sec:Xray_analysis} and the increase in the number of WISSH QSOs with measured {\sc Civ} velocity $\rm v_{CIV}$\footnote{$\rm v_{CIV}$ is defined as the velocity at the 50th percentile of the total flux.} (Vietri et al. in prep.). We did not consider BAL objects because of absorption features preventing a proper estimate of the {\sc Civ} emission line properties. As shown in Figure \ref{fig:vCIV_plots} (top panel), we confirm the presence of a strong $\rm (r_P = 0.77)$ and highly significant $\rm (p = 5 \times 10^{-6})$ correlation between $\rm Log(L_{2-10})$ and $\rm v_{CIV}$. In order to quantify the distribution in the $\rm Log(L_{2-10}) - v_{CIV}$ plane, we fitted a linear model to the data using the hierarchical Bayesian model \texttt{linmix} (see Section \ref{subsec:gamma_distributions}), resulting in $\rm Log(L_{2-10}) = (1.84 \pm 0.34) \times 10^{-4}v_{CIV} + (46.08 \pm 0.14)$.

We also confirm the presence of a robust correlation ($\rm p = 4\times 10^{-6}$; $\rm r_P = 0.82$) between $\rm \Delta(\alpha_{OX})$ and $\rm v_{CIV}$ for WISSH QSOs (Figure \ref{fig:vCIV_plots}, bottom panel). Using \texttt{linmix}, we obtain $\rm \Delta(\alpha_{OX}) = (7.01 \pm 1.19)\times 10^{-5}v_{CIV} + (0.22 \pm 0.05)$.

\section{Discussion}\label{sec:discussion} 

\subsection{Fraction of X-ray-weak sources among luminous QSOs}
WISSH QSOs are ideal sources for probing the extreme ends of $\rm L_{bol}$, $\rm L_{2500\,\AA}$ and $\rm \lambda L_{6\,\mu m}$ distributions, and they encompass the most extreme values of $\rm k_{bol}$, $\rm \alpha_{OX}$ and $\rm \lambda_{Edd}$ within the AGN population, as shown in Figures \ref{fig:Lbol_kBol}, \ref{fig:L2500_aox}, \ref{fig:eddRatio_gamma} and \ref{fig:Lmir_LX}. WISSH sources follow the $\rm k_{bol} - Log(L_{bol})$ and $\rm \alpha_{OX} - Log(L_{2500\,\AA})$ trends derived from samples of AGN at lower luminosities, confirming that $\rm L_{2-10}$ becomes relatively weak with increasing AGN luminosity. This suggests that the accretion disk $-$ X-ray corona coupling is not subject to a sudden change for the bulk of the most luminous AGN population, but instead evolves consistently with what predicted by the well-known $\rm k_{bol} - Log(L_{bol})$ and $\rm \alpha_{OX} - Log(L_{2500\,\AA})$ relations. Nevertheless, it is clear from Figures \ref{fig:Lbol_kBol}, \ref{fig:L2500_aox} and \ref{fig:Lmir_LX} that WISSH QSOs exhibit distinctive distributions. Due to their selection, these sources cover a narrow range of $\rm L_{bol}$, $\rm L_{2500\,\AA}$ and $\rm \lambda L_{6\,\mu m}$ values. Despite this, they show a highly dispersed distribution of $\rm k_{bol}$, $\rm \alpha_{OX}$ and $\rm \Delta_{6\,\mu m,X}$ values. It is worth noting that the complete X-ray coverage of WISSH has been crucial in revealing this feature, which was only suggested by \citet{Martocchia_2017} based on a smaller sample of sources with X-ray data. 

We adopted the threshold $\rm \Delta(\alpha_{OX}) \le -0.2$ from \citet{Luo_2015} to identify X-ray-weak QSOs. About 31\% of WISSH sources fall below the threshold: specifically, $\rm \approx 43\%$ of BAL and $\rm \approx 20\%$ of non-BAL QSOs (Figure \ref{fig:deltaAOX_distribution}). 
Therefore, although the fraction of intrinsically X-ray-weak sources is much higher in BAL QSOs, this phenomenon is also present in a sizeable fraction of normal QSOs, suggesting it must be considered relevant for the entire population of highly accreting and luminous AGN, in agreement to the results presented by \citet{Nardini_2019} and \citet{Laurenti_2022}. 

A large $\rm \lambda_{Edd}$ value seems to have a key role in triggering a weak state of the X-ray corona. The launch of powerful nuclear winds due to the increasing importance of radiation pressure can affect (and, possibly, remove) the innermost part of the accretion disk (i.e. the source of optical/UV seed photons) which is believed to be surrounded by the X-ray corona. X-ray weakness compared to UV emission is therefore fundamental to avoid nuclear gas over-ionisation, which would prevent radiation pressure from producing radiatively driven winds \citep[e.g. ][]{Proga_2003, Leighly_2004, Proga_2007, Xu_2024}. Figure \ref{fig:vCIV_plots} supports this scenario for WISSH objects, showing that the QSOs with the most negative $\rm \Delta(\alpha_{OX})$ values typically host the fastest nuclear ionised winds. 

The physical and geometrical properties of the inner accretion disk can be different in high-$\rm \lambda_{Edd}$ environments, as the enhanced radiation pressure driven by photon trapping favours the presence of slim (i.e. geometrically thick) accretion flow characterised by a low radiation efficiency \citep[e.g.][]{Abramowicz_1988, Cao_2022}. Furthermore, size and location of the X-ray corona itself could depend on the accretion rate and trigger intrinsic X-ray weakness due to a significant reduction of the active region and/or strong light bending, as suggested by \citet{Miniutti_2012}. Recent simulations of high-$\rm \lambda_{Edd}$ AGN presented by \citet{Pacucci_2024} show that the X-ray SED may depend on the SMBH spin and on the inclination of our line of sight, due to the presence of a puffed-up structure of the hot inner accretion flow in these sources. Slowly spinning SMBHs are associated with the X-ray-weakest sources with extremely steep (i.e. $\rm \Gamma > 3$) SEDs. In case of spinning SMBHs, the effect of the viewing angle is important for observing stronger X-ray emission and a continuum with flatter slopes as the inclination decreases. \citet{Inayoshi_2024} recently argued that X-ray weakness of AGN with high-$\rm \lambda_{Edd}$ accretion disks can be related to the presence of a warm X-ray corona with large optical depth which produces a significant softening of the X-ray continuum, leading to hard X-ray flux reduction. This steepening of the X-ray spectral slope is not observed in WISSH QSOs (see Figure \ref{fig:deltaAOX_gamma_both}). However, most of the X-ray-weak sources in the WISSH sample do not have a measured $\rm \Gamma$ and, therefore, it is not possible to completely rule out this scenario. Similarly, we cannot exclude that part of the X-ray continuum emission is scattered off our line of sight by a highly ionised medium with large $\rm N_H$ and small covering factor, as suggested by \citet{Laurenti_2022}. 

Finally, multiple X-ray observations of some WISSH QSOs offered the opportunity to detect possible transitions from an intrinsic X-ray-weak state to a normal state (and vice versa) over very different timescales (i.e. from a few days to years). However, none of those sources underwent such change, as reported in Appendix \ref{apdx:variability}. Future well-designed campaigns to monitor the X-ray emission of WISSH QSOs will be useful to shed light on the occurrence of these transitions, their timescales, the fraction of persistent, intrinsically X-ray-weak sources among the most luminous QSOs, and the duration of the X-ray-weak phase. These insights are essential for providing constraints, which are currently unavailable, to improve our understanding of the accretion disk $-$ corona system in AGN with high accretion rates on SMBHs with $\rm M_{BH} > 10^9 \,\,M_\odot$.

\subsection{The $ \Gamma - Log(\lambda_{Edd})$ relation}
The claim of a strong and tight correlation between $\rm \Gamma$ and $\rm \lambda_{Edd}$ has been reported by many papers in the last two decades \citep[e.g.][]{Shemmer_2008, Risaliti_2009, Brightman_2013, Liu_2021}. This has garnered great interest, as it offers the remarkable opportunity to estimate $\rm M_{BH}$ from high-quality X-ray observations for a large number of AGN and, potentially, even for narrow-line Type 2 objects, for which single-epoch relations cannot be used \citep[but see][]{Ricci_2022}. A popular interpretation for this relation is that the enhanced flux of UV seed photons from the accretion disk in highly accreting AGN causes a stronger cooling of the X-ray corona and, in turn, a softening of the X-ray spectrum. However, the existence of a strong dependence of $\rm \Gamma$ on $\rm \lambda_{Edd}$ has been questioned by \citet{Trakhtenbrot_2017} based on their study of a large sample of local AGN drawn from the BASS survey, which benefits from broad-band, high-quality X-ray spectral data. Specifically, they reported a weak correlation with a flatter slope compared to previous studies, and emphasised the significant amount of scatter in the $\rm \Gamma$ distribution. \citet{Laurenti_2022} confirmed such large scatter in their investigation of the X-ray spectral properties of a sample of $z \rm \approx 0.3-0.6$ QSOs with $\rm \lambda_{Edd} \approx 1$. 

We studied $\rm \Gamma$ as a function of $\rm \lambda_{Edd}$ for the 19 sources with both $\rm \ge 20$ X-ray counts and available H$\rm \beta$-derived $\rm M_{BH}$. It is evident from Figure \ref{fig:eddRatio_gamma} that $\rm \Gamma$ values around $\rm Log(\lambda_{Edd}) \approx 0$ are highly dispersed and no clear correlation is observed. Indeed, a substantial fraction $\rm (\approx 30\%)$ of WISSH QSOs with $\rm \ge 20$ counts exhibit intrinsically flatter $\rm \Gamma$ values $\rm (\Gamma \approx 1.2 - 1.7)$ than those predicted by previously published relations \citep[e.g. ][]{Risaliti_2009, Shemmer_2008, Brightman_2013}. These relations typically predict $\rm \Gamma \approx 2 - 2.3$ for AGN with $\rm Log(\lambda_{Edd}) \approx 0$. Our result extends the findings presented by \citet{Martocchia_2017} and is similar to those reported by \citet{Trefoloni_2023} for a sample of optically bright, luminous QSOs at Cosmic Noon. 

It is worth noting that recent studies have shown that, in case of luminous QSOs and highly accreting-AGN, the real size of the BLR could likely be smaller than the one derived from the popular BLR radius $-$ luminosity relationships \citep[e.g.][]{Du_2015, Du_2016, GRAVITY_2024, Li_2025}. Therefore, $\rm \lambda_{Edd}$ of WISSH QSOs in Figure \ref{fig:eddRatio_gamma} could be underestimated by a factor of a few, as they were derived using Equation \ref{eq:singleEpoch}. This would make the mismatch with the $\rm \Gamma - Log(\lambda_{Edd})$ relations in Figure \ref{fig:eddRatio_gamma} even more evident.

The wide range of $\rm \Gamma$ values may be a further indication of possible differences in the physical and geometrical properties of the inner accretion disk and the hot corona in the QSO population \citep[e.g.][]{Kubota_2018}, as well as differences in the micro-meso feeding mechanisms \citep[e.g. Bondi-like vs. chaotic cold accretion;][]{Gaspari_2017}. In this context, the findings presented by \citet{Laurenti_2024} are particularly relevant. This recent study considered several samples of AGN with high-quality X-ray spectra to populate the $\rm \Gamma -  Log(\lambda_{Edd})$ plane with sources spanning a wide range of redshifts, black hole masses and luminosities. By investigating a possible dependence on $\rm M_{BH}$ or $\rm L_{bol}$, \citet{Laurenti_2024} reported that the $\rm \Gamma -  Log(\lambda_{Edd})$ correlation is significant only for objects with either $\rm M_{BH} \lesssim 10^8\,\,M_\odot$ or $\rm L_{bol} \lesssim 10^{45.5}\,\,erg\,s^{-1}$, i.e. Seyfert-like AGN. This result, combined with the limited quality and spectral range of early studies on high-$z$ QSOs, may provide an explanation for the contradictory findings reported in these earlier works compared to those obtained for WISSH and other recent studies on luminous QSOs.

\subsection{Blue QSOs in the forbidden region of the $\rm Log(N_H) - Log(\lambda_{Edd})$ plane}
The $\rm Log(N_H) - Log(\lambda_{Edd})$ plane can be interpreted in the framework of AGN evolution. Consequently to a wet merger (i.e. between gas-rich galaxies), QSOs likely face a heavily obscured, high-Eddington accretion phase, which triggers intense feedback processes, eventually sweeping up most of nuclear gas \citep[e.g.][]{Gaspari_2014}. Then, sources enter an unobscured regime during which less-intense accretion and star formation occur, evolving towards a passive red galaxy phase \citep[e.g.][]{Hopkins_2008b, Hopkins_2008a}.

Dust-obscured red QSOs (see Section \ref{subsec:NH_Edd}), thought to be experiencing the obscured-accretion phase in the AGN evolutionary scenario, are expected to occupy the forbidden area in the $\rm Log(N_H) - Log(\lambda_{Edd})$ plane \citep[e.g. ][]{Stacey_2022, Glikman_2024}. We find that two WISSH red QSOs with available $\rm N_H$ and H$\rm \beta$-derived $\rm M_{BH}$ (highlighted with encircled symbols in Figure \ref{fig:eddRatio_nh}) are likely located in the blow-out region: in particular, we notice the presence of WISSH58, which is well constrained to lie in the forbidden region, and WISSH34, which shows an extreme $\rm N_H$ upper limit (i.e. $\rm N_H \le 5 \times 10^{23}\,\,cm^{-2}$).

Interestingly, all of the five blue WISSH sources (both BAL and non-BAL) with significant X-ray absorption also occupy the forbidden region. This number might be even larger since there are other blue QSOs with an upper limit on $\rm N_H$ consistent with $\rm 10^{22}\,\,cm^{-2}$ or higher. Future deeper X-ray observations will be able to shed light on their nuclear properties. These blue QSOs in the blow-out phase possibly share the feedback and environment properties of red objects, and may represent an intermediate phase of the transition between red obscured QSOs and blue unobscured sources. A different nature for obscuring material in red and blue QSOs has also been suggested: dust-rich large-scale gas would be responsible for the extinction of the former, while dust-free small-scale medium would be the cause of obscuration for the latter \citep[e.g. ][]{Maiolino_2001, Mizukoshi_2024}. Blue AGN in the forbidden region were also reported by \citet{Ballo_2014}. Unlike WISSH QSOs, they found narrow-line AGN at $z \rm \approx 0.5 - 1$, highlighting the fact that the forbidden region may be populated by sources with very different nuclear properties.

In particular, among the blue QSOs in the forbidden region, we find non-BAL QSOs. The non-BAL nature of these sources suggests a different origin for the X-ray absorber rather than being outflowing gas, as is typically the case in BAL objects. On the other hand, they possibly represent a late stage of the blow-out phase. Assuming the median $\rm \lambda_{Edd}$ value derived for WISSH QSOs with measured $\rm M_{BH}$ as representative for the sources with no $\rm M_{BH}$ (given the narrow $\rm Log(\lambda_{Edd})$ distribution of WISSH objects), in the HC-WISSH sample we find that seven (one of which is X-ray weak) and five more sources fall within and below the forbidden area, respectively.

\subsection{Comparison with $z$ > 6 QSOs}
The WISSH sample naturally bridges the gap between low-$z$ QSOs and the most distant luminous QSOs detected at the Epoch of Reionisation. Consequently, it is particularly interesting to compare their X-ray properties with those derived for $z \rm \ge 6$ QSOs. \citet{Tortosa_2024} recently published the results of the X-ray spectral analysis of the 21 QSOs at $z \rm > 6$ with the best X-ray coverage available so far (i.e. detected with at least 30 counts). These highly accreting objects show $\rm M_{BH} \approx 10^9 - 10^{10}\,\,M_\odot$ and $\rm L_{bol} \approx 10^{47} - 10^{48}\,\, erg \,s^{-1}$ and, therefore, can be meaningfully compared with WISSH objects.

\begin{figure*}
    \centering
    \subfloat[][\label{fig:L210_z6}]{\includegraphics[width=0.33\textwidth]{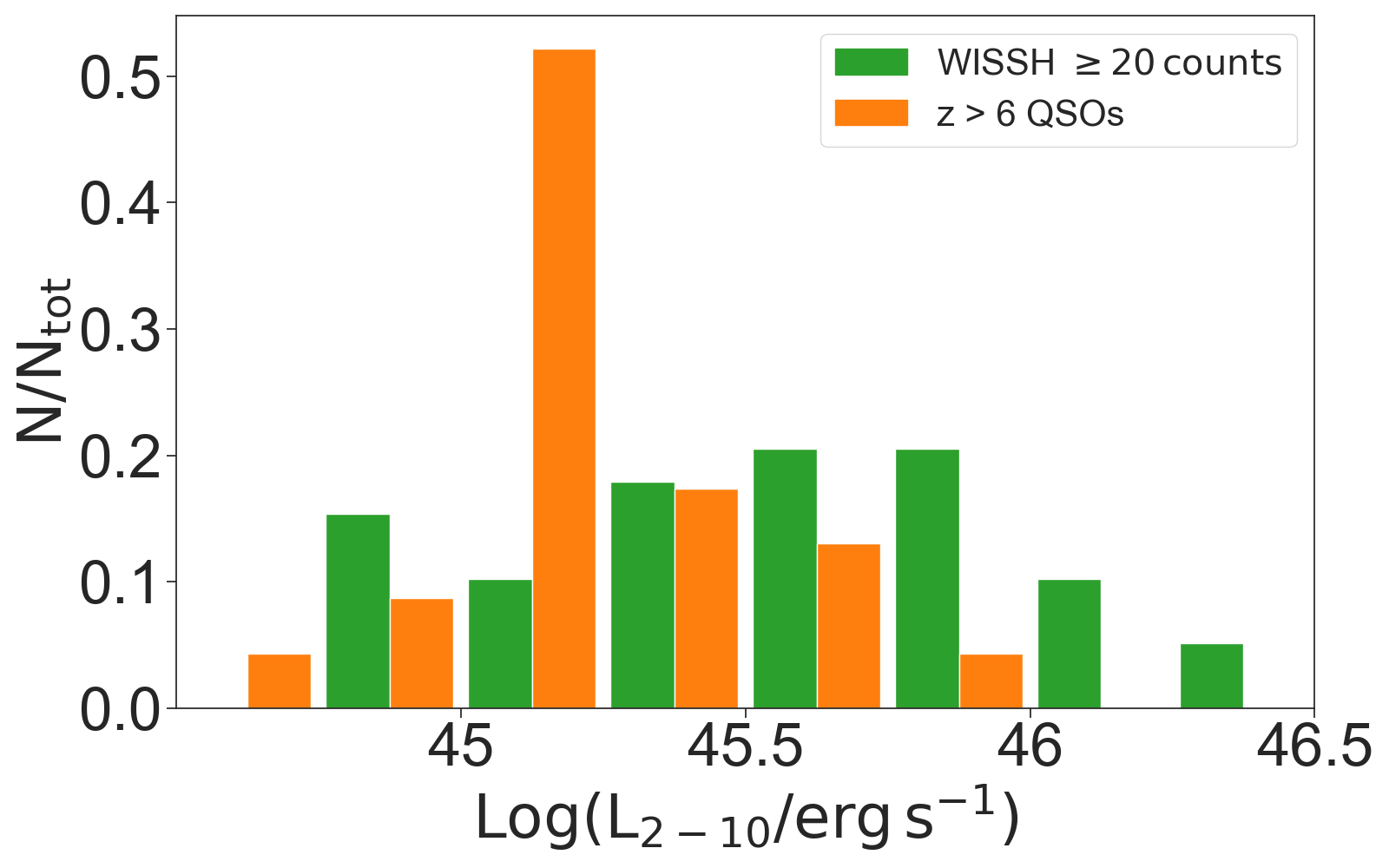}} \enspace
    \subfloat[][\label{fig:kBol_z6}]{\includegraphics[width=0.33\textwidth]{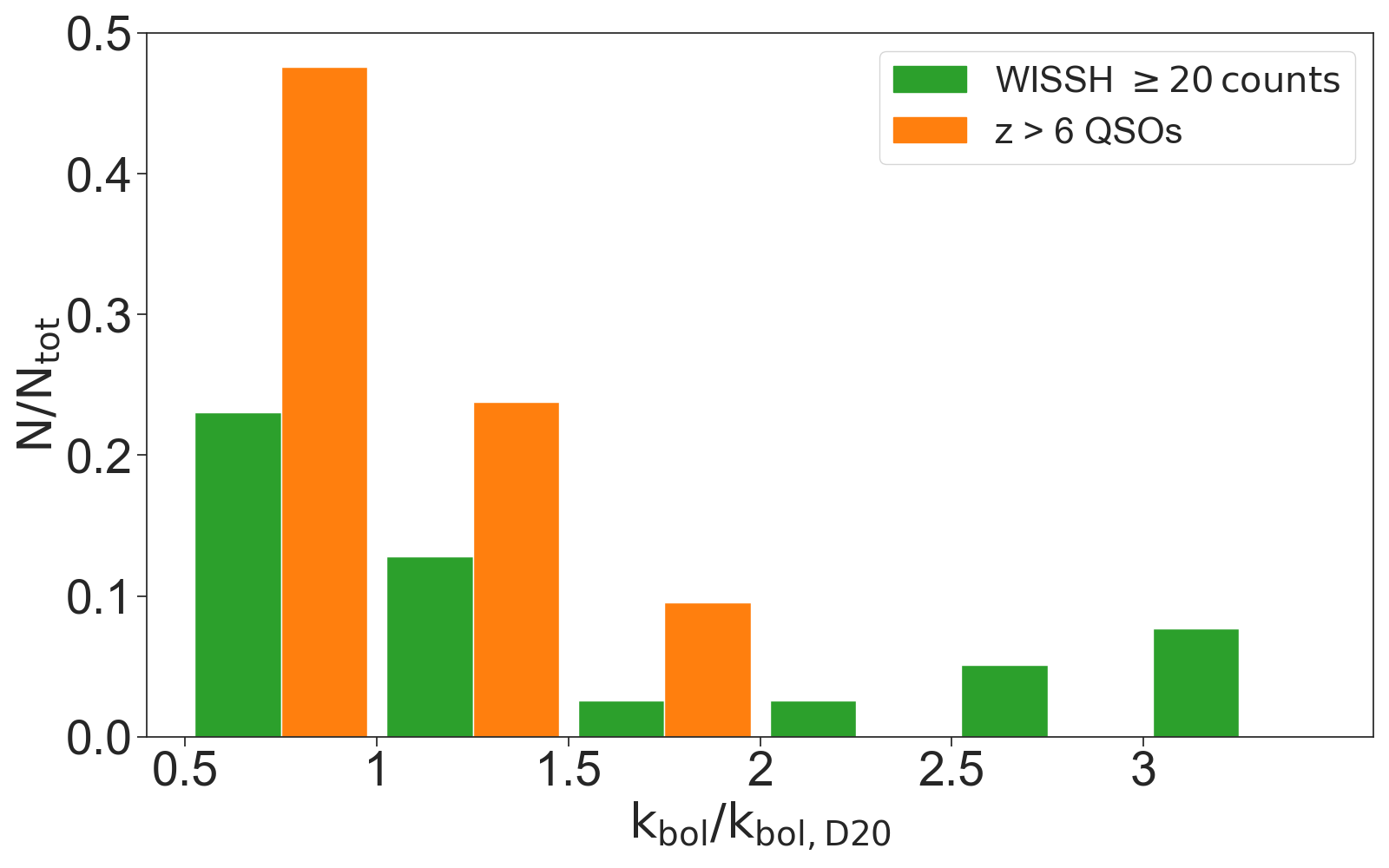}} \enspace
    \subfloat[][\label{fig:gamma_z6}]{\includegraphics[width=0.33\textwidth]{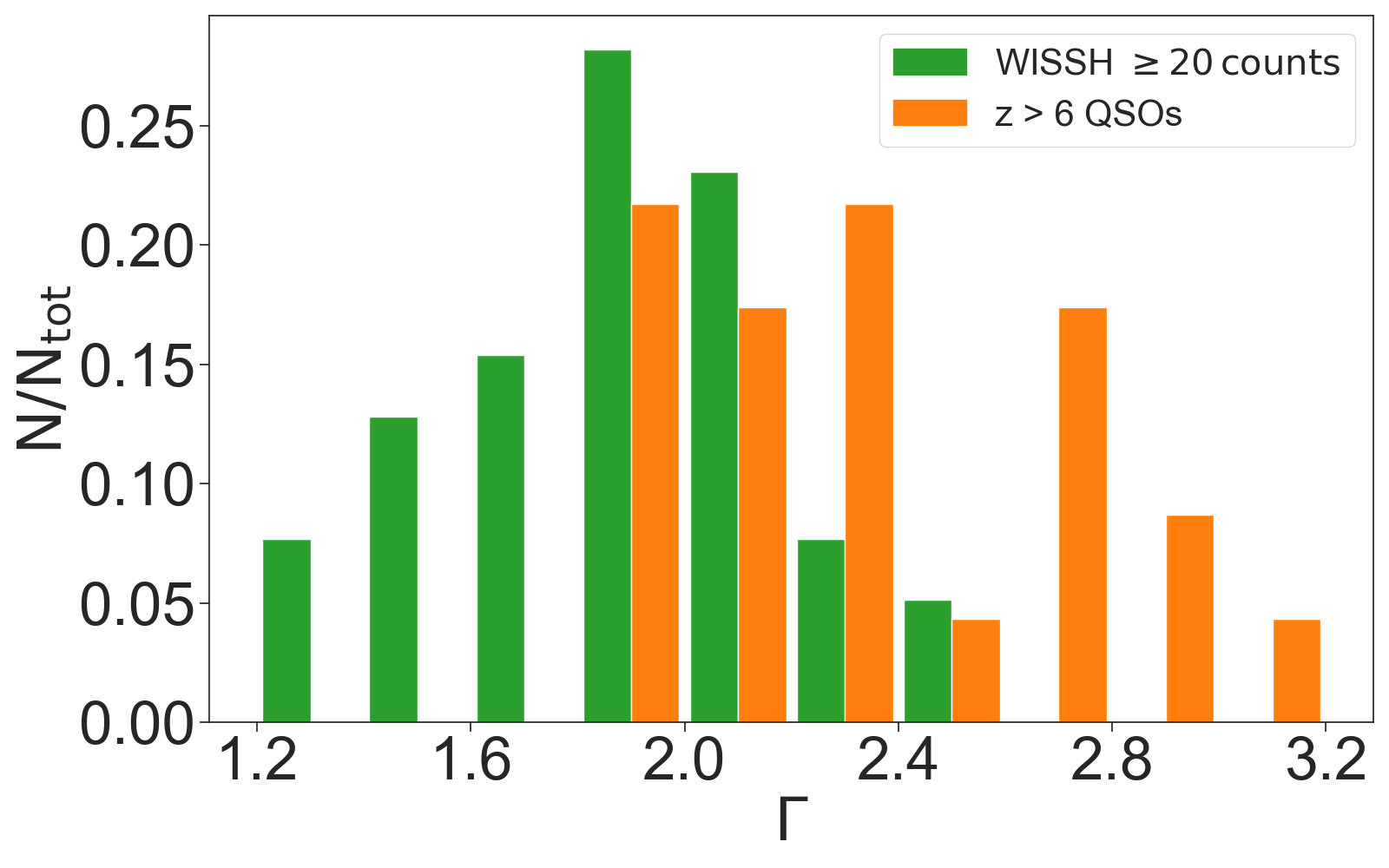}}
    \caption{(a) Intrinsic $\rm 2 - 10\,\,keV$ luminosity, (b) $\rm k_{bol}/k_{bol,D20}$, and (c) photon index distributions. The histograms show the comparison between WISSH sources with $\rm \ge 20$ counts (green) and a sample of QSOs at $z \rm > 6$ from \citet{Tortosa_2024} (orange).}
    \label{fig:hyperion}
\end{figure*}

Figure \ref{fig:hyperion} shows the distributions of $\rm L_{2-10}$, $\rm k_{bol}/k_{bol, D20}$, and $\rm \Gamma$ for the two samples. They approximately span the same range of intrinsic X-ray luminosity, i.e. $\rm 10^{45} \lesssim L_{2-10}/erg\,s^{-1} \lesssim 10^{46}$, although most of $z \rm > 6$ objects show $\rm L_{2-10} \approx (1-3) \times 10^{45}\,\, erg\, s^{-1}$ (Figure \ref{fig:L210_z6}). This results in a more clustered $\rm k_{bol}/k_{bol, D20}$ distribution for $z \rm > 6$ QSOs (Figure \ref{fig:kBol_z6}). Consequently, the large fraction of intrinsically X-ray-weak sources in the WISSH sample does not seem to be present in these QSOs shining in the early Universe (less than 1 Gyr old).

The distributions of the X-ray photon index are strikingly different: indeed, the distant QSOs typically exhibit $\rm \Gamma \ge 2.2$, with an average value of $\rm \Gamma \approx 2.40 \pm 0.42$\footnote{\label{nota:dispersion} The reported error is the dispersion of the distribution.} from \citet{Tortosa_2024} sample, which is not consistent with that derived for WISSH sources, showing a mean slope of $\rm 1.88 \pm 0.29$\textsuperscript{\ref{nota:dispersion}} (Figure \ref{fig:gamma_z6}). This confirms the intriguing presence of a significantly steeper X-ray continuum in $z \rm > 6$ QSOs compared to AGN at lower $z$, regardless of their bolometric luminosity, as pointed out by \citet{Zappacosta_2023}. \citet{Madau_2024} recently proposed peculiar properties of the X-ray corona to account for such steep $\rm \Gamma$ in distant QSOs. In particular, they suggested that in these high-$\rm \lambda_{Edd}$ sources undergoing a specific evolutionary phase, the corona may be embedded in a funnel-like geometry which enhances the density of seed UV photons, resulting in a cooler corona and a steeper X-ray slope than in normal AGN, for which the source of seed photon is only the underlying disk. However, another prediction of the \citet{Madau_2024} scenario is the intrinsic X-ray weakness of these high-$z$/high-$\rm \lambda_{Edd}$ objects, which is instead not observed for QSOs in the \citet{Tortosa_2024} sample. Interestingly, bearing in mind the limited number of sources, X-ray-weak WISSH QSOs shown in Figure \ref{fig:deltaAOX_gamma_both} are not characterised by a steep X-ray continuum.

\section{Conclusions and future perspectives}\label{sec:conclusions}
The present work constitutes one of the largest systematic investigations of the X-ray properties of sources at the brightest end of the AGN luminosity function. We presented the results of X-ray observations of all 85 QSOs in the WISSH sample. About $\rm \approx 90\%$ of these Type 1 sources were detected. We successfully performed X-ray spectral analysis on approximately one-half of the sample. For 16 sources, we used the hardness ratio analysis to derive the intrinsic column density $\rm N_H$, and thus the $\rm 2-10\,\,keV$ intrinsic luminosity $\rm L_{2-10}$. For the remaining objects, we estimated $\rm L_{2-10}$ assuming $\Gamma$ = 1.8 and $\rm N_H = 5 \times 10^{22}\,\,cm^{-2}$, which is the median value measured for the absorbed sources in the HC-WISSH sample (see Section \ref{sec:Xray_analysis}). Our main results can be summarised as follows: 

\begin{itemize}
    \item We estimated the presence of intrinsic absorption for the sources belonging to the HC-WISSH sample ($\rm \approx 65\%$ of the entire WISSH sample). The vast majority of them exhibit little to no obscuration (i.e. $\rm N_H \le 10^{22}\,\,cm^{-2}$). We populated the $\rm Log(N_H) - Log(\lambda_{Edd})$ plane with the HC-WISSH sources for which H$\beta$-based $\rm M_{BH}$ are estimated. Remarkably, we find all five blue WISSH QSOs, both broad absorption line (BAL) and non-BAL sources, showing significant X-ray absorption within the forbidden area of this plane. This region is indeed typically occupied by dust-reddened QSOs and is associated with an early short-lived blow-out evolutionary phase driven by intense feedback processes.
    \item The complete X-ray coverage of the WISSH sample allows us to perform a systematic investigation of X-ray variability at the highest bolometric luminosity $\rm (L_{bol})$ at Cosmic Noon. Variability in the observed $\rm 0.5-2\,\,keV$ and $\rm 2 - 10 \,\,keV$ flux ($\rm F_{0.5-2}$ and $\rm F_{2-10}$, respectively; with statistical significance $\rm > 3\sigma$) is found in three QSOs. $\rm F_{0.5-2}$-only variability ($\rm > 3\sigma$ significance level) is found in four QSOs. $\rm N_H$ variability ($\rm > 2\sigma$ significance level) down to 40-day rest-frame timescales is found in two QSOs, which may be considered good candidates for the changing look QSO population (thus far mostly sampled at lower luminosities and redshifts; e.g. \citealp{Risaliti_2009}; but see also \citealp{Vito_2025}).
    \item WISSH QSOs broadly follow the well-established trends in the $\rm k_{bol} - Log(L_{bol})$ and $\rm \alpha_{OX} - Log(L_{2500\,\AA})$ planes. They predict  progressively decreasing ratios of $\rm L_{2-10}$ to $\rm L_{bol}$, $\rm L_{2500\,\AA}$, and $\rm \lambda L_{6\,\mu m}$, which indicate a diminishing relative contribution of X-ray luminosity to the UV and bolometric values with respect to lower-luminosity samples. However, the distribution of $\rm k_{bol}$ and $\rm \alpha_{OX}$ values reported for WISSH sources is strikingly broad, suggesting that caution should be exercised when using $\rm L_{bol}$, $\rm L_{2500\,\AA}$, and $\rm \lambda L_{6\,\mu m}$ to estimate X-ray emission of individual luminous QSOs $\rm (L_{bol} \ge 10^{47}\,\,erg\,s^{-1})$.
    \item About one-third of  WISSH QSOs exhibit $\rm \Delta(\alpha_{OX}) \le -0.2$, thus classifying them as intrinsically X-ray-weak AGN (assuming that absorption is totally accounted for). Specifically, the fraction of X-ray-weak sources is $\rm \approx 47\%$ among our BAL QSOs and $\rm \approx 20\%$ among non-BAL QSOs (see Figure \ref{fig:deltaAOX_distribution}).
    \item Once we define $\rm \Delta_{6\,\mu m, X}$ as the difference between $\rm L_{2-10}$ and the X-ray luminosity expected from the \citetalias{Stern_2015} relation $\rm (L_{2-10, S15})$, a very tight correlation emerges in the $\rm \Delta(\alpha_{OX}) - \Delta_{6\mu m, X}$ plane (see Section \ref{subsec:MIR}). Therefore, if $\rm L_{2500\,\AA}$ and $\rm \lambda L_{6\,\mu m}$ are known, this can be used to robustly derive $\rm L_{2-10}$ from the X-ray$-$to$-$MIR comparison. Moreover, the comparison between $\rm L_{2-10}$ and the observed $\rm 2 - 10\,\,keV$ luminosity can provide an estimate of the intrinsic $\rm N_H$.
    \item The distribution of X-ray continuum slope of WISSH QSOs is similar for AGN at lower $z$ and lower $\rm L_{bol}$ (with an average value of $\rm \Gamma \approx 1.9$), and does not depend on X-ray weakness, suggesting that un-modelled extra-absorption is not the primary cause for X-ray weakness in our sample. Among objects at Cosmic Noon, we confirm the extreme rarity of $\rm \Gamma$ values as steep as those commonly found in the limited sample of luminous QSO at $z \rm  > 6$ observed with good X-ray statistics to date.
    \item WISSH QSOs, for which both $\rm \Gamma$ and H$\beta$-based $\rm M_{BH}$ have been measured, exhibit a significant dispersion in the $\rm \Gamma - Log(\lambda_{Edd})$ plane (see Figure \ref{fig:eddRatio_gamma}).
    \item We confirm the existence of a significant correlation between $\rm L_{2-10}$ and {\sc Civ} emission line velocity $\rm v_{CIV}$ (see Section \ref{subsec:CIV_shift}), as previously claimed by \citet{Zappacosta_2020} for a smaller sub-sample of WISSH QSOs.
\end{itemize} 
 
The diversity of X-ray emission (in $\rm L_{2-10}$ and in $\rm \Gamma$) shown by WISSH QSOs, in contrast to their narrow distributions in bolometric, UV, and MIR luminosity, clearly points to a range of physical and geometrical properties of the inner accretion flow in these luminous highly accreting AGN. Gaining a deeper understanding of the causes behind this diversity is therefore a priority for future studies on luminous QSOs. To shed new light on this, a promising  approach is performing high-quality X-ray spectroscopy and monitoring of the X-ray-weakest sources $\rm (k_{bol} \gtrsim 500-600)$ to investigate potential spectral and temporal behaviours characteristic of this peculiar AGN phase. It is important to note that recent models proposed to explain the X-ray weakness of high-$\rm \lambda_{Edd}$ sources predict very steep spectral slopes (i.e. $\rm \Gamma > 3$), which are not observed in our sample. However, these models typically consider AGN with $\rm M_{BH} \approx 10^{6-7}\,\,M_\odot$, i.e. several orders of magnitude lower than those reported for WISSH QSOs. As suggested by \citet{Laurenti_2024}, $\rm M_{BH}$ may be a key parameter in determining the X-ray spectral properties of high-$\rm \lambda_{Edd}$ AGN. Furthermore, the comparison with the sample of $z \rm >6$ QSOs  by \citet{Tortosa_2024}, which covers $\rm M_{BH}$ and $\rm L_{bol}$ ranges  similar to the WISSH QSOs ranges, indicates significant differences in their X-ray spectral properties and, in turn, a possible evolution in the inner accretion flow within the population of highly accreting QSOs powered by $\rm M_{BH} > 10^9\,\, M_\odot$ SMBHs along cosmic time. 

A deeper understanding of the nuclear properties driving the $\rm Log(L_{2-10}) - v_{CIV}$ relation shown in Figure \ref{fig:vCIV_plots} is highly desirable as it links accretion and ejection phenomena over accretion disk scales. The most powerful ionised winds are typically found in sources with the lowest $\rm L_{2-10}$ and $\rm \Delta(\alpha_{OX})$ values, which may indicate that a strong X-ray output could significantly impact the launch of UV winds due to over-ionisation of the gas surrounding the continuum source. Interestingly, \citet{Rankine_2020} and \citet{Yi_2020} recently proposed a scenario whereby blueshifted {\sc Civ} emission and BALs could represent different manifestations of the same outflow seen at different lines of sight. Our results seems to support this scenario as the fraction of X-ray-weak sources among WISSH QSOs with BAL or high $\rm v_{CIV}$ is similarly high.

The results emerging from our extensive investigation of WISSH QSO X-ray properties triggered the recently awarded \textit{XMM-Newton} Heritage and Multi-year `WISSH QSOs-Focused UFO Legacy sample (WISSHFUL)' programme (2.3 Ms; PI: G. Lanzuisi), which also benefits from (quasi-)simultaneous \textit{NuSTAR} coverage. Specifically, the WISSHFUL programme will target the 15 X-ray brightest WISSH QSOs with allocated long exposure times ranging from 60 to 220 ks per target. This project will allow us to place stronger constraints on the X-ray continuum (hence $\rm \Gamma$ and $\rm N_H$ distributions) of WISSH QSOs, detect X-ray outflowing winds using good photon-statistics data, and extend the variability studies over long timescales.

Finally, thanks to the ongoing multi-frequency radio campaign using JVLA and GMRT observations for the entire WISSH sample (PI: G. Bruni), it will be possible to conduct an innovative study of their nuclear properties by combining X-ray, optical, and radio data for a large sample of highly luminous AGN at Cosmic Noon.

\begin{acknowledgements}
The scientific results reported in this article are based to a significant degree on observations made by the \textit{Chandra} X-ray Observatory, and \textit{XMM-Newton}, an ESA science mission with instruments and contributions directly funded by ESA Member States and NASA. The research has made use of data obtained from the \textit{Chandra} Data Archive provided by the \textit{Chandra} X-ray Center (CXC). Data analysis was performed with the {\it XMM-Newton} SAS and CXC CIAO software packages. Support for this work was provided by the National Aeronautics and Space Administration through Chandra Award Number GO2-23087X issued by the Chandra X-ray Center, which is operated by the Smithsonian Astrophysical Observatory for and on behalf of the National Aeronautics Space Administration under contract NAS8-03060. We acknowledge financial support from the Bando Ricerca Fondamentale INAF Large Grant 2022 ``Toward an holistic view of the Titans: multi-band observations of $z \rm > 6$ QSOs powered by greedy supermassive black holes''. 

We thank the anonymous referee for the constructive comments that helped to improve the manuscript. We also acknowledge E. Ros for the useful suggestions.

FT and ML acknowledges funding from the European Union $-$ Next Generation EU, PRIN/MUR 2022 (2022K9N5B4). 

MG acknowledges support from the ERC Consolidator Grant \textit{BlackHoleWeather} (101086804).

EB acknowledges financial support from INAF under the Large Grant 2022 ``The metal circle: a new sharp view of the baryon cycle up to Cosmic Dawn with the latest generation IFU facilities''. 

FS acknowledges financial support from the PRIN MUR 2022 2022TKPB2P $-$ BIG-z, Ricerca Fondamentale INAF 2023 Data Analysis grant ``ARCHIE ARchive Cosmic HI \& ISM  Evolution'', Ricerca Fondamentale INAF 2024 under project ``ECHOS'' MINI-GRANTS RSN1.

MB acknowledges support from INAF project 1.05.12.04.01 $-$ MINI-GRANTS di RSN1 ``Mini-feedback'' and from UniTs under FVG LR 2/2011 project D55-microgrants23 ``Hyper-gal''.

GM is funded by Spanish MICIU/AEI/10.13039/501100011033 and ERDF/EU grant PID2023-147338NB-C21.

CP acknowledge funding by the European Union $-$ Next Generation EU, Mission 4 Component 1 CUP C53D23001330006.

EG acknowledges the generous support of the Cottrell Scholar Award through the Research Corporation for Science Advancement. EG is grateful to the Mittelman Family Foundation for their generous support.

LZ acknowledges support from the European Union – Next Generation EU, PRIN/MUR 2022 2022TKPB2P – BIG-z. LZ acknowledges partial support by grant NSF PHY-2309135 to the Kavli Institute for Theoretical Physics (KITP).

\end{acknowledgements}


\bibliographystyle{aa}
\bibliography{aa54943-25}

\begin{appendix}
\onecolumn

\section{Log of the X-ray observations}
\begin{table}[h!]
    \caption{Log of the X-ray observations considered in this study.}
    \label{tab:logObs}
    \centering
    \fontsize{9}{9}\selectfont
    \begin{tabular}{cccc}
    \toprule
    WISSH ID & Obs. ID & Exposure (ks) & Obs. date \\
    \midrule
WISSH01 &  6889\tablefootmark{a}  &  11.4   & 2006-07-24 \\
WISSH02 &  25318\tablefootmark{$\dagger$} &  4.02   & 2022-06-04 \\
WISSH03 &  25319\tablefootmark{$\dagger$} &  4.03   & 2022-05-22 \\
WISSH04 &  17078\tablefootmark{a} &  29.69  & 2014-11-18 \\
        &  6820\tablefootmark{a}  &  2.17   & 2005-12-02 \\
WISSH05 &  25320\tablefootmark{$\dagger$} &  3.89   & 2022-07-03 \\
WISSH06 &  25321\tablefootmark{$\dagger$} &  4.03   & 2022-11-23 \\
WISSH07 &  17077\tablefootmark{a} &  24.76  & 2015-10-02 \\
        &  6829\tablefootmark{a}  &  3.98   & 2005-12-06 \\
WISSH08 &  13308\tablefootmark{a} &  1.54   & 2012-01-01 \\
WISSH09 &  0844970601\tablefootmark{b} &  47.16  & 2020-03-29 \\
        &  3561\tablefootmark{a}  &  4.96   & 2002-12-03 \\
WISSH10 &  17081\tablefootmark{a} &  43.5   & 2014-12-11 \\
        &  10733\tablefootmark{a} &  4.1    & 2009-01-12 \\
WISSH11 &  25322\tablefootmark{$\dagger$} &  4.04   & 2022-01-27 \\
WISSH12 &  25323\tablefootmark{$\dagger$} &  8.97   & 2022-01-02 \\
WISSH13 &  0803950601\tablefootmark{b} &  10.4   & 2017-11-17 \\
        &  6810\tablefootmark{a}  &  3.91   & 2006-02-09 \\
WISSH14 &  0745010301\tablefootmark{b} &  9.1    & 2014-11-24 \\
        &  13336\tablefootmark{a} &  1.54   & 2011-12-08 \\
WISSH15 &  25324\tablefootmark{$\dagger$} &  8.94   & 2022-12-18 \\
WISSH16 &  25325\tablefootmark{$\dagger$} &  3.91   & 2022-01-29 \\
WISSH17 &  0803950801\tablefootmark{b} &  21.4   & 2017-04-28 \\
        &  13325\tablefootmark{a} &  1.56   & 2012-05-28 \\
WISSH18 &  25326\tablefootmark{$\dagger$} &  4.03   & 2022-05-22 \\
WISSH19 &  20444\tablefootmark{a} &  29.68  & 2018-09-30 \\
        &  21863\tablefootmark{a} &  11.08  & 2018-10-03 \\
WISSH20 &  25327\tablefootmark{$\dagger$} &  9.26   & 2023-01-03 \\
WISSH21 &  25328\tablefootmark{$\dagger$} &  4.03   & 2022-12-18 \\
WISSH22 &  0803950201\tablefootmark{b} &  21.6   & 2017-04-16 \\
        &  6809\tablefootmark{a}  &  4.13   & 2006-06-14 \\
WISSH23 &  25329\tablefootmark{$\dagger$} &  9.53   & 2022-01-19 \\
WISSH24 &  25330\tablefootmark{$\dagger$} &  4.04   & 2022-01-29 \\
WISSH25 &  23755\tablefootmark{$\dagger$} &  9.92   & 2021-03-02 \\
WISSH26 &  25331\tablefootmark{$\dagger$} &  4.03   & 2023-03-08 \\
WISSH27 &  0803950401\tablefootmark{b} &  16.2   & 2017-10-28 \\
        &  13312\tablefootmark{a} &  1.56   & 2012-04-02 \\
WISSH28 &  25332\tablefootmark{$\dagger$} &  8.82   & 2022-09-29 \\
WISSH29 &  25333\tablefootmark{$\dagger$} &  8.96   & 2022-02-28 \\
WISSH30 &  878\tablefootmark{a}   &  2.81   & 2000-06-14 \\
WISSH31 &  25334\tablefootmark{$\dagger$} &  4.13   & 2023-03-08 \\
WISSH32 &  25335\tablefootmark{$\dagger$} &  4.04   & 2022-01-23 \\
WISSH33 &  6811\tablefootmark{a}  &  3.65   & 2006-07-16 \\
        &  0553561401\tablefootmark{b} &  1.2    & 2008-11-29 \\
WISSH34 &  25336\tablefootmark{$\dagger$} &  3.89   & 2023-03-08 \\
WISSH35 &  0104861001\tablefootmark{b} &  24.7   & 2002-06-01 \\
        &  0059750401\tablefootmark{b} &  9.5    & 2002-04-25 \\
WISSH36 &  25337\tablefootmark{$\dagger$} &  4.03   & 2022-10-11 \\
WISSH37 &  17082\tablefootmark{a} &  43.06  & 2015-01-26 \\
        &  13321\tablefootmark{a} &  1.56   & 2012-01-31 \\
WISSH38 &  25338\tablefootmark{$\dagger$} &  4.03   & 2022-12-17 \\
WISSH39 &  25339\tablefootmark{$\dagger$} &  4.03   & 2022-12-28 \\
WISSH40 &  0865210601\tablefootmark{b} &  10.9   & 2020-11-24 \\
WISSH41 &  13323\tablefootmark{a} &  1.56   & 2012-06-29 \\
WISSH42 &  13309\tablefootmark{a} &  1.47   & 2012-03-18 \\
WISSH43 &  0803952201\tablefootmark{b} &  33.6   & 2017-06-06 \\
        &  13345\tablefootmark{a} &  1.56   & 2012-02-10 \\
WISSH44 &  13324\tablefootmark{a} &  1.56   & 2012-06-11 \\
WISSH45 &  25340\tablefootmark{$\dagger$} &  4.03   & 2023-06-04 \\
WISSH46 &  25341\tablefootmark{$\dagger$} &  4.03   & 2023-04-03 \\
        &  13366\tablefootmark{a} &  1.56   & 2012-07-09 \\
    \bottomrule
    \end{tabular}
    \quad
    \begin{tabular}{cccc}
    \toprule
    WISSH ID & Obs. ID & Exposure (ks) & Obs. date \\
    \midrule
WISSH47 &  4201\tablefootmark{a}  &  44.52  & 2003-11-14 \\
WISSH48 &  25342\tablefootmark{$\dagger$} &  4.03   & 2022-12-14 \\
WISSH49 &  25343\tablefootmark{$\dagger$} &  3.85   & 2023-03-06 \\
WISSH50 &  20443\tablefootmark{a} &  44.91  & 2018-07-29 \\
        &  6817\tablefootmark{a}  &  4.01   & 2006-08-29 \\
WISSH51 &  25344\tablefootmark{$\dagger$} &  4.03   & 2023-07-29 \\
WISSH52 &  2974\tablefootmark{a}  &  6.67   & 2002-05-03 \\
WISSH53 &  13335\tablefootmark{a} &  1.54   & 2011-12-03 \\
WISSH54 &  0143150201\tablefootmark{b} &  13.2   & 2003-06-18 \\
WISSH55 &  25345\tablefootmark{$\dagger$} &  3.71   & 2022-05-11 \\
WISSH56 &  25346\tablefootmark{$\dagger$} &  8.96   & 2022-06-08 \\
WISSH57 &  00016152001\tablefootmark{c}&  4.07   & 2023-07-27 \\
        &  00016152002\tablefootmark{c}&  4.96   & 2023-07-28 \\
        &  00016152003\tablefootmark{c}&  4.76   & 2023-07-30 \\
        &  25347\tablefootmark{$\dagger$} &  4.02   & 2022-07-03 \\ 
WISSH58 &  0804480101\tablefootmark{b} &  33.4   & 2017-12-30 \\
WISSH59 &  0921360101\tablefootmark{$\star$} &  82.1   & 2023-05-18 \\
        &  0921360201\tablefootmark{$\star$} &  39.5   & 2023-11-10 \\
	&  0405690501\tablefootmark{b} &  18.1   & 2006-11-25 \\
WISSH60 &  867\tablefootmark{a}   &  3.01   & 2000-04-03 \\
WISSH61 &  12859\tablefootmark{a} &  23.64  & 2011-06-20 \\
        &  6823\tablefootmark{a}  &  3.9    & 2006-09-15 \\
WISSH62 &  0865210401\tablefootmark{b} &  23.8   & 2021-01-07 \\
        &  13360\tablefootmark{a} &  1.54   & 2011-11-11 \\
WISSH63 &  0803950301\tablefootmark{b} &  19.0   & 2017-05-12 \\
        &  0402070101\tablefootmark{b} &  2.2    & 2006-11-12 \\
WISSH64 &  25348\tablefootmark{$\dagger$} &  8.96   & 2022-05-16 \\
        &  3959\tablefootmark{a}  &  3.5    & 2003-04-20 \\
WISSH65 &  12860\tablefootmark{a} &  21.46  & 2012-02-28 \\
WISSH66 &  25349\tablefootmark{$\dagger$} &  4.03   & 2022-01-08 \\
WISSH67 &  25350\tablefootmark{$\dagger$} &  4.02   & 2022-09-05 \\
WISSH68 &  25351\tablefootmark{$\dagger$} &  8.96   & 2022-08-16 \\
        &  4071\tablefootmark{a}  &  4.86   & 2012-10-24 \\
WISSH69 &  17079\tablefootmark{a} &  29.68  & 2016-04-06 \\
        &  13342\tablefootmark{a} &  1.56   & 2012-03-22 \\
WISSH70 &  15334\tablefootmark{a} &  37.39  & 2013-10-22 \\
        &  0840440101\tablefootmark{b} &  38.5   & 2019-07-26 \\
	&  0840440201\tablefootmark{b} &  33.2   & 2019-12-15 \\
	&  6808\tablefootmark{a}  &  4.05   & 2006-07-16 \\
WISSH71 &  25352\tablefootmark{$\dagger$} &  4.03   & 2022-07-01 \\
        &  13314\tablefootmark{a} &  1.56   & 2012-05-02 \\
WISSH72 &  25353\tablefootmark{$\dagger$} &  4.03   & 2023-07-05 \\
WISSH73 &  0763160201\tablefootmark{b} &  23.8   & 2016-02-04 \\
WISSH74 &  25354\tablefootmark{$\dagger$} &  4.0    & 2023-07-18 \\
WISSH75 &  25355\tablefootmark{$\dagger$} &  4.03   & 2022-06-30 \\
WISSH76 &  25356\tablefootmark{$\dagger$} &  3.85   & 2023-03-06 \\
WISSH77 &  25357\tablefootmark{$\dagger$} &  4.03   & 2022-10-10 \\
WISSH78 &  2184\tablefootmark{a}  &  1.57   & 2001-09-05 \\
WISSH79 &  25358\tablefootmark{$\dagger$} &  4.03   & 2023-07-21 \\
WISSH80 &  13315\tablefootmark{a} &  1.54   & 2011-11-24 \\
WISSH81 &  25359\tablefootmark{$\dagger$} &  4.03   & 2022-07-22 \\
WISSH82 &  0723700101\tablefootmark{b} &  23.2   & 2013-05-19 \\
        &  0723700301\tablefootmark{b} &  30.5   & 2013-08-04 \\
	&  0723700201\tablefootmark{b} &  30.7   & 2013-07-09 \\
	&  9756\tablefootmark{a}  &  32.26  & 2007-11-14 \\
WISSH83 &  17080\tablefootmark{a} &  39.55  & 2015-12-22 \\
	&  0745010401\tablefootmark{b} &  17.6   & 2014-11-14 \\
        &  6822\tablefootmark{a}  &  3.9    & 2006-03-30 \\
WISSH84 &  25360\tablefootmark{$\dagger$} &  4.03   & 2022-09-19 \\
WISSH85 &  25361\tablefootmark{$\dagger$} &  4.03   & 2022-05-07 \\
    \bottomrule
    \end{tabular}
    \tablefoot{For \textit{XMM-Newton} observations the after-cleaning exposure time is reported.\\
    \tablefoottext{a}{\textit{Chandra} observation.}
    \tablefoottext{b}{\textit{XMM-Newton} observation.}
    \tablefoottext{c}{\textit{Swift}-XRT observation.}
    \tablefoottext{$\dagger$}{Proprietary \textit{Chandra} data (PI: E. Piconcelli).}
    \tablefoottext{$\star$}{Proprietary \textit{XMM-Newton} data (PI: C. Pinto).}}
\end{table}

\section{Data analysis results}\label{apdx:dataAnalysis_results}

\begin{table}[h!]
    \caption{WISSH QSOs with $\rm \ge 20$ X-ray net (i.e. background-subtracted) counts.}
    \label{tab:20counts}
    \centering
    \begin{tabular}{cccccccc}
    \toprule
    WISSH ID & SDSS ID & $z$ & $\rm N_{H,gal}$ & $\rm \Gamma$ & $\rm N_H$ & $\rm Log(L_{2-10})$ & $\rm Log(F_{0.5-10})$ \\
    (1) & (2) & (3) & (4) & (5) & (6) & (7) & (8) \\
    \midrule
WISSH04 & J0209$-$0005 & 2.87   & 2.29 & $\rm 1.36 \pm 0.14$          & $\rm \le 2.0             $ & $\rm 45.20 \pm 0.04$ & $\rm -13.17 \pm 0.06$ \\ [0.75ex]
WISSH07 & J0735+2659   & 1.999  & 5.39 & $\rm 1.57 \pm 0.13$          & $\rm \le 2.6             $ & $\rm 45.13 \pm 0.03$ & $\rm -12.99 \pm 0.05$ \\ [0.75ex]
WISSH08 & J0745+4734   & 3.225  & 5.78 & $\rm 1.83_{-0.19}^{+0.20}$   & $\rm \le 3.4             $ & $\rm 46.38 \pm 0.05$ & $\rm -12.28 \pm 0.07$ \\ [0.75ex]
WISSH09 & J0747+2739   & 4.126  & 3.82 & $\rm 1.65 \pm 0.13$          & $\rm \le 4.5             $ & $\rm 45.09 \pm 0.04$ & $\rm -13.73 \pm 0.06$ \\ [0.75ex]
WISSH10 & J0801+5210   & 3.257  & 4.61 & $\rm 1.74 \pm 0.14$          & $\rm \le 0.9             $ & $\rm 45.31 \pm 0.04$ & $\rm -13.33 \pm 0.05$ \\ [0.75ex]
WISSH13 & J0900+4215   & 3.294  & 2.37 & $\rm 1.89_{-0.04}^{+0.05}$   & $\rm \le 0.7             $ & $\rm 46.24 \pm 0.01$ & $\rm -12.46 \pm 0.02$ \\ [0.75ex]
WISSH14 & J0904+1309   & 2.9765 & 2.58 & $\rm 2.04 \pm 0.07$          & $\rm \le 1.1            $ & $\rm 45.88 \pm 0.02$ & $\rm -12.77 \pm 0.03$ \\ [0.75ex]
WISSH17 & J0947+1421   & 3.031  & 2.95 & $\rm 1.98 \pm 0.11$          & $\rm 0.9 \pm 0.4         $ & $\rm 45.59 \pm 0.02$ & $\rm -13.03 \pm 0.03$ \\ [0.75ex]
WISSH18 & J0950+4329   & 1.7696 & 0.94 & $\rm 1.55 \pm 0.12$          & $\rm \le 5.8             $ & $\rm 46.16 \pm 0.03$ & $\rm -11.84 \pm 0.03$ \\ [0.75ex]
WISSH20 & J0959+1312   & 4.0781 & 2.62 & $\rm 2.56_{-0.46}^{+0.49}$   & $\rm \le 122.2            $ & $\rm 46.00 \pm 0.20$ & $\rm -13.16 \pm 0.10$ \\ [0.75ex]
WISSH22 & J1014+4300   & 3.1224 & 1.61 & $\rm 2.15 \pm 0.08$          & $\rm \le 1.3             $ & $\rm 45.47 \pm 0.02$ & $\rm -13.26 \pm 0.04$ \\ [0.75ex]
WISSH27 & J1027+3543   & 3.1182 & 0.81 & $\rm 1.99 \pm 0.06$          & $\rm \le 0.8             $ & $\rm 45.79 \pm 0.02$ & $\rm -12.89 \pm 0.03$ \\ [0.75ex]
WISSH30 & J1057+4555   & 4.1306 & 0.87 & $\rm 2.08 \pm 0.31$          & $\rm \le 1.2             $ & $\rm 45.75 \pm 0.08$ & $\rm -13.25 \pm 0.14$ \\ [0.75ex]
WISSH33 & J1106+6400   & 2.221  & 0.92 & $\rm 2.09 \pm 0.15$          & $\rm \le 0.5             $ & $\rm 45.73 \pm 0.04$ & $\rm -12.61 \pm 0.05$ \\ [0.75ex]
WISSH35 & J1110+4831   & 2.9741 & 1.62 & $\rm 1.92 \pm 0.07$          & $\rm \le 0.3             $ & $\rm 45.41 \pm 0.02$ & $\rm -13.20 \pm 0.03$ \\ [0.75ex]
WISSH36 & J1110+4305   & 3.8492 & 1.39 & $\rm 1.76 \pm 0.26$          & $\rm \le 26.4            $ & $\rm 46.36 \pm 0.10$ & $\rm -12.45 \pm 0.06$ \\ [0.75ex]
WISSH37 & J1111+1336   & 3.49   & 1.54 & $\rm 1.76 \pm 0.13$          & $\rm \le 6.4             $ & $\rm 45.39 \pm 0.04$ & $\rm -13.32 \pm 0.05$ \\ [0.75ex]
WISSH39 & J1130+0732   & 2.659  & 4.86 & $\rm 1.56 \pm 0.23$          & $\rm \le 36.0            $ & $\rm 45.98 \pm 0.07$ & $\rm -12.40 \pm 0.06$ \\ [0.75ex]
WISSH40 & J1157+2724   & 2.217  & 1.78 & $\rm 1.61_{-0.22}^{+0.25}$   & $\rm \le 1.1             $ & $\rm 44.89 \pm 0.08$ & $\rm -13.33 \pm 0.13$ \\ [0.75ex]
WISSH43 & J1201+0116   & 3.248  & 1.63 & $\rm 2.11_{-0.31}^{+0.37}$   & $\rm 4.4_{-2.3}^{+3.4}    $ & $\rm 44.80 \pm 0.05$ & $\rm -13.88 \pm 0.08$ \\ [0.75ex]
WISSH47 & J1215$-$0034 & 2.6987 & 1.69 & $\rm 1.53_{-0.29}^{+0.31}$   & $\rm 18.2_{-6.5}^{+7.7}  $ & $\rm 45.30 \pm 0.15$ & $\rm -13.21 \pm 0.07$ \\ [0.75ex]
WISSH48 & J1219+4940   & 2.6928 & 2.09 & $\rm 2.02_{-0.54}^{+0.55}$   & $\rm \le 21.8            $ & $\rm 45.52 \pm 0.14$ & $\rm -13.01 \pm 0.11$ \\ [0.75ex]
WISSH49 & J1220+1126   & 1.8962 & 2.20 & $\rm 2.21_{-0.36}^{+0.37}$   & $\rm \le 25.1             $ & $\rm 45.55 \pm 0.08$ & $\rm -12.64 \pm 0.07$ \\ [0.75ex]
WISSH50 & J1236+6554   & 3.424  & 2.31 & $\rm 2.45 \pm 0.17$          & $\rm \le 6.1             $ & $\rm 45.51 \pm 0.05$ & $\rm -13.40 \pm 0.04$ \\ [0.75ex]
WISSH54 & J1250+2631   & 2.0476 & 0.94 & $\rm 2.12 \pm 0.03$          & $\rm \le 0.04             $ & $\rm 45.97 \pm 0.01$ & $\rm -12.28 \pm 0.01$ \\ [0.75ex]
WISSH57 & J1310+4601   & 2.1423 & 1.17 & $\rm 2.10_{-0.40}^{+0.35} $  & $\rm \le 8.6             $ & $\rm 45.65 \pm 0.09$ & $\rm -12.65 \pm 0.13$ \\ [0.75ex]
WISSH58 & J1326$-$0005 & 3.303  & 1.69 & $\rm 1.79_{-0.13}^{+0.14}$   & $\rm 3.6_{-2.0}^{+1.1}   $ & $\rm 45.34 \pm 0.03$ & $\rm -13.25 \pm 0.04$ \\ [0.75ex]
WISSH59 & J1328+5818   & 3.14   & 1.94 & 1.8\tablefootmark{f}  & $\rm 56.6_{-17.3}^{+20.3}$ & $\rm 44.51 \pm 0.06$ & $\rm -13.52 \pm 0.06$ \\ [0.75ex]
WISSH60 & J1333+1649   & 2.099  & 1.60 & $\rm 1.38_{-0.11}^{+0.12}$   & $\rm \le 0.9             $ & $\rm 45.84 \pm 0.04$ & $\rm -12.27 \pm 0.06$ \\ [0.75ex]
WISSH61 & J1421+4633   & 3.454  & 1.10 & $\rm 1.23_{-0.26}^{+0.27}$   & $\rm \le 14.7             $ & $\rm 44.95 \pm 0.08$ & $\rm -13.51 \pm 0.11$ \\ [0.75ex]
WISSH62 & J1422+4417   & 3.648  & 0.81 & $\rm 1.99_{-0.39}^{+0.51}$   & $\rm 13.1_{-7.0}^{+12.7} $ & $\rm 44.96 \pm 0.06$ & $\rm -13.68 \pm 0.08$ \\ [0.75ex]
WISSH63 & J1426+6025   & 3.1972 & 2.08 & $\rm 1.91 \pm 0.05$          & $\rm \le 1.0             $ & $\rm 45.90 \pm 0.02$ & $\rm -12.78 \pm 0.02$ \\ [0.75ex]
WISSH65 & J1441+0454   & 2.08   & 2.81 & $\rm 1.83_{-0.39}^{+0.40}$   & $\rm 5.4_{-3.1}^{+3.4}   $ & $\rm 44.78 \pm 0.05$ & $\rm -13.38 \pm 0.08$ \\ [0.75ex]
WISSH69 & J1513+0855   & 2.8883 & 2.19 & $\rm 2.01_{-0.17}^{+0.18}$   & $\rm 5.3_{-1.9}^{+2.1}   $ & $\rm 45.74 \pm 0.07$ & $\rm -12.96 \pm 0.04$ \\ [0.75ex]
WISSH70 & J1521+5202   & 2.218  & 1.53 & $\rm 1.58_{-0.37}^{+0.38}$   & $\rm 11.4_{-4.4}^{+5.2}  $ & $\rm 44.91 \pm 0.06$ & $\rm -13.43 \pm 0.08$ \\ [0.75ex]
WISSH73 & J1549+1245   & 2.365  & 3.18 & $\rm 2.25_{-0.17}^{+0.18}$   & $\rm 4.7_{-0.9}^{+1.1}   $ & $\rm 45.20 \pm 0.02$ & $\rm -13.20 \pm 0.04$ \\ [0.75ex]
WISSH78 & J1621$-$0042 & 3.7285 & 6.49 & $\rm 1.81_{-0.37}^{+0.38}$   & $\rm \le 9.7             $ & $\rm 45.98 \pm 0.11$ & $\rm -12.81 \pm 0.13$ \\ [0.75ex]
WISSH82 & J1701+6412   & 2.753  & 2.11 & $\rm 2.20  \pm 0.05$         & $\rm 0.8 \pm 0.2         $ & $\rm 46.13 \pm 0.01$ & $\rm -12.46 \pm 0.01$ \\ [0.75ex]
WISSH83 & J2123$-$0050 & 2.282  & 3.65 & $\rm 1.97 \pm 0.07$          & $\rm \le 0.6             $ & $\rm 45.69 \pm 0.02$ & $\rm -12.66 \pm 0.02$ \\
    \bottomrule
    \end{tabular}
    \tablefoot{Columns: (1) WISSH ID; (2) SDSS ID; (3) Redshift \citep[from][]{Saccheo_2023}; (4) Galactic column density (in units of $\rm 10^{20}\,\,cm^{-2}$); (5) X-ray photon index; (6) Intrinsic column density (in units of $\rm 10^{22}\,\,cm^{-2}$). Upper limits indicate that the addition of an intrinsic absorption component is not significantly required (i.e. $\rm < 95\%$ c.l.); (7) Intrinsic $\rm 2 - 10\,\,keV$ luminosity (in units of $\rm Log(L_{2-10}/erg\,s^{-1})$); (8) Observed $\rm 0.5 - 10 \,\,keV$ flux (in units of $\rm Log(F_{0.5-10}/erg\,s^{-1}\,cm^{-2})$).
    \tablefoottext{f}{Fixed.}}
\end{table}

\begin{table}[h!]
    \caption{WISSH QSOs with $\rm 5 - 20$ X-ray net (i.e. background-subtracted) counts.}
    \label{tab:520counts}
    \centering
    \begin{tabular}{cccccccc}
    \toprule
    WISSH ID & SDSS ID & $z$ & $\rm N_{H,gal}$ & HR & $\rm N_H$ & $\rm Log(L_{2-10})$ & $\rm Log(F_{0.5-10})$ \\
    (1) & (2) & (3) & (4) & (5) & (6) & (7) & (8) \\
    \midrule
WISSH01 & J0045+1438   & 1.9897 & 5.50 & $\rm -0.16 \pm 0.30$           & $ 15.0_{-13.0}^{+15.0} $ & $ 44.31 \pm 0.27 $ & $\rm -13.91 \pm 0.11$ \\ [0.75ex] 
WISSH02 & J0124+0044   & 3.822  & 3.13 & $\rm -0.72_{-0.27}^{+0.07}$    & $ \le 0.1             $ & $\rm 45.42 \pm 0.12 $ & $\rm -13.43 \pm 0.12$ \\ [0.75ex]
WISSH03 & J0125$-$1027 & 3.3588 & 3.20 & $\rm 0.25_{-0.24}^{+0.31}$     & $ 50.0_{-35.0}^{+50.0} $ & $ 45.43 \pm 0.14 $ & $\rm -13.30 \pm 0.11$ \\ [0.75ex]
WISSH16 & J0941+3257   & 3.454  & 1.44 & $\rm -0.30_{-0.44}^{+0.28}$    & $ \le 16.5             $ & $\rm 45.14 \pm 0.15 $ & $\rm -13.59 \pm 0.15$ \\ [0.75ex]
WISSH19 & J0958+2827   & 3.434  & 1.46 & $\rm -0.06 \pm 0.24$           & $ \le 99.0             $ & $\rm 44.63 \pm 0.07 $ & $\rm -14.10 \pm 0.07$ \\ [0.75ex]
WISSH21 & J1013+5615   & 3.6507 & 0.91 & $\rm 0.00 \pm 0.38$         & $ \le 132.0             $ & $\rm 45.20 \pm 0.15 $ & $\rm -13.59 \pm 0.15$ \\ [0.75ex]
WISSH23 & J1015+0020   & 4.407  & 2.90 & $\rm -0.16_{-0.22}^{+0.10}$    & $ \le 49.5             $ & $\rm 45.52 \pm 0.10 $ & $\rm -13.47 \pm 0.10$ \\ [0.75ex]
WISSH29 & J1051+3107   & 4.2742 & 2.12 & $\rm -0.63_{-0.30}^{+0.10}$    & $ \le 0.1             $ & $\rm 44.96 \pm 0.18 $ & $\rm -13.99 \pm 0.18$ \\ [0.75ex]
WISSH34 & J1110+1930   & 2.502  & 1.73 & $\rm 0.07_{-0.26}^{+0.27}$     & $ \le 82.5             $ & $\rm 45.14 \pm 0.10 $ & $\rm -13.25 \pm 0.10$ \\ [0.75ex]
WISSH42 & J1200+3126   & 2.9947 & 1.49 & $\rm -0.53 \pm 0.23$           & $ \le 24.8             $ & $\rm 45.55 \pm 0.09 $ & $\rm -13.04 \pm 0.09$ \\ [0.75ex]
WISSH44 & J1201+1206   & 3.512  & 1.89 & $\rm -0.35_{-0.26}^{+0.19}$    & $ \le 66.0             $ & $\rm 45.64 \pm 0.10 $ & $\rm -13.11 \pm 0.10$ \\ [0.75ex]
WISSH51 & J1237+0647   & 2.7891 & 1.53 & $\rm -0.01 \pm 0.28$           & $ \le 82.5             $ & $\rm 45.22 \pm 0.11 $ & $\rm -13.29 \pm 0.11$ \\ [0.75ex]
WISSH52 & J1245+0105   & 2.8068 & 1.42 & $\rm -0.18_{-0.32}^{+0.27}$    & $ 30.0_{-23.0}^{+30.0} $ & $ 44.82 \pm 0.27 $ & $\rm -13.75 \pm 0.12$ \\ [0.75ex]
WISSH56 & J1305+0521   & 4.101  & 1.88 & $\rm -0.06_{-0.26}^{+0.24}$    & $ \le 82.5             $ & $\rm 45.37 \pm 0.09 $ & $\rm -13.54 \pm 0.09$ \\ [0.75ex]
WISSH79 & J1633+3629   & 3.5747 & 1.14 & $\rm -0.14_{-0.38}^{+0.31}$    & $ \le 66.0             $ & $\rm 45.23 \pm 0.15 $ & $\rm -13.54 \pm 0.15$ \\ [0.75ex]
WISSH80 & J1639+2824   & 3.846  & 2.93 & $\rm 0.00 \pm 0.38$ & $ 70.0_{-55.0}^{+80.0} $ & $45.61 \pm 0.33$ & $\rm -13.28 \pm 0.15$ \\
    \bottomrule
    \end{tabular}
    \tablefoot{Columns: (1) WISSH ID; (2) SDSS ID; (3) Redshift \citep[from][]{Saccheo_2023}; (4) Galactic column density (in units of $\rm 10^{20}\,\,cm^{-2}$); (5) Hardness Ratio; (6) Intrinsic column density (in units of $\rm 10^{22}\,\,cm^{-2}$); (7) Intrinsic $\rm 2 - 10\,\,keV$ luminosity (in units of $\rm Log(L_{2-10}/erg\,s^{-1})$); (8) Observed $\rm 0.5 - 10 \,\,keV$ flux (in units of $\rm Log(F_{0.5-10}/erg\,s^{-1}\,cm^{-2})$).}
\end{table}

\begin{table}[h!]
    \caption{WISSH QSOs with $\rm \le 5$ X-ray net (i.e. background-subtracted) counts.}
    \label{tab:05counts}
    \centering
    \begin{tabular}{cccccc}
    \toprule
    WISSH ID & SDSS ID & $z$ & $\rm N_{H,gal}$ & $\rm Log(L_{2-10})$ & $\rm Log(F_{0.5-10})$ \\
    (1) & (2) & (3) & (4) & (5) & (6) \\
    \midrule
WISSH05 & J0216$-$0921 & 3.7387 & 2.91 & $\rm 45.06 $ & $\rm -13.78 \pm 0.20 $ \\
WISSH06 & J0414+0609   & 2.6324 & 10.2 & $\rm 44.44 $ & $\rm -14.09 \pm 0.30 $ \\
WISSH11 & J0818+0958   & 3.6943 & 2.79 & $\rm 45.05 $ & $\rm -13.78 \pm 0.20 $ \\
WISSH12 & J0846+2411   & 4.7218 & 3.05 & $\rm < 44.75        $ & $\rm < -14.30        $ \\
WISSH15 & J0928+5340   & 4.466  & 1.66 & $\rm < 45.00        $ & $\rm < -14.00        $ \\
WISSH24 & J1020+0922   & 3.6584 & 2.89 & $\rm 44.88 $ & $\rm -13.94 \pm 0.23 $ \\
WISSH25 & J1025+2454   & 2.3917 & 1.16 & $\rm < 43.89        $ & $\rm < -14.53        $ \\
WISSH26 & J1026+0329   & 3.8808 & 3.25 & $\rm < 44.87        $ & $\rm < -14.00        $ \\
WISSH28 & J1048+4407   & 4.408  & 1.16 & $\rm 44.91 $ & $\rm -14.08 \pm 0.20 $ \\
WISSH31 & J1103+1004   & 3.6004 & 2.33 & $\rm < 44.80        $ & $\rm < -14.00        $ \\
WISSH32 & J1106$-$1731 & 2.572  & 4.27 & $\rm 44.79 $ & $\rm -13.71 \pm 0.18 $ \\
WISSH38 & J1122+1645   & 3.0398 & 1.47 & $\rm 44.94 $ & $\rm -13.71 \pm 0.20 $ \\
WISSH41 & J1159+1337   & 4.0048 & 2.07 & $\rm 45.09 $ & $\rm -13.81 \pm 0.24 $ \\
WISSH45 & J1204+3309   & 3.638  & 1.42 & $\rm < 44.81        $ & $\rm < -14.00        $ \\
WISSH46 & J1210+1741   & 3.831  & 3.35 & $\rm < 44.86        $ & $\rm < -14.00        $ \\
WISSH53 & J1249$-$0159 & 3.6286 & 1.66 & $\rm 45.01 $ & $\rm -13.80 \pm 0.24 $ \\
WISSH55 & J1250+2046   & 3.543  & 2.08 & $\rm 44.70 $ & $\rm -14.08 \pm 0.30 $ \\
WISSH64 & J1433+0227   & 4.728  & 2.58 & $\rm 44.98 $ & $\rm -14.08 \pm 0.20 $ \\
WISSH66 & J1447+1038   & 3.7042 & 1.56 & $\rm 45.05 $ & $\rm -13.78 \pm 0.20 $ \\
WISSH67 & J1451+1441   & 3.094  & 1.52 & $\rm 44.96 $ & $\rm -13.71 \pm 0.20 $ \\
WISSH68 & J1506+5220   & 4.0995 & 1.84 & $\rm 44.62 $ & $\rm -14.30 \pm 0.22 $ \\
WISSH71 & J1538+0855   & 3.567  & 3.09 & $\rm 44.86 $ & $\rm -13.94 \pm 0.23 $ \\
WISSH72 & J1544+4120   & 3.5513 & 1.86 & $\rm 44.85 $ & $\rm -13.94 \pm 0.23 $ \\
WISSH74 & J1554+1109   & 2.93   & 3.46 & $\rm < 44.61        $ & $\rm < -14.01        $ \\
WISSH75 & J1555+1003   & 3.529  & 3.90 & $\rm 44.70 $ & $\rm -14.08 \pm 0.30 $ \\
WISSH76 & J1559+4828   & 3.419  & 1.69 & $\rm 44.87 $ & $\rm -13.88 \pm 0.24 $ \\
WISSH77 & J1559+1923   & 3.9532 & 2.72 & $\rm 44.81 $ & $\rm -14.08 \pm 0.30 $ \\
WISSH81 & J1650+2507   & 3.337  & 4.63 & $\rm 44.80 $ & $\rm -13.94 \pm 0.23 $ \\
WISSH84 & J2238$-$0808 & 3.1422 & 3.56 & $\rm 44.74 $ & $\rm -13.94 \pm 0.23 $ \\
WISSH85 & J2346$-$0016 & 3.511  & 3.71 & $\rm 45.08 $ & $\rm -13.70 \pm 0.20 $ \\
    \bottomrule
    \end{tabular}
    \tablefoot{Columns: (1) WISSH ID; (2) SDSS ID; (3) Redshift \citep[from][]{Saccheo_2023}; (4) Galactic column density (in units of $\rm 10^{20}\,\,cm^{-2}$); (5) Intrinsic $\rm 2 - 10\,\,keV$ luminosity (in units of $\rm Log(L_{2-10}/erg\,s^{-1})$); (6) Observed $\rm 0.5 - 10 \,\,keV$ flux (in units of $\rm Log(F_{0.5-10}/erg\,s^{-1}\,cm^{-2})$). $\rm F_{0.5-10}$ and $\rm L_{2-10}$ are calculated assuming $\rm \Gamma = 1.8$ and $\rm N_H = 5 \times 10^{22}\,\,cm^{-2}$. Errors on  $\rm Log(L_{2-10}/erg\,s^{-1})$ are the same as those reported for $\rm Log(F_{0.5-10}/erg\,s^{-1}\,cm^{-2})$.}
\end{table}

\clearpage
\section{Variable QSOs}\label{apdx:variability}
\begin{figure}[h!]
    \centering
    \resizebox{0.4\hsize}{!}{\includegraphics{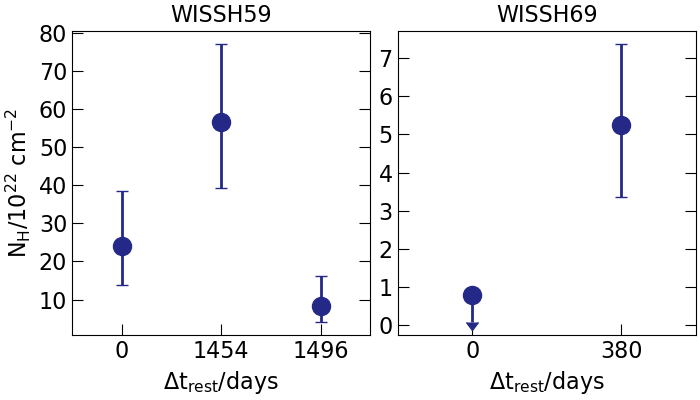}}
    \caption{Intrinsic column density as a function of rest-frame time interval for the WISSH QSOs with multiple X-ray observations exhibiting $\rm > 2\sigma$ $\rm N_H$ variability.}\label{fig:nH_variability}
\end{figure}

\begin{figure}[h!]
    \centering
    \includegraphics[width=17cm]{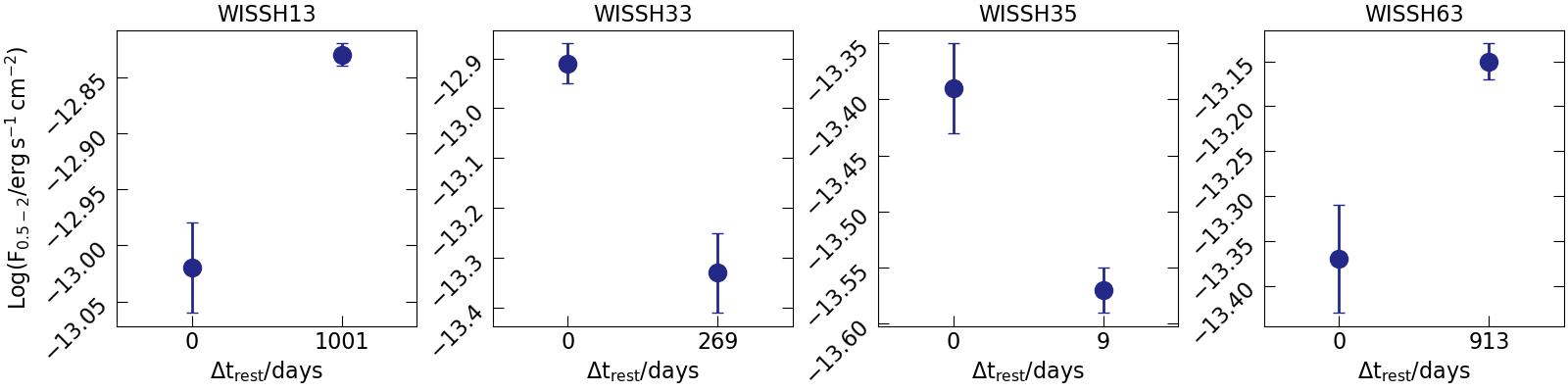}
    \caption{$\rm 0.5 - 2 \,\,keV$ flux as a function of rest-frame time interval for the four WISSH QSOs with multiple X-ray observations exhibiting $\rm >3\sigma$ $\rm F_{0.5-2}$ variability.}\label{fig:F052_variability}
\end{figure}

\begin{figure}[h!]
    \centering
    \includegraphics[width=12cm]{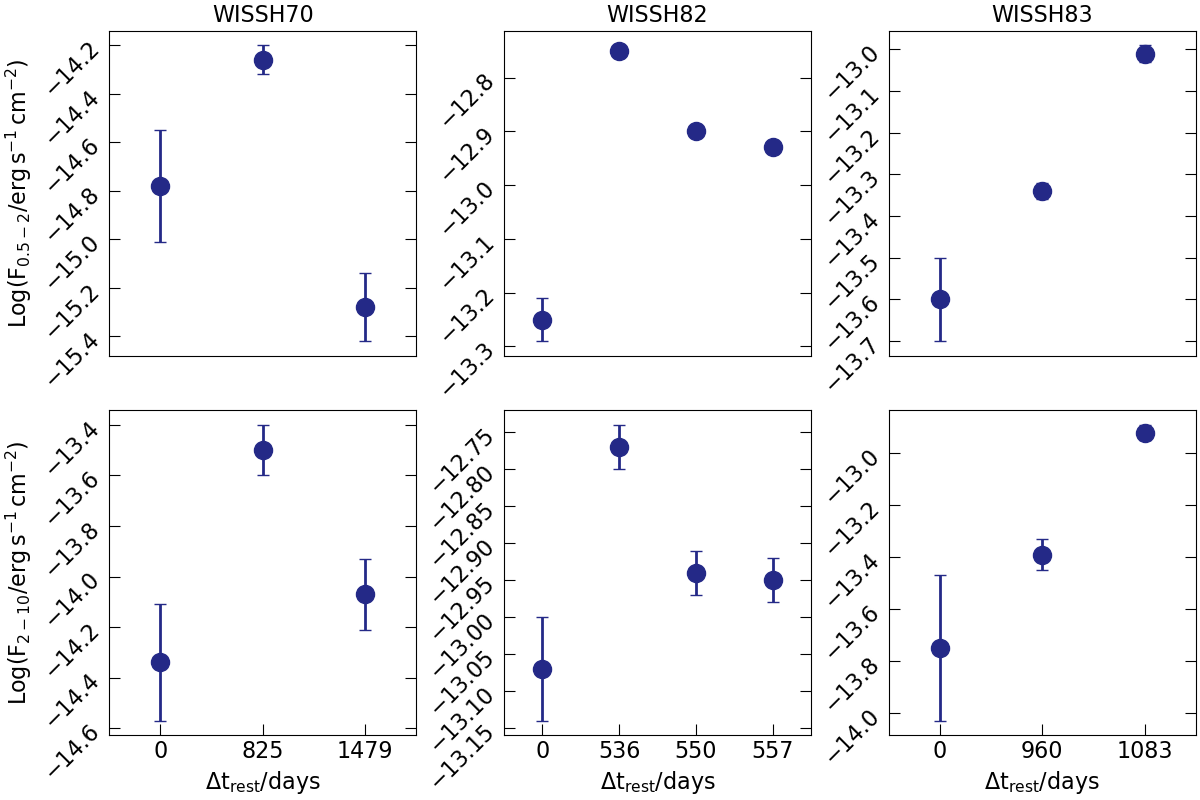}
    \caption{$\rm 0.5 - 2 \,\,keV$ (top panels) and $\rm 2 - 10\,\,keV$ (bottom panels) flux as a function of rest-frame time interval for the three WISSH QSOs with multiple X-ray observations exhibiting $\rm >3\sigma$ $\rm F_{0.5-2}$ and $\rm F_{2-10}$ variability.}\label{fig:F052_F210_variability}
\end{figure}

As described in Section \ref{subsec:multi_obs}, 28 sources have multiple observations, thus allowing us to search for possible variations in terms of $\rm N_H$ and flux. For what concerns $\rm N_H$ variability, we find a possible $\rm > 2\sigma$ variations in WISSH59 and WISSH69 over a rest-frame time interval ranging from $\rm \approx 380$ to $\rm \approx 40$ days. Figure \ref{fig:nH_variability} and Table \ref{tab:Nh_variability} summarise the X-ray parameters from different epochs. A $\rm > 3\sigma$ variability in the soft $\rm (0.5 - 2\,\,keV)$ flux occurs in WISSH13, WISSH33, WISSH35 and WISSH63 over a rest-frame time interval ranging from $\rm \approx 1000$ to 9 days. Figure \ref{fig:F052_variability} and Table \ref{tab:F052_variability} summarise X-ray parameters from different epochs. A $\rm > 3\sigma$ variability in the soft and hard $\rm (2 - 10\,\,keV)$ flux occurs in WISSH70, WISSH82 and WISSH83 over a rest-frame time interval ranging from $\rm \approx 1000$ to 7 days. Figure \ref{fig:F052_F210_variability} and Table \ref{tab:F052_F210_variability} summarise X-ray parameters from different epochs.

\begin{table}[h!]
  \caption{WISSH QSOs with multiple X-ray observations showing $\rm >2\sigma$ $\rm N_H$ variability.}\label{tab:Nh_variability}
  \centering
  \begin{tabular}{cccccccccc}
    \toprule
    Obs. ID & $\rm \Gamma$ & $\rm N_H$ & $\rm Log( L_{2-10})$ & $\rm Log( F_{0.5 - 2})$ & $\rm Log(F_{2 - 10})$ & $\rm \alpha_{OX}$ & $\rm \Delta (\alpha_{OX})$ & $\rm \Delta t_{rest}$ \\
    (1) & (2) & (3) & (4) & (5) & (6) & (7) & (8) & (9) \\
    \midrule
    \multicolumn{9}{c}{\textbf{WISSH59 $-$ J1328+5818}} \\[0.75ex]
    0405690501 & 1.8\tablefootmark{f} & $\rm 23.9_{-10.0}^{+14.5}$ & $\rm 44.80 \pm 0.11$ & $\rm -14.30 \pm 0.11$ & $\rm -13.62 \pm 0.11$ & $\rm -1.9 \pm 0.04$ & $\rm -0.18$ & \multirow{2}{*}{1454}\\ [0.75ex]
    \textbf{0921360101} & 1.8\tablefootmark{f} & $\rm 56.6_{-17.3}^{+20.3}$ & $\rm 44.51 \pm 0.06$ & $\rm -14.68 \pm 0.07$ & $\rm -13.66 \pm 0.07$ & $\rm -1.72 \pm 0.03$ & $\rm -0.003$ & \multirow{2}{*}{43}\\ [0.75ex] 
    0921360201 & 1.8\tablefootmark{f} & $\rm 8.3_{-4.1}^{+7.8}$ & $\rm 44.87 \pm 0.10$ & $\rm -14.42 \pm 0.12$ & $\rm -13.98 \pm 0.12$ & $\rm -1.87 \pm 0.04$ & $\rm -0.15$ \\
    \midrule
    \multicolumn{9}{c}{\textbf{WISSH69 $-$ J1513+0855}} \\[0.75ex]
    13342 & 1.8\tablefootmark{f} & $\rm \le 1.3$ & $\rm 45.35 \pm 0.11$ & $\rm -13.62 \pm 0.11$ & $\rm -13.41 \pm 0.11$ & $\rm -1.89 \pm 0.04$ & $\rm -0.09$ & \multirow{2}{*}{380}\\ [0.75ex] 
    \textbf{17079} & $\rm 2.01_{-0.17}^{+0.18}$ & $\rm 5.3_{-1.9}^{+2.1}$ & $\rm 45.74 \pm 0.07$ & $\rm -13.41 \pm 0.03$ & $\rm -13.15 \pm 0.06$ & $\rm -1.71 \pm 0.04$ & $\rm 0.09$ & \\
    \bottomrule
  \end{tabular}
  \tablefoot{Columns: (1) Observation ID. The ID in boldface refers to the dataset from which the reference results for the tables and figures in the study are derived; (2) X-ray photon index; (3) Intrinsic column density (in units of $\rm 10^{22}\,\,cm^{-2}$); (4) Intrinsic $\rm 2 - 10\,\,keV$ luminosity (in units of $\rm Log (L_{2-10}/erg\,s^{-1})$); (5) $\rm 0.5 - 2\,\,keV$ flux (in units of $\rm Log (F_{0.5-2}/erg\,s^{-1}\,cm^{-2})$); (6) $\rm 2 - 10\,\,keV$ flux (in units of $\rm Log (F_{2-10}/erg\,s^{-1}\,cm^{-2})$); (7) X-ray$-$to$-$optical index; (8) $\rm \Delta(\alpha_{OX}) = \alpha_{OX} - \alpha_{OX, L10}$; (9) Rest-frame time interval (in units of days).\\
  \tablefoottext{f}{Fixed.}}
\end{table}

\newpage

\begin{table}[h!]
  \caption{WISSH QSOs with multiple X-ray observations showing $\rm >3\sigma$ $\rm F_{0.5 - 2}$ variability.}\label{tab:F052_variability}
  \centering
  \begin{tabular}{ccccccccc}
    \toprule
    Obs. ID & $\rm \Gamma$ & $\rm N_H$ & $\rm Log(L_{2-10})$ & $\rm Log(F_{0.5 - 2})$ & $\rm Log( F_{2 - 10})$ & $\rm \alpha_{OX}$ & $\rm \Delta (\alpha_{OX})$ & $\rm \Delta t_{rest}$ \\
    (1) & (2) & (3) & (4) & (5) & (6) & (7) & (8) & (9) \\
    \midrule
    \multicolumn{8}{c}{\textbf{WISSH13 $-$ J0900+4215}} \\[0.75ex]
    6810 & $\rm 1.93 \pm 0.16$ & $\rm \le 0.7$ & $\rm 46.10 \pm 0.04$ & $\rm -13.02 \pm 0.04$ & $\rm -12.90 \pm 0.10$ & $\rm -1.60 \pm 0.03$ & $\rm 0.20$ & \multirow{2}{*}{1001} \\ [0.75ex]
    \textbf{0803950601} & $\rm 1.89_{-0.04}^{+0.05}$ & $\rm \le 0.8$ & $\rm 46.24 \pm 0.01$ & $\rm -12.83 \pm 0.01$ & $\rm -12.70 \pm 0.03$ & $\rm -1.55 \pm 0.01$ & $\rm 0.25$ & \\
    \midrule
    \multicolumn{8}{c}{\textbf{WISSH33 $-$ J1106+6400}} \\[0.75ex]
    \textbf{6811} & $\rm 2.09 \pm 0.15$ & $\rm \le 0.3$ & $\rm 45.73 \pm 0.04$ & $\rm -12.91 \pm 0.04$ & $\rm -12.91 \pm 0.10$ & $\rm -1.69 \pm 0.03$ & $\rm 0.10$ & \multirow{2}{*}{269}\\ [0.75ex]
    0553561401 & $\rm 2.12_{-0.27}^{+0.30}$ & $\rm \le 0.5$ & $\rm 45.31 \pm 0.09$ & $\rm -13.33 \pm 0.08$ & $\rm -13.34 \pm 0.22$ & $\rm -1.85 \pm 0.05$ & $\rm -0.05$ & \\
    \midrule
    \multicolumn{8}{c}{\textbf{WISSH35 $-$ J1110+4831}} \\[0.75ex]
    0059750401 & $\rm 1.56_{-0.15}^{+0.16}$ & $\rm \le 2.0$ & $\rm 45.61 \pm 0.04$ & $\rm -13.39 \pm 0.04$ & $\rm -13.04 \pm 0.12$ & $\rm -1.81 \pm 0.03$ & $\rm -0.01$ & \multirow{2}{*}{9}\\ [0.75ex]
    \textbf{0104861001} & $\rm 1.92 \pm 0.07$ & $\rm \le 0.3$ & $\rm 45.41 \pm 0.02$ & $\rm -13.57 \pm 0.02$ & $\rm -13.45 \pm 0.05$ & $\rm -1.83 \pm 0.01$ & $\rm -0.04$ & \\
    \midrule
    \multicolumn{8}{c}{\textbf{WISSH63 $-$ J1426+6025}} \\[0.75ex]
    0402070101 & $\rm 1.63_{-0.18}^{+0.19}$ & $\rm \le 4.3$ & $\rm 45.69 \pm 0.06$ & $\rm -13.37 \pm 0.06$ & $\rm -13.06 \pm 0.13$ & $\rm -1.85 \pm 0.04$ & $\rm -0.02$ & \multirow{2}{*}{913}\\ [0.75ex]
    \textbf{0803950301} & $\rm 1.91 \pm 0.05$ & $\rm \le 0.8$ & $\rm 45.90 \pm 0.02$ & $\rm -13.15 \pm 0.02$ & $\rm -13.03 \pm 0.04$ & $\rm -1.73 \pm 0.01$ & $\rm 0.10$ & \\
    \bottomrule
  \end{tabular}
  \tablefoot{Columns: (1) Observation ID. The ID in boldface refers to the dataset from which the reference results for the tables and figures in the study were derived; (2) X-ray photon index; (3) Intrinsic column density (in units of $\rm 10^{22}\,\,cm^{-2}$); (4) Intrinsic $\rm 2 - 10\,\,keV$ luminosity (in units of $\rm Log (L_{2-10}/erg\,s^{-1})$); (5) $\rm 0.5 - 2\,\,keV$ flux (in units of $\rm Log (F_{0.5-2}/erg\,s^{-1}\,cm^{-2})$); (6) $\rm 2 - 10\,\,keV$ flux (in units of $\rm Log (F_{2-10}/erg\,s^{-1}\,cm^{-2})$); (7) X-ray$-$to$-$optical index; (8) $\rm \Delta(\alpha_{OX}) = \alpha_{OX} - \alpha_{OX, L10}$; (9) Rest-frame time interval (in units of days).}
\end{table}

\begin{table}[h!]
  \caption{WISSH QSOs with multiple X-ray observations showing $\rm >3\sigma$ $\rm F_{0.5 - 2}$ and $\rm F_{2 - 10}$ variability.}\label{tab:F052_F210_variability}
  \centering
  \begin{tabular}{ccccccccc}
    \toprule
    Obs. ID & $\rm \Gamma$ & $\rm N_H$ & $\rm Log( L_{2-10})$ & $\rm Log( F_{0.5 - 2})$ & $\rm Log( F_{2 - 10})$ & $\rm \alpha_{OX}$ & $\rm \Delta (\alpha_{OX})$ & $\rm \Delta t_{rest}$ \\
    (1) & (2) & (3) & (4) & (5) & (6) & (7) & (8) & (9) \\
    \midrule
    \multicolumn{8}{c}{\textbf{WISSH70 $-$ J1521+5202}} \\[0.75ex]
    6808 & 1.8\tablefootmark{f} & 5\tablefootmark{f} & $\rm 44.16 \pm 0.23$ & $\rm -14.78 \pm 0.23$ & $\rm -14.34 \pm 0.23$ & $\rm -2.41 \pm 0.09$ & $\rm -0.59$ & \multirow{2}{*}{825} \\ [0.75ex]
    \textbf{15334} & $\rm 1.58_{-0.37}^{+0.38}$ & $\rm 11.4_{-4.4}^{+5.2}$ & $\rm 44.91 \pm 0.06$ & $\rm -14.26 \pm 0.06$ & $\rm -13.50 \pm 0.10$ & $\rm -2.16 \pm 0.06$ & $\rm -0.33$ & \multirow{2}{*}{654} \\ [0.75ex]
    0840440101/ & \multirow{2}{*}{1.8\tablefootmark{f}} & \multirow{2}{*}{$\rm 152.4_{-63.3}^{+115.3}$} & \multirow{2}{*}{$\rm 43.56 \pm 0.17$} & \multirow{2}{*}{$\rm -15.28 \pm 0.14$} & \multirow{2}{*}{$\rm -14.07 \pm 0.14$} & \multirow{2}{*}{$\rm -2.64 \pm 0.07$} & \multirow{2}{*}{$\rm -0.82$} & \\ [0.75ex]
    0840440201 & & & & & & & & \\ 
    \midrule
    \multicolumn{8}{c}{\textbf{WISSH82 $-$ J1701+6412}} \\[0.75ex]
    9756 & $\rm 1.83 \pm 0.12$ & $\rm \le 3.1$ & $\rm 45.65 \pm 0.03$ & $\rm -13.25 \pm 0.04$ & $\rm -13.07 \pm 0.07$ & $\rm -1.84 \pm 0.02$ & $\rm -0.01$ & \multirow{2}{*}{536}\\ [0.75ex]
    \textbf{0723700101} & $\rm 2.20 \pm 0.05$ & $\rm 0.8 \pm 0.2$ & $\rm 46.13 \pm 0.01$ & $\rm -12.75 \pm 0.01$ & $\rm -12.77 \pm 0.03$ & $\rm -1.61 \pm 0.01$ & $\rm 0.22$ & \multirow{2}{*}{14} \\ [0.75ex]
    0723700201 & $\rm 2.21 \pm 0.06$ & $\rm 0.6 \pm 0.2$ & $\rm 45.98 \pm 0.01$ & $\rm -12.90 \pm 0.01$ & $\rm -12.94 \pm 0.03$ & $\rm -1.66 \pm 0.01$ & $\rm 0.17$ & \multirow{2}{*}{7} \\ [0.75ex]
    0723700301 & $\rm 2.20_{-0.15}^{+0.16}$ & $\rm 0.8 \pm 0.2$ & $\rm 45.95 \pm 0.01$ & $\rm -12.93 \pm 0.01$ & $\rm -12.95 \pm 0.03$ & $\rm -1.67 \pm 0.01$ & $\rm 0.15$ & \\
    \midrule
    \multicolumn{8}{c}{\textbf{WISSH83 $-$ J2123-0050}} \\ [0.75ex]
    6822 & $\rm 2.34_{-0.45}^{+0.47}$ & $\rm \le 1.5$ & $\rm 45.05 \pm 0.10$ & $\rm -13.60 \pm 0.10$ & $\rm -13.75 \pm 0.28$ & $\rm -1.89 \pm 0.07$ & $\rm -0.11$ & \multirow{2}{*}{960}\\ [0.75ex]
    0745010401 & $\rm 2.27 \pm 0.14$ & $\rm 0.7 \pm 0.3$ & $\rm 45.34 \pm 0.02$ & $\rm -13.34 \pm 0.02$ & $\rm -13.39 \pm 0.06$ & $\rm -1.79 \pm 0.02$ & $\rm -0.004$ & \multirow{2}{*}{123}\\ [0.75ex]
    \textbf{17080} & $\rm 1.97 \pm 0.07$ & $\rm \le 0.7$ & $\rm 45.69 \pm 0.02$ & $\rm -13.01 \pm 0.02$ & $\rm -12.92 \pm 0.03$ & $\rm -1.69 \pm 0.01$ & $\rm 0.09$ &  \\ 
    \bottomrule
  \end{tabular}
  \tablefoot{Columns: (1) Observation ID. The ID in boldface refers to the dataset from which the reference results for the tables and figures in the study were derived; (2) X-ray photon index; (3) Intrinsic column density (in units of $\rm 10^{22}\,\,cm^{-2}$); (4) Intrinsic $\rm 2 - 10\,\,keV$ luminosity (in units of $\rm Log (L_{2-10}/erg\,s^{-1})$); (5) $\rm 0.5 - 2\,\,keV$ flux (in units of $\rm Log (F_{0.5-2}/erg\,s^{-1}\,cm^{-2})$); (6) $\rm 2 - 10\,\,keV$ flux (in units of $\rm Log (F_{2-10}/erg\,s^{-1}\,cm^{-2})$); (7) X-ray$-$to$-$optical index; (8) $\rm \Delta(\alpha_{OX}) = \alpha_{OX} - \alpha_{OX, L10}$; (9) Rest-frame time interval (in units of days).\\
  \tablefoottext{f}{Fixed.}}
\end{table}

\clearpage
\section{Corner plots for the best fit parameters}

\begin{figure}[h!]
    \centering
    \subfloat[][$\rm \Delta(\alpha_{OX}) = \textit{m}\, \Delta_{6\,\mu m,X} + \textit{q}$\label{fig:deltaMIR_deltaAOX_corner}]{\includegraphics[width=0.32\textwidth]{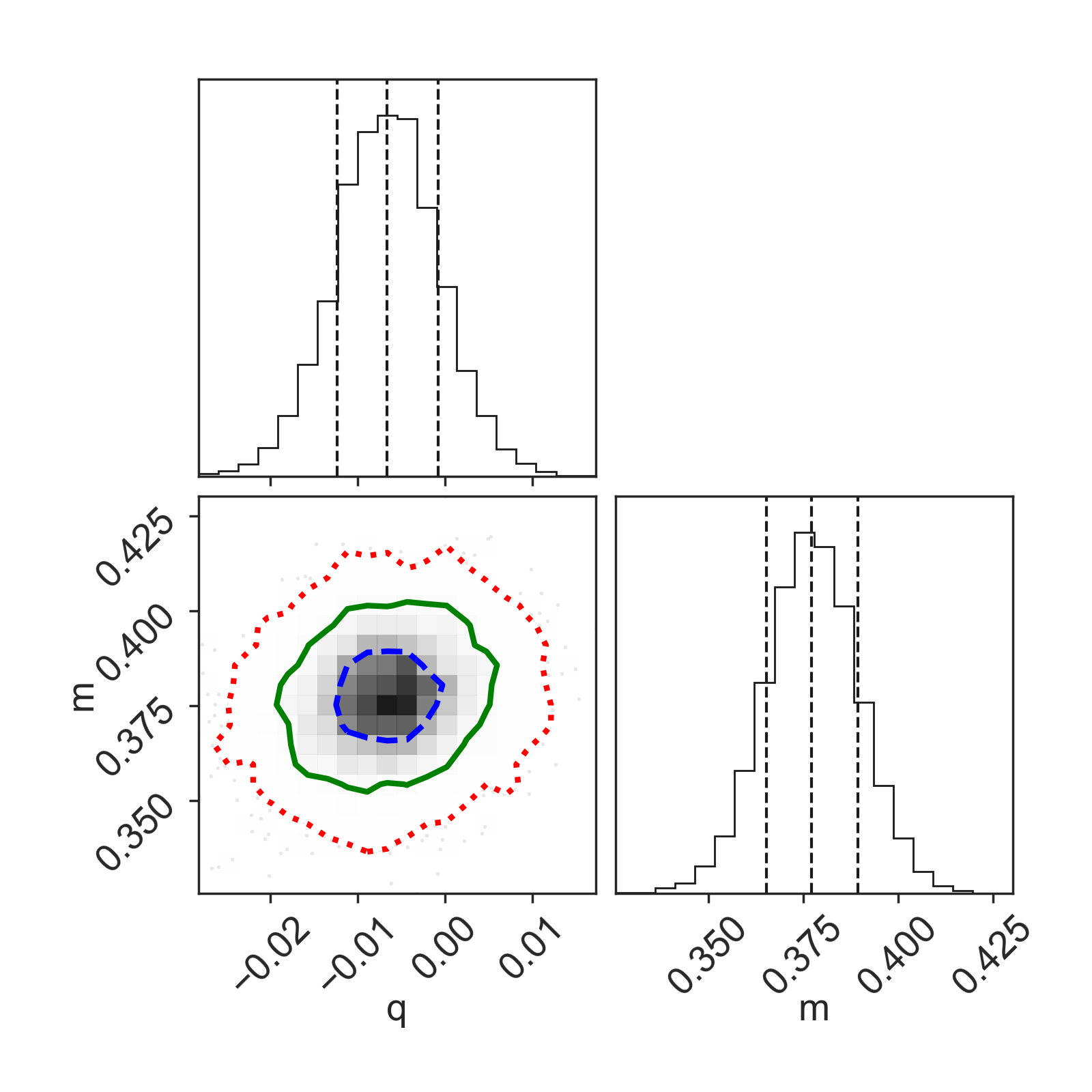}} \quad
    \subfloat[][$\rm Log(L_{2-10}/erg\,s^{-1}) = \textit{m}\, (v_{CIV}/km\,s^{-1}) + \textit{q}$\label{fig:vCIV_L210_corner}]{\includegraphics[width=0.32\textwidth]{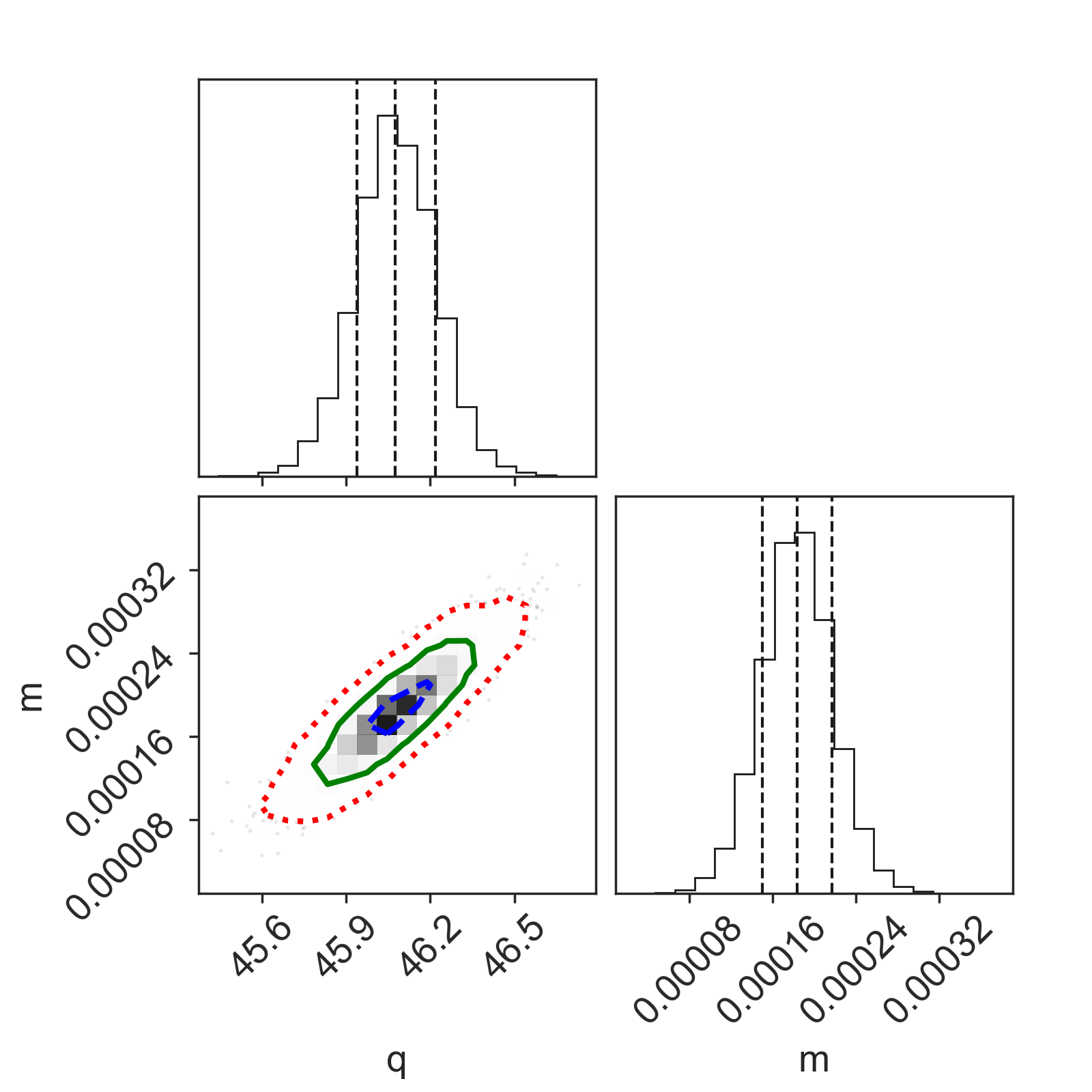}} \quad
    \subfloat[][$\rm \Delta(\alpha_{OX}) = \textit{m}\, (v_{CIV}/km\,s^{-1}) + \textit{q}$\label{fig:vCIV_deltaAOX_corner}]{\includegraphics[width=0.32\textwidth]{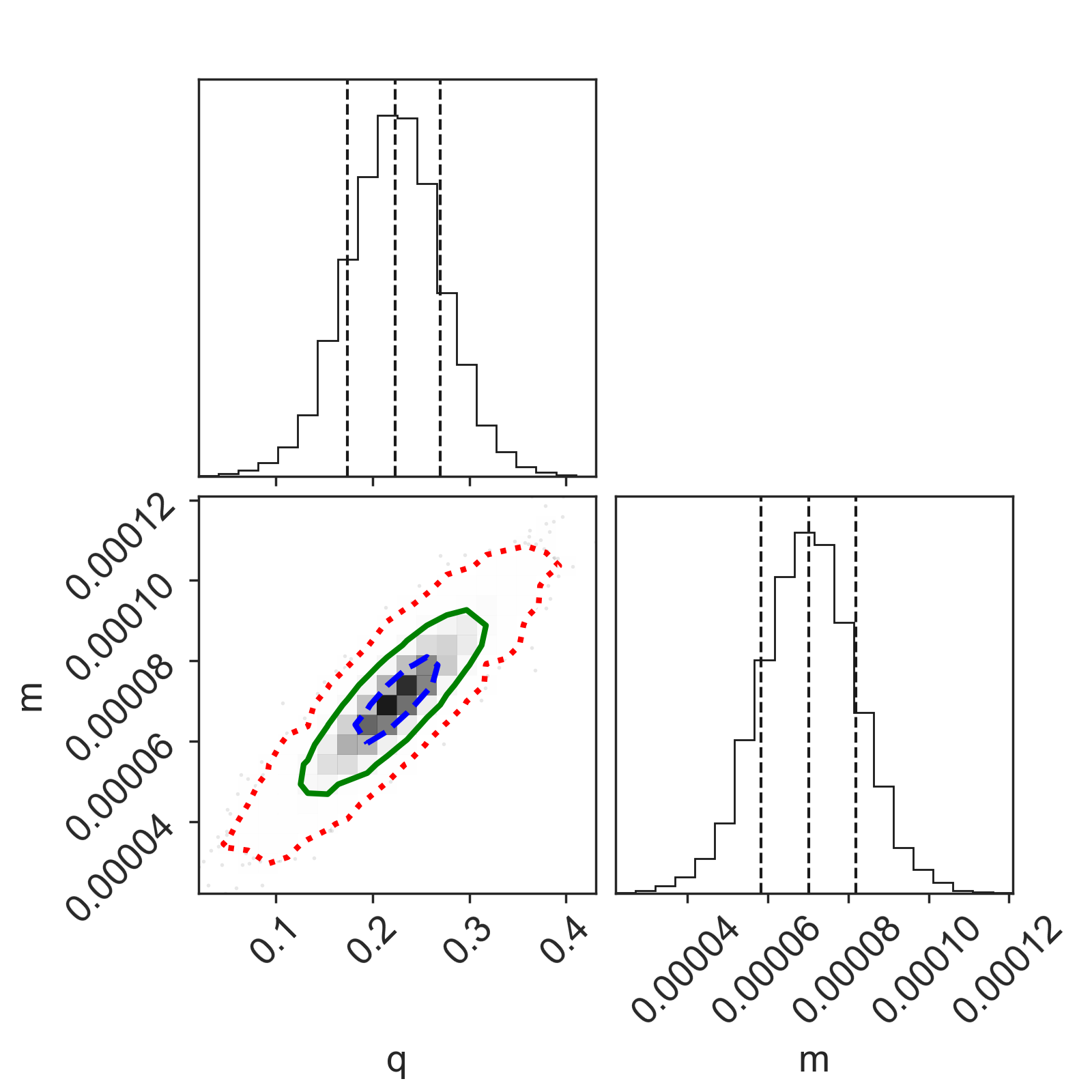}}
    \caption{Corner plot of the best fit parameters in Sections \ref{subsec:MIR} and \ref{subsec:CIV_shift}. The correlation parameters are reported in Table \ref{tab:relations_parameters}. The blue, green, and red contours show the 1$\rm \sigma$, 2$\rm \sigma$, and 3$\rm \sigma$ confidence level, respectively.}
    \label{fig:corner_plots}
\end{figure}

\begin{table}[h!]
    \caption{Correlation parameters of the best fit relations in Sections \ref{subsec:MIR} and \ref{subsec:CIV_shift}.}
    \label{tab:relations_parameters}
    \centering
    \begin{tabular}{cccccc}
    \toprule
    & Relation & p\tablefootmark{a} & $\rm r_P$\tablefootmark{b} & m\tablefootmark{c} & q\tablefootmark{c} \\
    \midrule
    (a) & $\rm \Delta_{6\,\mu m, X} - \Delta(\alpha_{OX})$ & $\rm 3 \times10^{-50}$ & 0.97 & $\rm 0.38\pm 0.01$ & $\rm -0.01 \pm 0.01$ \\ [0.75ex]
    (b) & $\rm v_{CIV}/km\,s^{-1} - Log(L_{2-10}/erg\,s^{-1})$ & $\rm 5 \times 10^{-6}$ & 0.77 & $\rm (1.84 \pm 0.34)\times 10^{-4}$ & $\rm 46.08 \pm 0.14$ \\ [0.75ex]
    (c) & $\rm v_{CIV}/km\,s^{-1} - \Delta(\alpha_{OX})$ & $\rm 4 \times 10^{-6}$ & 0.82 & $\rm (7.01 \pm 1.19)\times 10^{-5}$ & $\rm 0.22 \pm 0.05$ \\
    \bottomrule
    \end{tabular}
    \tablefoot{Figure \ref{fig:corner_plots} shows the corresponding corner plots.\\
    \tablefoottext{a}{Null-hypothesis probability.}
    \tablefoottext{b}{Pearson correlation coefficient.}
    \tablefoottext{c}{Correlation parameters ($ y = \text{m}x + \text{q}$) derived using \texttt{linmix} \citep{Kelly_2007}.}}
\end{table}

\newpage
\section{Summary of the results}

\begin{longtable}[t]{ccccc}
\caption{$\rm \lambda L_{6\,\mu m}$, $\rm k_{bol}$, $\rm \alpha_{OX}$ and $\rm \Delta(\alpha_{OX})$ for each QSO in the WISSH sample.}
\label{tab:results} \\

\toprule
WISSH ID & $\rm Log(\lambda L_{6\,\mu m}/erg\,s^{-1})$ & $\rm k_{bol}$ & $\rm \alpha_{OX}$ & $\rm \Delta(\alpha_{OX})$ \\
(1) & (2) & (3) & (4) & (5) \\
\midrule
\endfirsthead

\caption{continued}\\
\midrule
WISSH ID & $\rm Log(\lambda L_{6\,\mu m}/erg\,s^{-1})$ & $\rm k_{bol}$ & $\rm \alpha_{OX}$ & $\rm \Delta(\alpha_{OX})$ \\
(1) & (2) & (3) & (4) & (5) \\
\midrule
\endhead

\midrule
\endfoot
\bottomrule
\endlastfoot
WISSH01 & $\rm 47.00 \pm 0.01 $ & $\rm 1380 \pm 541 $ & $\rm -2.14 \pm 0.04 $ & $\rm -0.39 $ \\
WISSH02 & $\rm 46.96 \pm 0.05 $ & $\rm 135  \pm 42  $ & $\rm -1.72 \pm 0.05 $ & $\rm 0.02  $ \\
WISSH03 & $\rm 46.99 \pm 0.03 $ & $\rm 79   \pm 25  $ & $\rm -1.63 \pm 0.04 $ & $\rm 0.1   $ \\
WISSH04 & $\rm 47.09 \pm 0.01 $ & $\rm 295  \pm 49  $ & $\rm -1.93 \pm 0.03 $ & $\rm -0.16 $ \\
WISSH05 & $\rm 47.02 \pm 0.03 $ & $\rm 295  \pm 142 $ & $\rm -1.87 \pm 0.08 $ & $\rm -0.12 $ \\
WISSH06 & $\rm 47.02 \pm 0.01 $ & $\rm 1413 \pm 1002$ & $\rm -2.15 \pm 0.12 $ & $\rm -0.39 $ \\
WISSH07 & $\rm 47.04 \pm 0.01 $ & $\rm 339  \pm 45  $ & $\rm -1.94 \pm 0.02 $ & $\rm -0.17 $ \\
WISSH08 & $\rm 47.18 \pm 0.02 $ & $\rm 38   \pm 14  $ & $\rm -1.56 \pm 0.03 $ & $\rm 0.27  $ \\
WISSH09 & $\rm 47.04 \pm 0.06 $ & $\rm 257  \pm 43  $ & $\rm -1.89 \pm 0.04 $ & $\rm -0.14 $ \\
WISSH10 & $\rm 47.23 \pm 0.02 $ & $\rm 316  \pm 47  $ & $\rm -1.89 \pm 0.03 $ & $\rm -0.1  $ \\
WISSH11 & $\rm 46.99 \pm 0.05 $ & $\rm 269  \pm 129 $ & $\rm -1.87 \pm 0.08 $ & $\rm -0.12 $ \\
WISSH12 & $\rm 47.04 \pm 0.12 $ & $\rm > 331 $ & $\rm < -1.93 $ & $\rm < -0.21 $ \\
WISSH13 & $\rm 47.13 \pm 0.02 $ & $\rm 45   \pm 8  	$ & $\rm -1.55 \pm 0.01 $ & $\rm 0.25 $ \\
WISSH14 & $\rm 47.37 \pm 0.01 $ & $\rm 93   \pm 18 	$ & $\rm -1.63 \pm 0.01 $ & $\rm 0.16 $ \\
WISSH15 & $\rm 47.01 \pm 0.07 $ & $\rm > 295 $ & $\rm < -1.94 $ & $\rm < -0.17$ \\
WISSH16 & $\rm 46.98 \pm 0.04 $ & $\rm 295  \pm 112 $ & $\rm -1.88 \pm 0.06 $ & $\rm -0.12 $ \\
WISSH17 & $\rm 47.07 \pm 0.02 $ & $\rm 120  \pm 15  $ & $\rm -1.71 \pm 0.02 $ & $\rm 0.07  $ \\
WISSH18 & $\rm 47.01 \pm 0.01 $ & $\rm 32   \pm 6   $ & $\rm -1.54 \pm 0.02 $ & $\rm 0.23  $ \\
WISSH19 & $\rm 47.21 \pm 0.02 $ & $\rm 912  \pm 256 $ & $\rm -2.07 \pm 0.03 $ & $\rm -0.31 $ \\
WISSH20 & $\rm 47.05 \pm 0.06 $ & $\rm 65   \pm 33  $ & $\rm -1.50  \pm 0.10  $ & $\rm 0.28  $ \\
WISSH21 & $\rm 47.08 \pm 0.03 $ & $\rm 257  \pm 98  $ & $\rm -1.83 \pm 0.06 $ & $\rm -0.08 $ \\
WISSH22 & $\rm 46.99 \pm 0.02 $ & $\rm 219  \pm 66  $ & $\rm -1.83 \pm 0.01 $ & $\rm -0.02 $ \\
WISSH23 & $\rm 46.84 \pm 0.21 $ & $\rm 59   \pm 17  $ & $\rm -1.59 \pm 0.06 $ & $\rm 0.12  $ \\
WISSH24 & $\rm 46.75 \pm 0.08 $ & $\rm 240  \pm 135 $ & $\rm -1.87 \pm 0.09 $ & $\rm -0.15 $ \\
WISSH25 & $\rm 46.97 \pm 0.01 $ & $\rm > 2512 $ & $\rm < -2.21 $ & $\rm < -0.51 $ \\
WISSH26 & $\rm 46.88 \pm 0.09 $ & $\rm > 324  $ & $\rm < -1.86 $ & $\rm < -0.15 $ \\
WISSH27 & $\rm 47.41 \pm 0.01 $ & $\rm 158  \pm 27  $ & $\rm -1.73 \pm 0.01 $ & $\rm 0.08  $ \\
WISSH28 & $\rm 47.14 \pm 0.06 $ & $\rm 676  \pm 355 $ & $\rm -2.03 \pm 0.08 $ & $\rm -0.24 $ \\
WISSH29 & $\rm 47.03 \pm 0.08 $ & $\rm 372  \pm 169 $ & $\rm -1.91 \pm 0.08 $ & $\rm -0.16 $ \\
WISSH30 & $\rm 47.13 \pm 0.05 $ & $\rm 138  \pm 36  $ & $\rm -1.73 \pm 0.05 $ & $\rm 0.08  $ \\
WISSH31 & $\rm 46.96 \pm 0.04 $ & $\rm > 513 $ & $\rm < -1.97 $ & $\rm < -0.22 $ \\
WISSH32 & $\rm 47.10 \pm 0.01 $ & $\rm 427  \pm 230 $ & $\rm -1.92 \pm 0.07 $ & $\rm -0.2  $ \\
WISSH33 & $\rm 47.06 \pm 0.01 $ & $\rm 95   \pm 30  $ & $\rm -1.69 \pm 0.03 $ & $\rm 0.1   $ \\
WISSH34 & $\rm 47.20 \pm 0.01 $ & $\rm 407  \pm 387 $ & $\rm -1.92 \pm 0.04 $ & $\rm -0.14 $ \\
WISSH35 & $\rm 47.32 \pm 0.01 $ & $\rm 269  \pm 39  $ & $\rm -1.83 \pm 0.01 $ & $\rm -0.04 $ \\
WISSH36 & $\rm 47.26 \pm 0.03 $ & $\rm 40   \pm 17  $ & $\rm -1.55 \pm 0.05 $ & $\rm 0.27  $ \\
WISSH37 & $\rm 47.11 \pm 0.03 $ & $\rm 214  \pm 32  $ & $\rm -1.84 \pm 0.03 $ & $\rm -0.06 $ \\
WISSH38 & $\rm 47.21 \pm 0.01 $ & $\rm 468  \pm 257 $ & $\rm -1.92 \pm 0.08 $ & $\rm -0.17 $ \\
WISSH39 & $\rm 47.28 \pm 0.01 $ & $\rm 52   \pm 20  $ & $\rm -1.59 \pm 0.04 $ & $\rm 0.17  $ \\
WISSH40 & $\rm 46.98 \pm 0.01 $ & $\rm 490  \pm 120 $ & $\rm -1.99 \pm 0.05 $ & $\rm -0.23 $ \\
WISSH41 & $\rm 47.11 \pm 0.05 $ & $\rm 457  \pm 260 $ & $\rm -1.98 \pm 0.09 $ & $\rm -0.19 $ \\
WISSH42 & $\rm 47.09 \pm 0.02 $ & $\rm 174  \pm 70  $ & $\rm -1.82 \pm 0.03 $ & $\rm -0.02 $ \\
WISSH43 & $\rm 47.13 \pm 0.03 $ & $\rm 692  \pm 137 $ & $\rm -1.98 \pm 0.05 $ & $\rm -0.21 $ \\
WISSH44 & $\rm 47.06 \pm 0.03 $ & $\rm 129  \pm 40  $ & $\rm -1.74 \pm 0.04 $ & $\rm 0.04  $ \\
WISSH45 & $\rm 47.15 \pm 0.03 $ & $\rm > 891 $ & $\rm < -2.07 $ & $\rm < -0.28 $ \\
WISSH46 & $\rm 47.19 \pm 0.03 $ & $\rm > 912 $ & $\rm < -2.13 $ & $\rm < -0.31 $ \\
WISSH47 & $\rm 47.12 \pm 0.02 $ & $\rm 219  \pm 81  $ & $\rm -1.86 \pm 0.07 $ & $\rm -0.1  $ \\
WISSH48 & $\rm 46.99 \pm 0.01 $ & $\rm 126  \pm 45  $ & $\rm -1.72 \pm 0.09 $ & $\rm 0.05  $ \\
WISSH49 & $\rm 47.18 \pm 0.01 $ & $\rm 182  \pm 45  $ & $\rm -1.73 \pm 0.06 $ & $\rm 0.06  $ \\
WISSH50 & $\rm 47.12 \pm 0.02 $ & $\rm 148  \pm 27  $ & $\rm -1.69 \pm 0.03 $ & $\rm 0.09  $ \\
WISSH51 & $\rm 46.87 \pm 0.02 $ & $\rm 151  \pm 47  $ & $\rm -1.74 \pm 0.04 $ & $\rm -0.02 $ \\
WISSH52 & $\rm 47.01 \pm 0.02 $ & $\rm 316  \pm 146 $ & $\rm -1.87 \pm 0.05 $ & $\rm -0.14 $ \\
WISSH53 & $\rm 47.11 \pm 0.03 $ & $\rm 380  \pm 219 $ & $\rm -1.92 \pm 0.09 $ & $\rm -0.16 $ \\
WISSH54 & $\rm 47.13 \pm 0.01 $ & $\rm 78   \pm 36  $ & $\rm -1.65 \pm 0.01 $ & $\rm 0.16  $ \\
WISSH55 & $\rm 47.00 \pm 0.03 $ & $\rm 562  \pm 418 $ & $\rm -1.95 \pm 0.12 $ & $\rm -0.22 $ \\
WISSH56 & $\rm 47.26 \pm 0.03 $ & $\rm 145  \pm 58  $ & $\rm -1.77 \pm 0.04 $ & $\rm -0.02 $ \\
WISSH57 & $\rm 46.98 \pm 0.01 $ & $\rm 87   \pm 26  $ & $\rm -1.63 \pm 0.06 $ & $\rm 0.13  $ \\
WISSH58 & $\rm 47.05 \pm 0.02 $ & $\rm 214  \pm 100 $ & $\rm -1.83 \pm 0.02 $ & $\rm -0.05 $ \\
WISSH59 & $\rm 47.08 \pm 0.02 $ & $\rm 129  \pm 63  $ & $\rm -1.72 \pm 0.03 $ & $\rm -0.0  $ \\
WISSH60 & $\rm 46.96 \pm 0.01 $ & $\rm 69   \pm 20  $ & $\rm -1.73 \pm 0.02 $ & $\rm 0.06  $ \\
WISSH61 & $\rm 47.12 \pm 0.02 $ & $\rm 537  \pm 117 $ & $\rm -2.05 \pm 0.05 $ & $\rm -0.28 $ \\
WISSH62 & $\rm 47.38 \pm 0.02 $ & $\rm 1000 \pm 249 $ & $\rm -2.08 \pm 0.07 $ & $\rm -0.25 $ \\
WISSH63 & $\rm 47.44 \pm 0.01 $ & $\rm 138  \pm 20  $ & $\rm -1.73 \pm 0.01 $ & $\rm 0.1   $ \\
WISSH64 & $\rm 47.19 \pm 0.06 $ & $\rm 389  \pm 190 $ & $\rm -1.97 \pm 0.08 $ & $\rm -0.2  $ \\
WISSH65 & $\rm 46.83 \pm 0.01 $ & $\rm 389  \pm 100 $ & $\rm -1.91 \pm 0.06 $ & $\rm -0.19 $ \\
WISSH66 & $\rm 47.00 \pm 0.03 $ & $\rm 316  \pm 157 $ & $\rm -1.91 \pm 0.08 $ & $\rm -0.15 $ \\
WISSH67 & $\rm 47.17 \pm 0.02 $ & $\rm 501  \pm 269 $ & $\rm -1.94 \pm 0.08 $ & $\rm -0.18 $ \\
WISSH68 & $\rm 47.15 \pm 0.03 $ & $\rm 1096 \pm 621 $ & $\rm -2.07 \pm 0.09 $ & $\rm -0.31 $ \\
WISSH69 & $\rm 47.39 \pm 0.01 $ & $\rm 132  \pm 35  $ & $\rm -1.71 \pm 0.04 $ & $\rm 0.09  $ \\
WISSH70 & $\rm 47.14 \pm 0.01 $ & $\rm 871  \pm 217 $ & $\rm -2.16 \pm 0.06 $ & $\rm -0.33 $ \\
WISSH71 & $\rm 47.08 \pm 0.03 $ & $\rm 933  \pm 517 $ & $\rm -2.08 \pm 0.09 $ & $\rm -0.28 $ \\
WISSH72 & $\rm 47.05 \pm 0.02 $ & $\rm 525  \pm 298 $ & $\rm -1.99 \pm 0.09 $ & $\rm -0.23 $ \\
WISSH73 & $\rm 47.04 \pm 0.01 $ & $\rm 347  \pm 105 $ & $\rm -1.86 \pm 0.02 $ & $\rm -0.07 $ \\
WISSH74 & $\rm 47.29 \pm 0.01 $ & $\rm > 1549 $ & $\rm < -2.12 $ & $\rm < -0.34 $ \\
WISSH75 & $\rm 47.00 \pm 0.03 $ & $\rm 513  \pm 391 $ & $\rm -2.03 \pm 0.12 $ & $\rm -0.27 $ \\
WISSH76 & $\rm 46.94 \pm 0.02 $ & $\rm 427  \pm 241 $ & $\rm -1.95 \pm 0.09 $ & $\rm -0.2  $ \\
WISSH77 & $\rm 47.08 \pm 0.05 $ & $\rm 676  \pm 488 $ & $\rm -2.03 \pm 0.12 $ & $\rm -0.26 $ \\
WISSH78 & $\rm 46.96 \pm 0.05 $ & $\rm 63   \pm 21  $ & $\rm -1.63 \pm 0.07 $ & $\rm 0.16  $ \\
WISSH79 & $\rm 47.28 \pm 0.02 $ & $\rm 372  \pm 145 $ & $\rm -1.91 \pm 0.06 $ & $\rm -0.12 $ \\
WISSH80 & $\rm 47.55 \pm 0.02 $ & $\rm 204  \pm 78  $ & $\rm -1.81 \pm 0.06 $ & $\rm 0.02  $ \\
WISSH81 & $\rm 47.17 \pm 0.02 $ & $\rm 661  \pm 410 $ & $\rm -1.99 \pm 0.09 $ & $\rm -0.23 $ \\
WISSH82 & $\rm 47.27 \pm 0.01 $ & $\rm 68   \pm 20  $ & $\rm -1.61 \pm 0.01 $ & $\rm 0.22  $ \\
WISSH83 & $\rm 47.01 \pm 0.01 $ & $\rm 100  \pm 15  $ & $\rm -1.69 \pm 0.01 $ & $\rm 0.09  $ \\
WISSH84 & $\rm 46.94 \pm 0.03 $ & $\rm 537  \pm 294 $ & $\rm -1.97 \pm 0.09 $ & $\rm -0.23 $ \\
WISSH85 & $\rm 46.91 \pm 0.04 $ & $\rm 282  \pm 134 $ & $\rm -1.83 \pm 0.08 $ & $\rm -0.1  $ \\
\end{longtable}
\tablefoot{Columns: (1) WISSH ID; (2) $\rm 6\,\mu m$ luminosity (in units of $\rm Log(\lambda L_{6\,\mu m}/erg\,s^{-1})$); (3) hard X-ray bolometric correction; (4) X-ray$-$to$-$optical index; (5) $\rm \Delta(\alpha_{OX}) = \alpha_{OX} - \alpha_{OX, L10}$.}

\end{appendix}

\end{document}